\documentclass[12pt]{article}
\pdfoutput=1

\usepackage{tabu}
\usepackage{geometry}

\usepackage{putex}
\usepackage{float}
\usepackage{caption}
\usepackage{psfrag}
\usepackage{amssymb}
\usepackage{amsmath}
\usepackage{empheq}
\usepackage{amsthm}
\usepackage{epsf}
\usepackage{graphicx}
\usepackage{epstopdf}

\usepackage{enumerate}
\usepackage{cite}
\usepackage{caption}
\usepackage{subcaption}

\usepackage[utf8]{inputenc}
\usepackage{braket}
\usepackage{titlesec}

\usepackage{psfrag}
\usepackage{color}
\definecolor{darkblue}{rgb}{0.1,0.1,.7}
\usepackage[colorlinks, linkcolor=darkblue, citecolor=darkblue, urlcolor=darkblue, linktocpage]{hyperref}

\hypersetup{breaklinks}
\usepackage{comment}

\def\1{{\rm 1-loop}}
\newcommand{\dDisc}{\text{dDisc}}
\newcommand{\qDisc}{\text{qDisc}}

\newcommand\numberthis{\addtocounter{equation}{1}\tag{\theequation}}
\usepackage{braket}

\def\t{\tilde}
\def\un{\underline}
\def\Cut{{\mathbf{Cut}}}
\def\hCut{\widehat\Cut}
\def\bx{{\rm box}}
\def\cA{\mathcal{A}}
\def\Dt{\widetilde{\Delta}}
\def\zb{\bar z}
\def\cG{\mathcal{G}}

\newcommand{\bc}{}
\def\<{\langle}
\def\>{\rangle}

\newcommand   \f  {\phi}

\newcommand{\bea}{\begin{eqnarray}}
\newcommand{\eea}{\end{eqnarray}}

\def\c{\cite}

\def\foot{\footnote}

\def\o{\over}

\def\eqr{\eqref}
\def\sec{\section}
\def\subsec{\subsection}

\def\rar{\rightarrow}

\def\la{\langle}
\def\ra{\rangle}
\def\O{{\cal O}}

\def\ssec{\subsection}
\def\sssec{\subsubsection}
\def\sec{\section}

\def\D{\Delta}

\def\g{\gamma}

\newcommand {\be} {\begin {equation}}
\newcommand {\ee} {\end {equation}}

\newcommand {\bes} {\begin {equation*}}
\newcommand {\ees} {\end {equation*}}

\newcommand{\es}[2] {\begin{equation} \label{#1} \begin{split} #2 \end{split} \end{equation}}
\newcommand{\e}[2] {\begin{equation} \label{#1} #2 \end{equation}}

\newcommand\sixj[6]{\ensuremath{\left\{\begin{array}{ccc} #1 & #2 & #6 \\ #3 & #4 & #5\end{array}\right\}}}
\newcommand\sixjBlock[6]{\ensuremath{\left(\begin{array}{ccc} #1 & #2 & #6 \\ #3 & #4 & #5\end{array}\right)}}

\DeclareMathOperator*{\res}{Res}

\newcommand{\beq}{\begin{equation}}
\newcommand{\eeq}{\end{equation}}

\def\be{ \begin{equation} }
\def\ee{ \end{equation} }

\def\subsec{\subsection}

 \newmuskip\pFqmuskip

\newcommand*\pFq[6][8]{%
  \begingroup 
  \pFqmuskip=#1mu\relax
  \mathcode`\,=\string"8000
  \begingroup\lccode`\~=`\,
  \lowercase{\endgroup\let~}\pFqcomma
  {}_{#2}F_{#3}{\left[\genfrac..{0pt}{}{#4}{#5};#6\right]}%
  \endgroup
}
\newcommand{\pFqcomma}{\mskip\pFqmuskip}

\makeatletter
\renewcommand{\@maketitle}{
\newpage
 \begin{center}%
  {\large\bfseries \@title \par}%
 \end{center}%
 \par} \makeatother

\renewcommand{\baselinestretch}{1.1}

\numberwithin{equation}{section}

\institution{CT}{Walter Burke Institute for Theoretical Physics, California Institute of Technology, \cr Pasadena, CA, 91125}
\institution{KY}{Department of Physics and Astronomy, University of Kentucky, Lexington, KY, 40506}

\begin{document}
\preprint{CALT-TH-2019-052}

\title{Unitarity Methods in AdS/CFT}

\authors{David Meltzer\worksat{\CT}
, Eric Perlmutter\worksat{\CT}
, Allic Sivaramakrishnan\worksat{\CT,\KY} 
\let\thefootnote\relax\footnote{\texttt{${}$dmeltzer@caltech.edu }}
\let\thefootnote\relax\footnote{\texttt{${}$perl@caltech.edu}}
\let\thefootnote\relax\footnote{\texttt{${}$allicsiva@uky.edu}}
 }
 
\abstract{We develop a systematic unitarity method for loop-level AdS scattering amplitudes, dual to non-planar CFT correlators, from both bulk and boundary perspectives. We identify cut operators acting on bulk amplitudes that put virtual lines on shell, and show how the conformal partial wave decomposition of the amplitudes may be efficiently computed by gluing lower-loop amplitudes. A central role is played by the double discontinuity of the amplitude, which has a direct relation to these cuts. We then exhibit a precise, intuitive map between the diagrammatic approach in the bulk using cutting and gluing, and the algebraic, holographic unitarity method of \c{Aharony:2016dwx} that constructs the non-planar correlator from planar CFT data. Our analysis focuses mostly on four-point, one-loop diagrams -- we compute cuts of the scalar bubble, triangle and box, as well as some one-particle reducible diagrams -- in addition to the five-point tree and four-point double-ladder. Analogies with S-matrix unitarity methods are drawn throughout. 

}

\date{ }

\maketitle
\setcounter{tocdepth}{3}

\renewcommand{\baselinestretch}{0.7}\normalsize

\tableofcontents
\renewcommand{\baselinestretch}{1.1}\normalsize

\section{Introduction}
Unitarity in quantum field theory (QFT) is most commonly stated as a condition on the $S$-matrix:
\e{}{\mathcal{S}^{\dagger} \mathcal{S}=1.}
Separating out the trivial part of the $S$-matrix, $\mathcal{S}=1+i \mathcal{T}$, this becomes the optical theorem,
\begin{align}
2\Im(\mathcal{T})=\mathcal{T}^{\dagger}\mathcal{T}.
\end{align} 
In perturbation theory, the imaginary part of the transition matrix $\mathcal{T}$ at $L$ loops is determined by the same matrix $\mathcal{T}$ at $L-1$ loops. The combination of the optical theorem with dispersion relations forms the backbone of unitarity methods for scattering amplitudes in QFT. 

In conformal field theories (CFTs), dual to theories of AdS quantum gravity via the  AdS/CFT correspondence \cite{Maldacena:1997re,Witten:1998qj,Gubser:1998bc}
, $\mathcal{S}$ is not well-defined. One may nevertheless attempt to define unitarity methods for AdS scattering amplitudes and conformal correlators. Though a theory in AdS lacks asymptotic states necessary to define a standard $S$-matrix, AdS amplitudes (with appropriate boundary conditions) compute dual CFT correlation functions of local operators. Therefore, one may expect that unitarity conditions for AdS can be formulated in terms of Witten diagrams or dual CFT data, and that the two descriptions will be simply related. 

This perspective formed the crux of the unitarity methods introduced in \cite{Aharony:2016dwx} which, building on elements of \cite{Heemskerk:2009pn, Penedones:2010ue,Fitzpatrick:2011dm}, proposed a new approach to perturbative AdS amplitudes and their dual non-planar CFT correlators. The idea was to compute AdS loop amplitudes holographically by developing the $1/N$ expansion of the CFT correlator. One main achievement of this work was to make precise the sense in which one can ``square'' tree-level OPE data -- i.e. anomalous dimensions and OPE coefficients -- to generate one-loop OPE data. Aside from providing conceptual insight, this approach bypasses the difficulty of computing directly in AdS, where results have been mostly restricted to tree level. 
The loop expansion in AdS has been further studied along these lines in  \cite{Alday:2017xua,Alday:2017vkk,Alday:2018pdi,Alday:2018kkw,Meltzer:2018tnm,Ponomarev:2019ltz,Shyani:2019wed,Alday:2019qrf,Alday:2019nin,Meltzer:2019pyl}. Other subsequent work on AdS loops from diverse perspectives includes \c{Aprile:2017bgs,Aprile:2017xsp,Aprile:2017qoy,Giombi:2017hpr,Cardona:2017tsw,Yuan:2017vgp,Yuan:2018qva,Bertan:2018afl,Bertan:2018khc,Liu:2018jhs,Carmi:2018qzm,Aprile:2018efk,Ghosh:2018bgd,Mazac:2018ycv,Beccaria:2019stp, Chester:2019pvm,Beccaria:2019dju,Carmi:2019ocp,Aprile:2019rep,Drummond:2019hel}.\foot{Early work on AdS loops can be found in \cite{Cornalba:2007zb,Penedones:2010ue,Fitzpatrick:2011hu,Fitzpatrick:2011dm}.}

The current paper aims to provide a more systematic approach to unitarity methods for correlators in AdS/CFT, cast in {\it both} bulk and boundary language. Stated concisely, the main results of this paper are the following:

\begin{itemize}

\item We develop an ``AdS unitarity method'' directly in the bulk, which uses bulk operations, such as the cutting and gluing of Witten diagrams, and makes conformal symmetry manifest. The method is best phrased in terms of the double discontinuity (dDisc) of the amplitudes. The dDisc is analogous to $\Im(\mathcal{T})$ in flat space and implements internal line cuts of loop-level bulk diagrams by putting virtual lines on shell. From here, the Lorentzian inversion formula \cite{Caron-Huot:2017vep} then reconstructs the full diagram.

\item We apply this method to a variety of one-loop diagrams, computing cuts of the scalar bubble, triangle, and box diagrams, along with studying one-particle reducible diagrams for mass and vertex corrections.

\item We give a simple and elegant dictionary between the {\it diagrammatic} bulk unitarity method, which involves cutting and gluing AdS graphs, and the {\it algebraic} holographic unitarity method of \cite{Aharony:2016dwx}, which involves summing over products of OPE data. Using CFT tools \cite{Liu:2018jhs}, we show that the two methods are equivalent. In doing so we recast the method of \cite{Aharony:2016dwx} in the language of the dDisc. 
 
\item We begin an extension to higher loops by analyzing the AdS double-ladder diagram. The diagram admits several distinct cuts, one of which involves five-point tree diagrams whose crossing properties we also study. The methods developed at one loop persist to higher loops and cleanly identify the multi-trace states.

\end{itemize}

\noindent The next subsection is an extended, self-contained conceptual summary of our work, followed by an outline of the remainder of the paper.

\subsec{Summary of AdS/CFT Unitarity}
\label{sec:UnitarityWittenDiagrams}

An overarching point of this paper is that one should not compute AdS loop amplitudes by brute force. However, to set the stage, suppose one does. An efficient algorithm for this purpose that manifests the simplifying role of conformal symmetry was developed in \cite{Liu:2018jhs}. First, by manipulating AdS bulk-to-bulk propagators, one can recast Witten diagrams as spectral integrals over conformally-invariant gluings of CFT three-point functions. Second, one employs techniques from harmonic analysis to perform crossing transformations \cite{Gadde:2017sjg,Caron-Huot:2017vep} and position-space integrals \cite{Kravchuk:2018htv,Karateev:2018oml, SimmonsDuffin:2012uy} of these conformal structures. In this way, any $n$-point, loop-level diagram can be expressed in a conformal partial wave (CPW) expansion. The OPE data is then determined by a set of non-trivial spectral integrals.

While a significant improvement over direct bulk integration, this method has certain limitations. Although it makes bulk integrations trivial, we are still left with spectral integrals which are difficult to compute. We also know from the existence of the Lorentzian inversion formula in the CFT, and anticipate based on S-matrix techniques in QFT, that we should be able to reconstruct the diagram from a simpler, more minimal object. 

In this work, we pursue an alternative approach using unitarity. Let us start with a general Witten diagram $\cA^{1234}(x_i) \equiv \cA(x_1,x_2,x_3,x_4)$, drawn below, with four external points and two sub-diagrams $\mathcal{A}_{L,R}$ connected by two propagators.\foot{To make it manifest which operators are being isolated and how external cuts differ from internal cuts, we will choose all internal and external lines to correspond to different operators/fields. For convenience we label fields in AdS by their dual operator in the CFT.} 

\begin{figure}[h!]
\centering
\includegraphics[scale=.3]{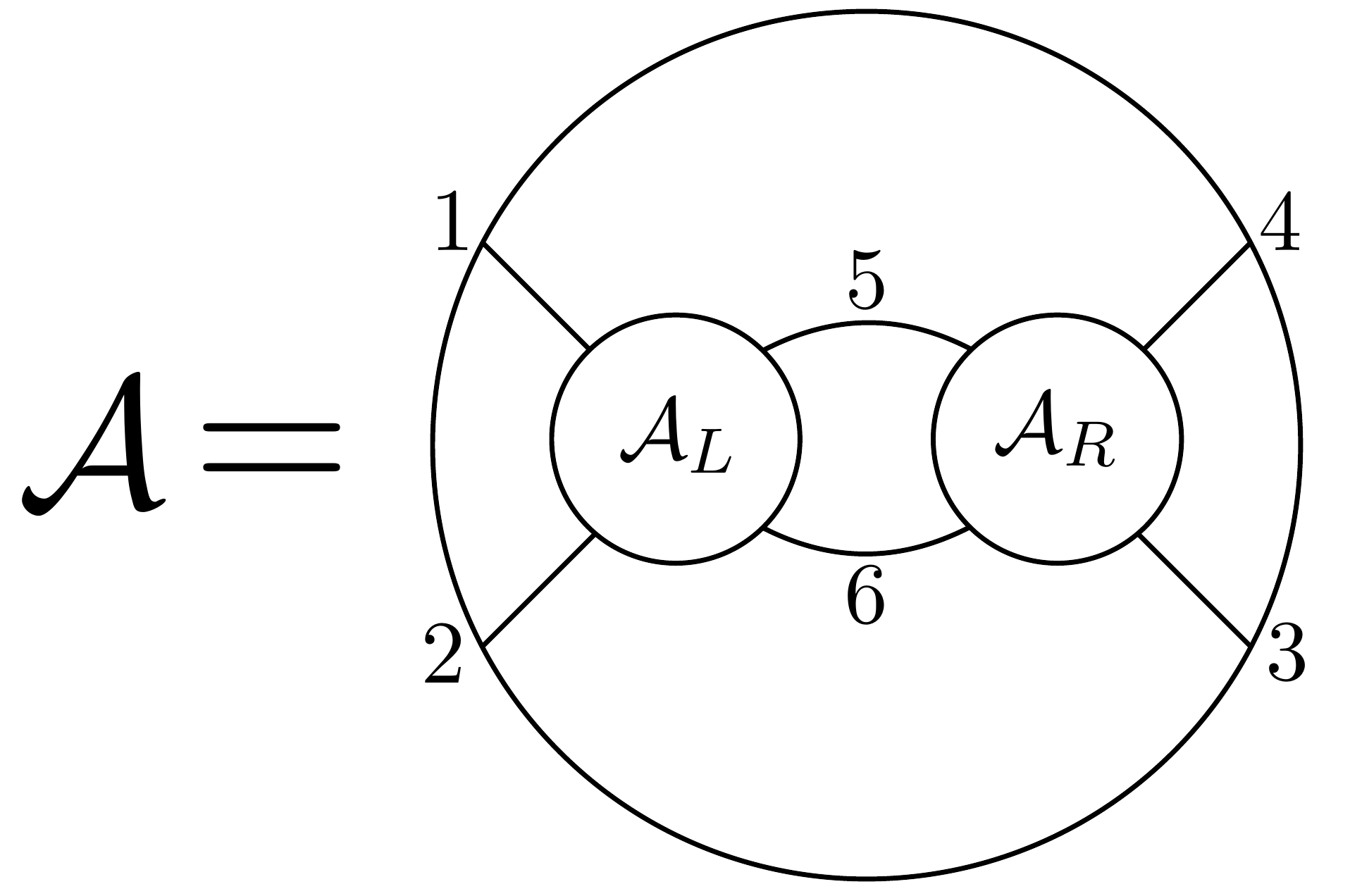}
\label{fig:Two_Prop_Witten}
\end{figure}

\noindent One expects that a useful AdS unitarity method should make the following explicit: 

{\bf i)} How to implement a ``cut" of the two propagators.

{\bf ii)} How this cut factorizes $\cA$ into the lower-loop diagrams $\mathcal{A}_{L,R}$.

{\bf iii)} How this relates to operator exchanges in the CPW decomposition of $\cA$.

\noindent Let us draw a picture of three possible ``vertical line cuts'' of this Witten diagram: 
\begin{figure}[h!]
\centering
\includegraphics[scale=.3]{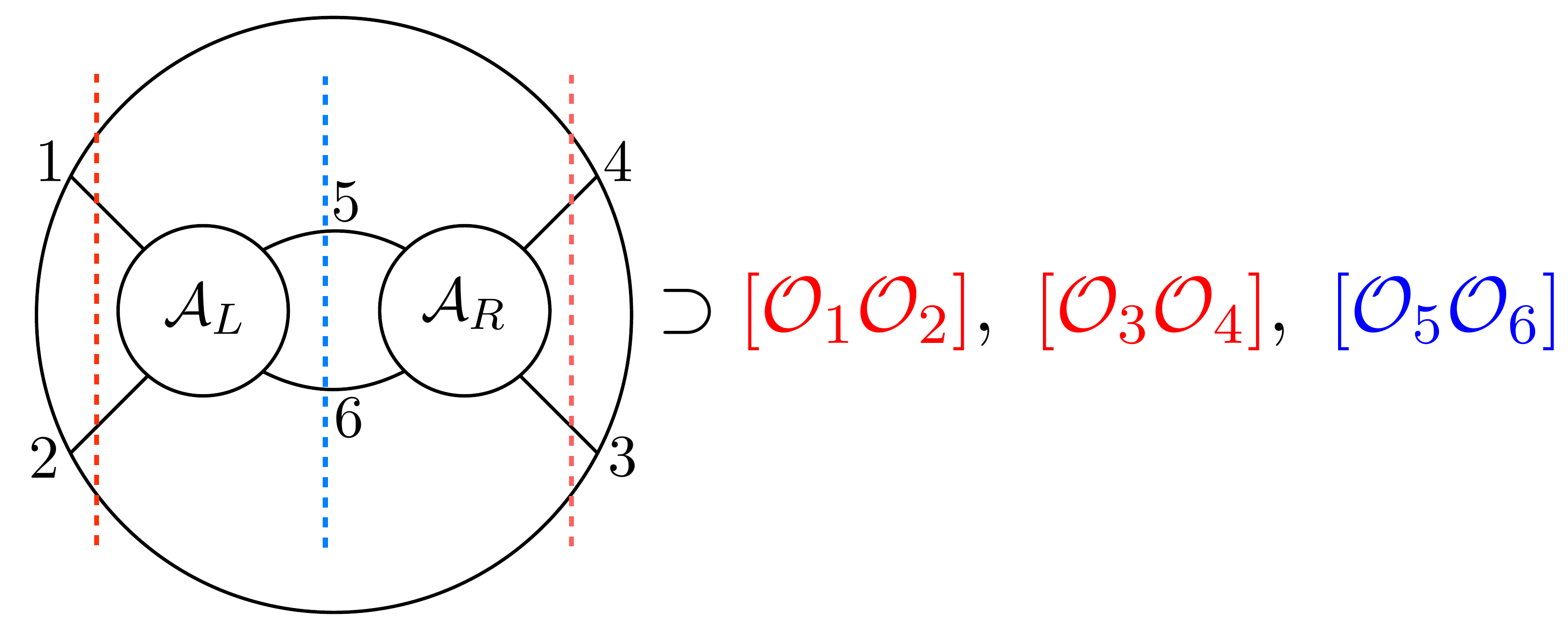}
\label{fig:Witten_diagram_Cut}
\end{figure}

\noindent Since the lines always cut through two propagators, AdS/CFT intuition suggests that those cuts should isolate the respective double-trace families in the $s$-channel CPW decomposition as shown above \cite{Fitzpatrick:2011dm}. In general, we may expect that vertical line cuts should tell us about operators exchanged in the $s$-channel, $\O_1\O_2\rightarrow\O_3\O_4$, while horizontal cuts should tell us about exchanges in the $t$-channel, $\O_3\O_2\rightarrow \O_1\O_4$. We realize these ideas concretely as follows.

To start, we use the split representation of the bulk-to-bulk propagator $G_{\Delta}(y_1,y_2)$, with bulk coordinates $y_i$ and boundary coordinates $x_i$:
\begin{align}\label{Gsplit}
G_{\Delta}(y_1,y_2)=\int\limits_{-\infty}^{\infty}{d\nu}\,P(\nu,\Delta)\int\limits_{\partial \rm AdS}d^{d}x K_{\frac{d}{2}+i\nu}(x,y_1)K_{\frac{d}{2}-i\nu}(x,y_2),
\end{align}
where $K_{\Delta}(x,y)$ is the bulk-to-boundary propagator for a CFT operator of dimension $\Delta$, and 
\begin{align}
P(\nu,\Delta)&\equiv\frac{1}{\nu^{2}+(\Delta-\frac{d}{2})^{2}}\frac{\nu^{2}}{\pi}. \label{eq:PDef}
\end{align}
The dimensions of $K_{\Delta}(x,y)$ in \eqref{Gsplit} sit on the Euclidean principal series, $\Delta={d\o 2}+i\nu$. The measure $P(\nu,\Delta)$ has poles at $\nu=\pm i(\D-{d\o 2})$. For reasons that will become clearer soon, we call the pole at $\nu=-i(\D-{d\o 2})$ the {\bf single-trace pole}: it lies at the value of $\nu$ corresponding to the physical, single-trace operator of dimension $\D$ propagating on the original line. This split is shown in the following figure, where we introduce the notation $\widetilde{\Delta}\equiv d-\Delta$, so that $\widetilde{\nu}=-\nu$ represents the ``shadow pole'' at $\nu=+i(\D-{d\o 2})$: 
\begin{figure}[H]
\centering
\includegraphics[scale=.32]{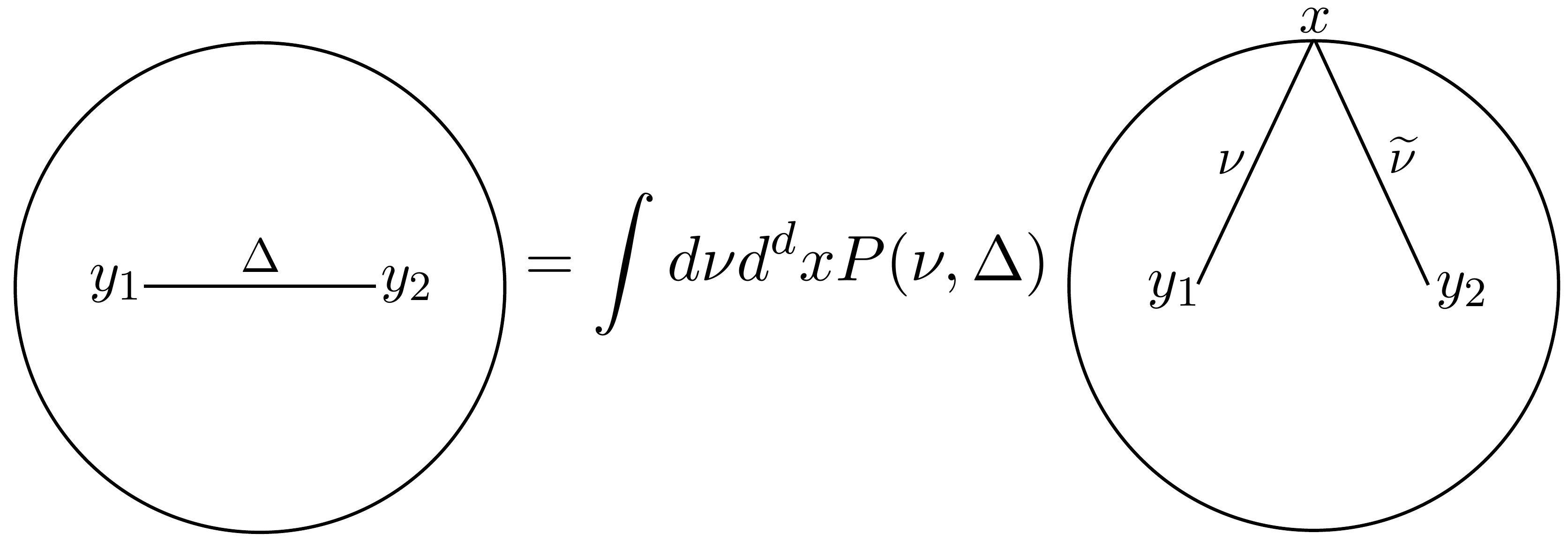}
\label{fig:Split_Rep_BB}
\end{figure}
\noindent If the bulk-to-boundary propagator sits on the principal series, we say it is \textbf{off shell}. On the other hand, if the bulk-to-boundary propagator has the scaling dimension of a physical single-trace operator, we say it is \textbf{on shell}. In the language above, {\it a propagator goes on shell when localized onto the single-trace pole of $P(\nu,\D)$}. This captures the essential distinction, central to the unitarity method, between the original diagram and its split representation. 

Since $\nu_\O$ (an off-shell quantity) and $\Delta_\O$ (an on-shell quantity) will often appear in the same expression, to keep the presentation unambiguous we define:
\begin{align}
\underline{\Delta}\equiv \frac{d}{2}+i\nu,
\end{align}
so that $\underline{\O}$ will denote an off-shell operator with dimension $\underline{\Delta}_{\O}$ and spin $J_{\O}$. This will be convenient for distinguishing the scaling dimensions of bulk-to-boundary propagators $K_{\frac{d}{2}+i\nu}(x,y)$, which come from splitting a bulk-to-bulk propagator $G_{\Delta_\O}(y_1,y_2)$, from propagators of the original diagram with physical scaling dimensions $\Delta_\O$. 

It will be convenient to define two notions of conformal gluing for four-point amplitudes.\foot{To use identities involving conformal integrals and, later, the 6j symbol, we must take all operators to lie on the principal series \cite{Gadde:2017sjg,Liu:2018jhs} and only analytically continue to real, physical dimensions at the end. We henceforth take this to be understood.} 

First, we define {\bf on-shell conformal gluing:}
\begin{align}\label{circdef}
\cA_L^{1265}\circ \cA_R^{\t5\t634}\equiv \int\limits_{\partial \rm AdS}d^{d}x_{5}d^{d}x_{6}\mathcal{A}_{L}^{1265}(x_1,x_2,x_6,x_5)\mathcal{A}_{R}^{\tilde5\tilde634}(x_5,x_6,x_3,x_4). 
\end{align}
This is a conformally-invariant gluing of two four-point amplitudes along their common boundary points that generates another four-point amplitude. 

Next, we define {\bf off-shell conformal gluing:}
\begin{align}\label{crossdef}
\cA_L^{12\un6\un5}\otimes \cA_R^{\underline{\tilde5}\underline{\tilde6}34}\equiv \int\limits_{-\infty}^{\infty}d\nu_{5}d\nu_{6}\,P(\nu_5,\Delta_5)P(\nu_6,\Delta_6)\mathcal{A}_{L}^{12\underline{6}\underline{5}}\circ \mathcal{A}_{R}^{\underline{\tilde5}\underline{\tilde6}34}.
\end{align}
This is also a conformally-invariant gluing, but it includes two spectral integrals weighted by the factors from the split representation. By construction, {\it the off-shell gluing defines a full Witten diagram} because $G_{\D}(y_1,y_2)$ splits onto off-shell propagators, cf. \eqr{Gsplit}.\footnote{The on-shell gluing is related to the ``pre-amplitude", defined as a Witten diagram with bulk-to-bulk propagators replaced by harmonic functions \cite{Liu:2018jhs,Yuan:2018qva}. The on-shell gluing defines a pre-amplitude in which we replace only the 5 and 6 propagators.}

\begin{figure}[t]
\centering
\includegraphics[scale=.44]{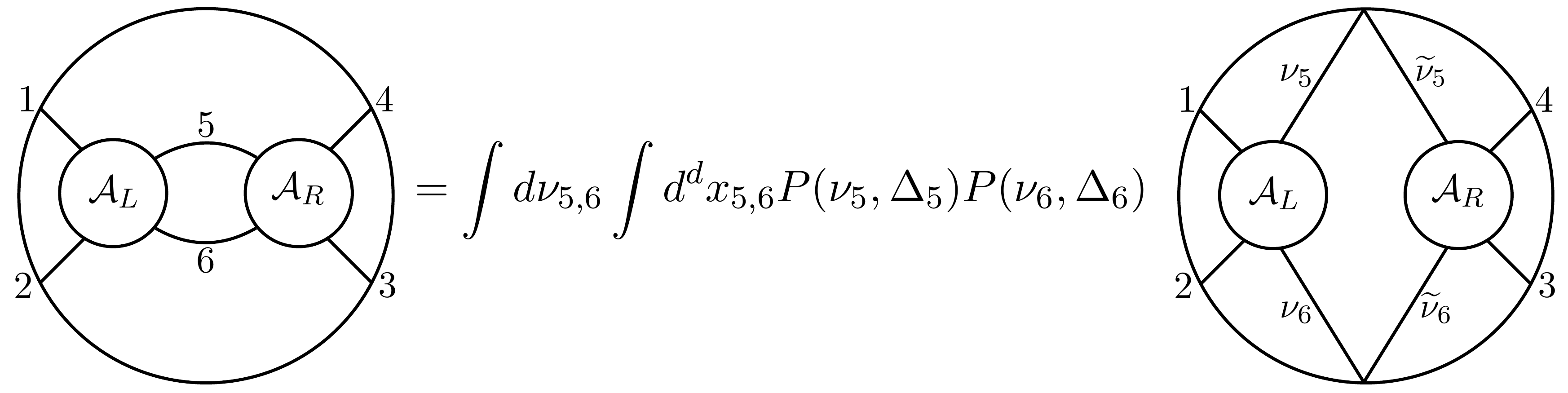}
\caption{General Witten diagram with an internal, double-trace cut.}
\label{fig:Two_Prop_Witten_Split}
\end{figure}

Returning to our amplitude $\cA$, using the split representation twice gives figure \ref{fig:Two_Prop_Witten_Split}, i.e.
\e{}{\cA^{1234}(x_i) = \cA_L^{12\un6\un5}\otimes \cA_R^{\underline{\tilde5}\underline{\tilde6}34}.}
This is an exact equality for the full diagram, and it is a non-trivial problem to perform the spectral integrals. To go further, we use unitarity.

\sssec*{\large Main results}
Let us define the following {\bf cut operator}:
\begin{align}\label{cutdef}
\Cut_{56}[\mathcal{A}^{1234}] \equiv  (2\D_5-d) (2\D_6-d)\,\cA_L^{1265}\circ \cA_R^{\t5\t634},
\end{align}
where we note that 
\e{}{2\D_i-d= \res\limits_{\nu_i=-i\left(\D_i-{d\o 2}\right)}2P(\nu_i,\D_i)~.}
That is, this cut operation places the virtual 5 and 6 lines on shell by picking up the single-trace poles of the measure. This vertical cut factorizes the diagram into position integrals of two on shell, lower-loop four-point functions. (One can define an $n$-fold cut operator $\Cut_{i_1 i_2\cdots i_n}$ which puts $n$ virtual lines on-shell in the same way.) To make contact with the CFT, it is convenient to further introduce
\e{eq:HatCut}{\widehat\Cut_{56}[\mathcal{A}^{1234}] \equiv  \Cut_{56}[\mathcal{A}^{1234}]\Big|_{[\O_5\O_6]_{n,\ell}},}
The symbol $|_{[\O_5\O_6]_{n,\ell}}$ means ``project onto the $[\O_5\O_6]_{n,\ell}$ family in the conformal block decomposition'',\footnote{In Mellin space \cite{Mack:2009mi,Penedones:2010ue} this corresponds to localizing onto certain poles of the Mellin amplitude.} where $[\O_5\O_6]_{n,\ell}$ are double-trace primary operators with the schematic form
\begin{align}
[\O_5\O_6]_{n,\ell}=\O_5\partial_{\mu_1}...\partial_{\mu_\ell}\partial^{2n}\O_6 - \text{traces}
\end{align}
This projection is non-trivial because the decomposition of $\Cut_{56}[\mathcal{A}^{1234}]$ includes not just the $[\O_5\O_6]_{n,\ell}$ family, but also the shadow families where $\O_{5,6} \rar \widetilde \O_{5,6}$. We will discuss this in more detail in Section \ref{sec:AdS1Loop}.\foot{One can define an external line cut, as simply projecting onto double-trace operators composed of external operators. For example, for $\mathcal{A}^{1234}$ an external, vertical line cut means we project onto the $[\O_1\O_2]_{n,\ell}$ or $[\O_3\O_4]_{n,\ell}$ families. These cuts do not factorize the diagram: external lines are already on shell and cutting them provides no simplification, unlike internal cuts which place virtual lines on shell.}

We can now summarize our results.
\begin{quotation}
\noindent {\bf Result 1:} {\it $\widehat\Cut_{56}$ isolates the full contribution of the $[\O_5\O_6]_{n,\ell}$ operators to the conformal block decomposition of the amplitude. As an equation,}
\begin{align}\nonumber
\widehat\Cut_{56}[\mathcal{A}^{1234}] = \cA^{1234}\Big|_{[\O_5\O_6]_{n,\ell}}.
\end{align}
\end{quotation}
\noindent We prove this statement by direct computation of one-loop diagrams. That is, we show that all  poles in the spectral integrals besides the single-trace pole contribute only to external line cuts.

To establish an AdS unitarity method, we need to reconstruct the diagram from its internal line cuts. To do this, we relate these cuts to the dDisc, defined in \eqr{dDiscdef}; the determination of the full diagram then follows from Lorentzian inversion/CFT dispersion \cite{Caron-Huot:2017vep, Carmi:2019cub}. As is now well-known, the dDisc annihilates conformal blocks of double-trace composites comprised of the external operators in that channel, as stated in \eqr{dDisctch}. Acting on the amplitude $\cA^{1234}$ considered here, the $s$-channel dDisc$_s$ will remove the operators $[\O_1\O_2]_{n,\ell}$ and $[\O_{3}\O_{4}]_{n,\ell}$ from the conformal block expansion. Therefore, taking dDisc$_{s}$ projects $\cA^{1234}$ onto a sum over all internal, vertical line cuts, weighted by some trigonometric prefactors: 
\begin{quotation}
\noindent 
{\bf Result 2:} {\it ${\rm dDisc}$ is a weighted sum of the $\widehat\Cut$ operator acting on internal lines.} 
\end{quotation}
\noindent In the body of the paper, we will classify what the allowed internal cuts are, i.e. what operators can appear in the CPW decomposition of a Witten diagram. Cuts are to be understood as acting in a given channel. At one loop -- i.e. taking $\cA_L$ and $\cA_R$ to be tree diagrams in figure \ref{fig:Two_Prop_Witten_Split} -- Result 2 implies the following: 
\begin{quotation}
\noindent 
{\bf Result 3:} {\it For one-loop 1PI diagrams, ${\rm dDisc}$ is proportional to a single $\widehat\Cut$.}
\end{quotation}
\noindent This also follows straightforwardly from Result 1. For generic scalar operators, for example, 
\e{ddcutintro}{\dDisc_s(\mathcal{A}_\1^{1234}) = 2\sin\left({\pi\o2}(\D_5+\D_6-\D_1-\D_2)\right)\sin\left({\pi\o2}(\D_5+\D_6-\D_3-\D_4)\right)\widehat\Cut_{56}[\mathcal{A}_\1^{1234}] .}
The dDisc of non-1PI diagrams is still a weighted sum of $\widehat\Cut$s, but there can be more than one such cut. This is intuitive from looking at the diagrams, and is discussed in more detail for mass and vertex renormalization diagrams in Section \ref{sec:AdS1Loop}. 

\begin{figure}
\begin{center}
\includegraphics[scale=.62]{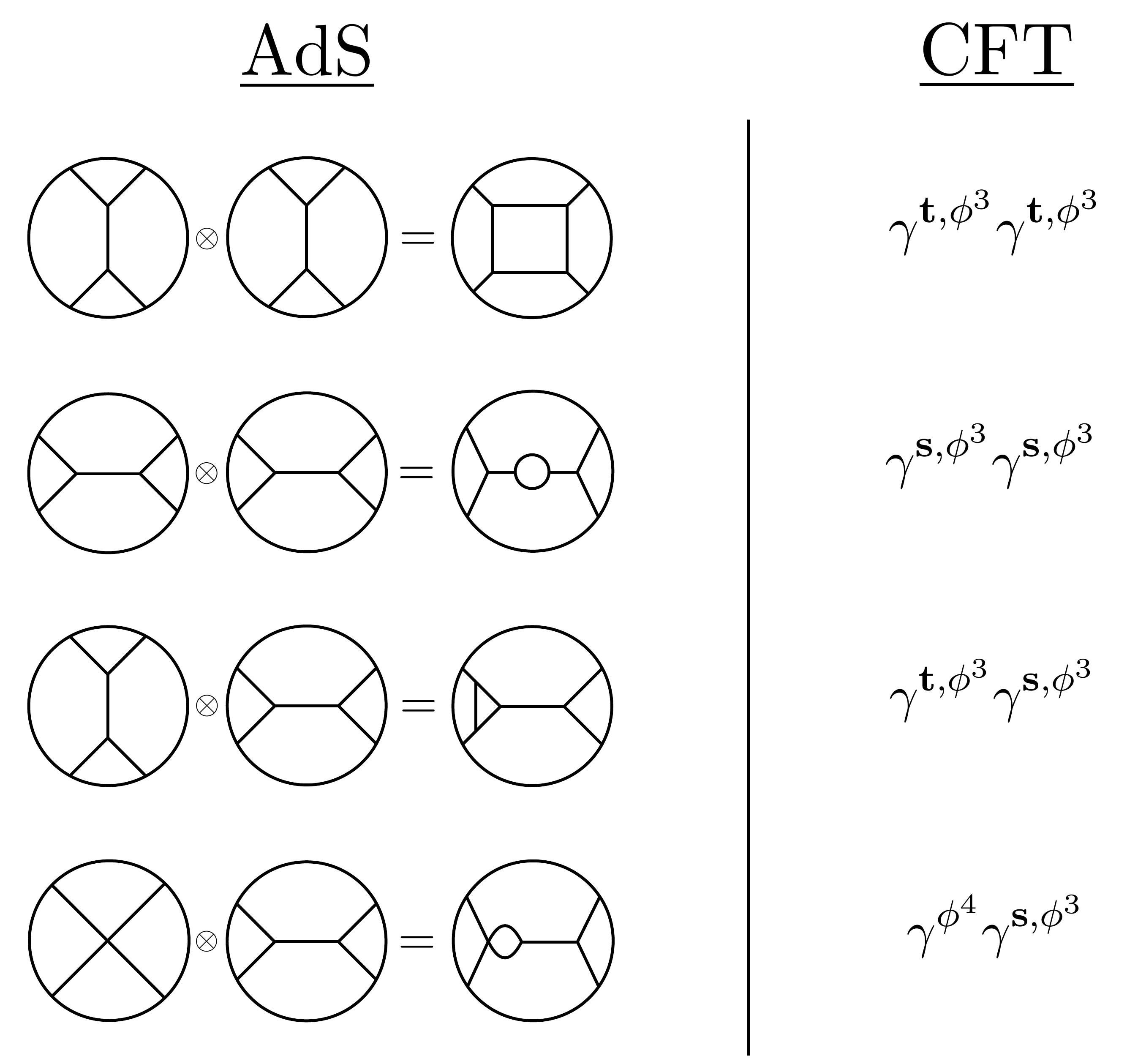}
\end{center}
\caption{A summary of the map between bulk and boundary unitarity methods for one-loop amplitudes in AdS $\phi^3+\phi^4$ theory. The left column shows the off-shell, $s$-channel conformal gluing of Witten diagrams, the $\hCut$s of which map to the dDisc of the correlator, $\dDisc_{s}(\cG_{\rm 1-loop})$, as explained in the text. The right column shows the CFT data -- namely, the product of tree-level anomalous dimensions -- that are inserted into the $s$-channel expansion of $\dDisc_{s}(\cG_{\rm 1-loop})$. The two ways of computing $\dDisc_{s}(\cG_{\rm 1-loop})$ match. We have suppressed the tree-level ``$(1)$'' superscripts on the anomalous dimensions for clarity. The fourth line may be obtained from the third by contraction of the $t$-channel exchange sub-diagram; this is neatly implemented on the CFT side by replacing a $\phi^3$ anomalous dimension with its $\phi^4$ counterpart.}
\label{fig:Gluing_Dictionary}
\end{figure}

With these ideas in hand, we can now directly relate the bulk and boundary unitarity methods -- that is, we seek a map between the {\it diagrammatic} conformal gluing defined by the $\widehat\Cut$ operator, and the {\it algebraic} holographic unitarity method that constructs the amplitude from OPE data. We exhibit this in figure \ref{fig:Gluing_Dictionary} for some diagrams in AdS $\phi^3+\phi^4$ theory.

The boundary unitarity method addressed the following question: given the planar OPE data of a large $N$ CFT, how do we generate leading non-planar corrections? Consider a one-loop, four-point function of identical scalars $\<\O\O\O\O\>$, as in \cite{Aharony:2016dwx}. Double-trace operators $[\O\O]_{n,\ell}$ have anomalous dimensions that admit an expansion
\e{}{\g_{n,\ell} = {\g^{(1)}_{n,\ell}\o N^2} + {\g^{(2)}_{n,\ell}\o N^4}+\ldots}
where $\g^{(1)}_{n,\ell}$ is the tree-level piece. The upshot of \cite{Aharony:2016dwx} phrased in the present language is that the dDisc of the one-loop correlator is completely determined by the product of tree-level anomalous dimensions
\e{cftunitintro}{{\rm dDisc}_t(\cA_{\rm 1-loop}) \supset {\pi^2\o 2}\sum_{n,\ell} p_{n,\ell}^{(0)} (\g^{(1)}_{n,\ell})^2 g_{n,\ell}(1-z,1-\zb).}
Here $p_{n,\ell}^{(0)}$ are the squared OPE coefficients of Mean Field Theory \cite{Heemskerk:2009pn,Fitzpatrick:2011dm} and $g_{n,\ell}(1-z,1-\zb)$ are $t$-channel conformal blocks for the exchange of $[\O\O]_{n,\ell}$.

Now, while the above reconstructs the (dDisc of the) {\it full, crossing-symmetric} one-loop amplitude,  it can be refined to reconstruct {\it individual} bulk diagrams as well. The sum of all bulk tree-level amplitudes gives the total tree-level anomalous dimension $\g_{n,\ell}^{(1)}$. Let us write it as
\e{}{\g_{n,\ell}^{(1)} = \g_{n,\ell}^{(1),\mathbf{s}}+\g_{n,\ell}^{(1),\mathbf{t}}+\g_{n,\ell}^{(1),\mathbf{u}}}
where $\g_{n,\ell}^{(1),\mathbf{x}}$ denotes the contribution from diagrams in channel $\mathbf{x}$. Inserting this into the right-hand side of \eqr{cftunitintro} and expanding, the bulk-boundary relation is clear: 
\begin{quotation}
\noindent 
{\bf Result 4:} {\it Replacing $(\gamma_{n,\ell}^{(1)})^{2}$ in \eqref{cftunitintro} with $\g_{n,\ell}^{(1),\mathbf{x}}\g_{n,\ell}^{(1),\mathbf{y}}$ computes the piece of \\${\rm dDisc}_{t}(\cA_{\rm 1-loop})$ obtained by gluing the bulk diagrams in the $\mathbf{x}$ and $\mathbf{y}$ channels. }
\end{quotation}
\noindent This is a simple and straightforward map between geometric and algebraic unitarity methods. The bulk-boundary map is explained further in Section \ref{sec:UnitaritiesUnified}, where we also give a CFT prescription for implementing contraction of bulk lines. 

Let us recap the analogy to the unitarity method in flat space scattering amplitudes. The integration over $\nu$ is analogous to the integration over momenta in standard Feynman diagrams, and the measure factor $P(\nu,\Delta)$ is analogous to a propagator. The $\dDisc$ of a diagram plays the role of the imaginary piece and is proportional to a sum over the action of the AdS $\widehat\Cut$ operator. Localizing onto the single-trace poles \eqref{eq:PDef} is analogous to localizing onto on-shell momentum eigenstates. The Lorentzian inversion formula then functions as a dispersion relation for the CFT data, allowing us to reconstruct the full CFT correlator from its cuts.

In closing, we note that the ideas we present here are valid for any perturbative field theory in AdS. For simplicity we will focus mostly on AdS scalar field theories, though we emphasize (and hope to convey) that the AdS unitarity method does not rely on that simplification. In Section \ref{sec:UnitaritiesUnified}, when making contact with the CFT unitarity method, we do not make such a restriction. The generalization to external spinning operators is possible using weight-shifting operators \cite{Karateev:2017jgd,Costa:2018mcg} and readily available identities for AdS bulk-to-bulk propagators \cite{Costa:2014kfa,Bekaert:2014cea,Bekaert:2015tva,Sleight:2017fpc}.

\ssec{Outline}

In {\bf Section \ref{sec:AdSUnitarity}}, we summarize the necessary CFT technology, including conformal integrals and the inversion formula. In {\bf Section \ref{sec:tree-level}}, we review how to compute tree diagrams, which will be the basic building blocks for loop amplitudes. In {\bf Section \ref{sec:AdS1Loop}} we present our central computational results, explicating the AdS unitarity method for one-loop, four-point scalar amplitudes, and computing their dDiscs. We study both 1PI and non-1PI diagrams. In {\bf Section \ref{sec:UnitaritiesUnified}} we tie things together by exhibiting the map between the bulk and boundary unitarity methods. In {\bf Section \ref{sec:AdSHigherPoint}} we initiate an extension to higher loops by applying the unitarity method to the two-loop double-ladder amplitude and, in conjunction with that, the five-point tree amplitude. Finally in {\bf Section \ref{sec:Conclusion}} we discuss some future directions. The appendices collect various conventions and technical details used in the body of the paper. Appendix \ref{app:Details} summarizes our conventions. Appendix \ref{app:MoreBox} contains further discussions on identities for the 6j symbol and AdS box, and some comments about the box Mellin amplitude and its relation to the results of \c{Aharony:2016dwx}. In Appendix \ref{app:Checks_Bulk_Vs_Bdy} we perform explicit comparisons between the bulk and boundary unitarity methods.

\textit{Note: after this work was completed, \cite{Ponomarev:2019ofr} appeared, which has some overlap with this work.}

\sec{CFT Ingredients}
\label{sec:AdSUnitarity}
\ssec{Rudiments}

We will be studying conformal four-point functions of real, scalar operators,
\begin{align}
\<\O_1(x_1)\O_2(x_2)\O_3(x_3)\O_4(x_4)\> &=T_{s}(x_i)\mathcal{G}(z,\bar{z}), \label{eq:CorrtoG}
\end{align} 
where $T_{s}(x_i)$ is a kinematic prefactor
\begin{align}\label{Tsfactor}
T_{s}(x_i)=\frac{1}{(x_{12}^2)^{(\Delta_1+\Delta_2)/2} (x_{34}^2)^{(\Delta_3+\Delta_4)/2} }
\left( \frac{x_{14}^2}{x_{24}^2}  \right)^{\frac{\Delta_2-\Delta_1}{2}}
\left( \frac{x_{14}^2}{x_{13}^2}  \right)^{\frac{\Delta_3-\Delta_4}{2}},
\end{align}
and the cross-ratios $(z,\zb)$ are defined by:
\begin{align}
z\bar{z}=\frac{x_{12}^{2}x_{34}^{2}}{x_{13}^{2}x_{24}^{2}}, \qquad (1-z)(1-\bar{z})=\frac{x_{14}^{2}x_{23}^{2}}{x_{13}^{2}x_{24}^{2}}.
\end{align}
In our convention, ``$s$-channel'' means the $\O_1\O_2 \rar \O_3\O_4$ OPE and ``$t$-channel'' means the $\O_1\O_4 \rar \O_3\O_2$ OPE. The crossing transformation maps $(z,\bar{z})\rightarrow (1-\bar{z},1-z)$. We will sometimes refer to four-point objects, like amplitudes, as $\cA^{1234}(x_i)$, where $x_i$ is implicit for $(x_1,x_2,x_3,x_4)$ and $1\ldots 4$ denotes the operators $\O_1\ldots \O_4$. 


The four-point function may be expanded equivalently in conformal blocks or conformal partial waves (CPWs). The $s$-channel CPW expansion is
\begin{align}
\<\O_1(x_1)\O_2(x_2)\O_3(x_3)\O_4(x_4)\>&=\sum\limits_{J=0}^{\infty}\int\limits_{\frac{d}{2}}^{\frac{d}{2}+i\infty}\frac{d\Delta}{2\pi i}\rho^{1234}(\Delta,J)\Psi^{1234}_{\Delta,J}(x_i).\label{eq:fourPt-CPW-V1}
\end{align}
The CPW, $\Psi^{1234}_{\Delta,J}(x_i)$, is related to the $s$-channel conformal block as
\e{eq:CPWtoCB}{\Psi^{1234}_{\Delta,J}(x_i)=K^{34}_{\widetilde{\Delta},J}g^{1234}_{\Delta,J}(x_i)+K^{12}_{\Delta,J}g^{1234}_{\widetilde{\Delta},J}(x_i),}
where $\widetilde{\Delta}=d-\Delta$ is the dimension of the shadow operator $\widetilde{\O}$, and the kinematic factors of $K^{ij}_{\Dt,J}$ are defined in Appendix \ref{app:Details}, although we will not need their explicit form. The conformal block as a function of all four points, $g_{{\Delta},J}(x_i)$, may be written as a function of the cross-ratios, $g_{\D,J}(z,\zb)$, multiplied by the $T_s(x_i)$ factor in \eqr{Tsfactor}. The important, non-kinematic object here is the {\bf OPE function} $\rho^{1234}(\Delta,J)$: this contains all data of the CPW decomposition in the $\O_1\O_2 \rightarrow \O_3\O_4$ channel.\foot{Our notation is that $\rho^{1234}(\D,J)$ is the OPE function in the $s$-channel, and likewise for the CPWs and conformal blocks.} 
To study the physical OPE of our theory and remove the shadow operator contributions, we need to go from the CPW expansion to the conformal block decomposition. Using the identity
\begin{align}
\rho^{1234}(\widetilde{\Delta},J)=\rho^{1234}(\Delta,J)\frac{K^{34}_{\widetilde{\Delta},J}}{K^{12}_{\widetilde{\Delta},J}},
\end{align}
we can extend the $\Delta$ contour along the entire axis and replace the CPW with a block:
\begin{align}
{\<\O_1(x_1)\O_2(x_2)\O_3(x_3)\O_4(x_4)\>=\sum\limits_{J=0}^{\infty}\int\limits_{\frac{d}{2}-i\infty}^{\frac{d}{2}+i\infty}\frac{d\Delta}{2\pi i}\rho^{1234}(\Delta,J)K^{34}_{\widetilde{\Delta},J}g^{1234}_{\Delta,J}(x_i)~.}
\label{eq:fourPt-CPW-V2}
\end{align}
Since each block decays for large real $\Delta > 0$, we can close the contour to the right, pick up the poles in the OPE function, and reproduce the physical conformal block expansion.\footnote{There are non-normalizable terms, coming from operators with $\Delta \leq d/2$, which are not captured by this expansion\cite{Caron-Huot:2017vep,ssw,Costa:2012cb}. In this work we can ignore such terms. Note that the function $\rho^{1234}(\Delta,J)$ can have families of poles extending to the left, so we define the contour such that it passes to the right of these poles.}

We emphasize again that \eqr{eq:fourPt-CPW-V1} and \eqr{eq:fourPt-CPW-V2} are equivalent. As is evident from comparison, the OPE function for the conformal block decomposition \eqr{eq:fourPt-CPW-V2} is just $\rho^{1234}(\D,J)$ times the kinematic factor $K^{34}_{\Dt,J}$. We will often just quote the result for $\rho^{1234}(\D,J)$. 

The CPWs have several useful properties which are not inherited by the blocks, some of which we mention here. If we restrict the scaling dimension to the Euclidean principal series, $\Delta=\frac{d}{2}+i\nu$ with $\nu\in\mathbb{R}_{\geq0}$, then the CPWs form a normalizable, complete basis for CFT four-point functions, modulo non-normalizable terms. The CPWs are orthogonal with respect to the following inner product:
\begin{align}
\left(\Psi^{1234}_{\frac{d}{2}+i\nu_{1},J_1},\Psi^{\tilde{1}\tilde{2}\tilde{3}\tilde{4}}_{\frac{d}{2}-i\nu_2,J_{2}}\right)&=2\pi\delta(\nu_1-\nu_2)\delta_{J_1,J_2}n_{\frac{d}{2}+i\nu_1,J_1},
\\
\left(F,G\right)&=\int \frac{d^{d}x_{1}...d^{d}x_{4}}{\text{vol(SO(d+1,1))}}F(x_i)G(x_i),
\end{align}
where $n_{\Delta,J}$ is the normalization of the CPW, see (\ref{eq:CPW_Norm}). 

\begin{figure*}
\centering
\begin{subfigure}[t]{.4\textwidth}
\centering
\includegraphics[scale=.35]{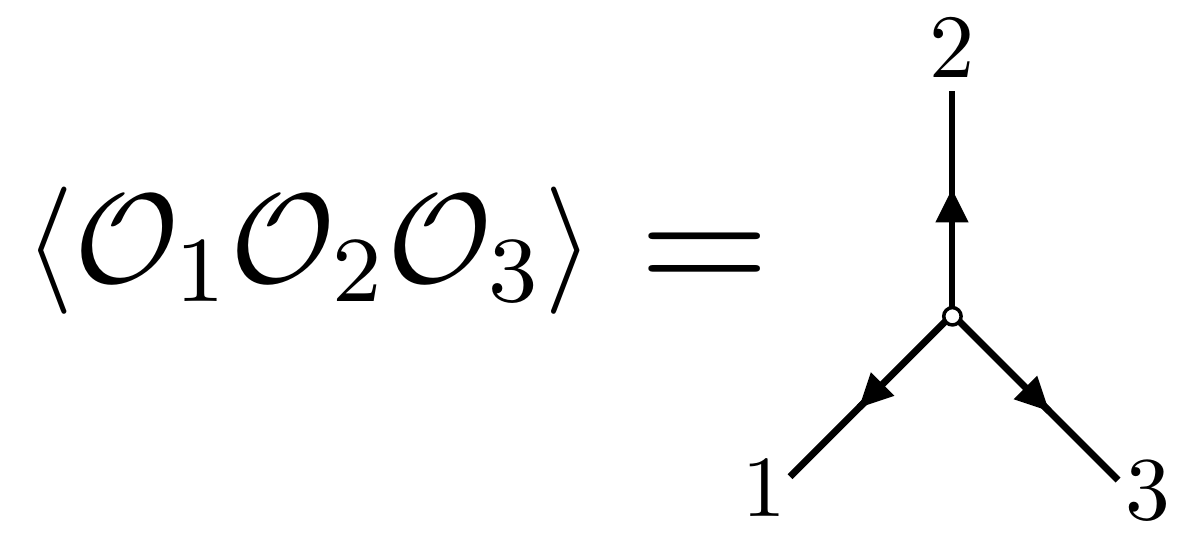}
\end{subfigure}%
~
\begin{subfigure}[t]{.4\textwidth}
\centering
\includegraphics[scale=.3]{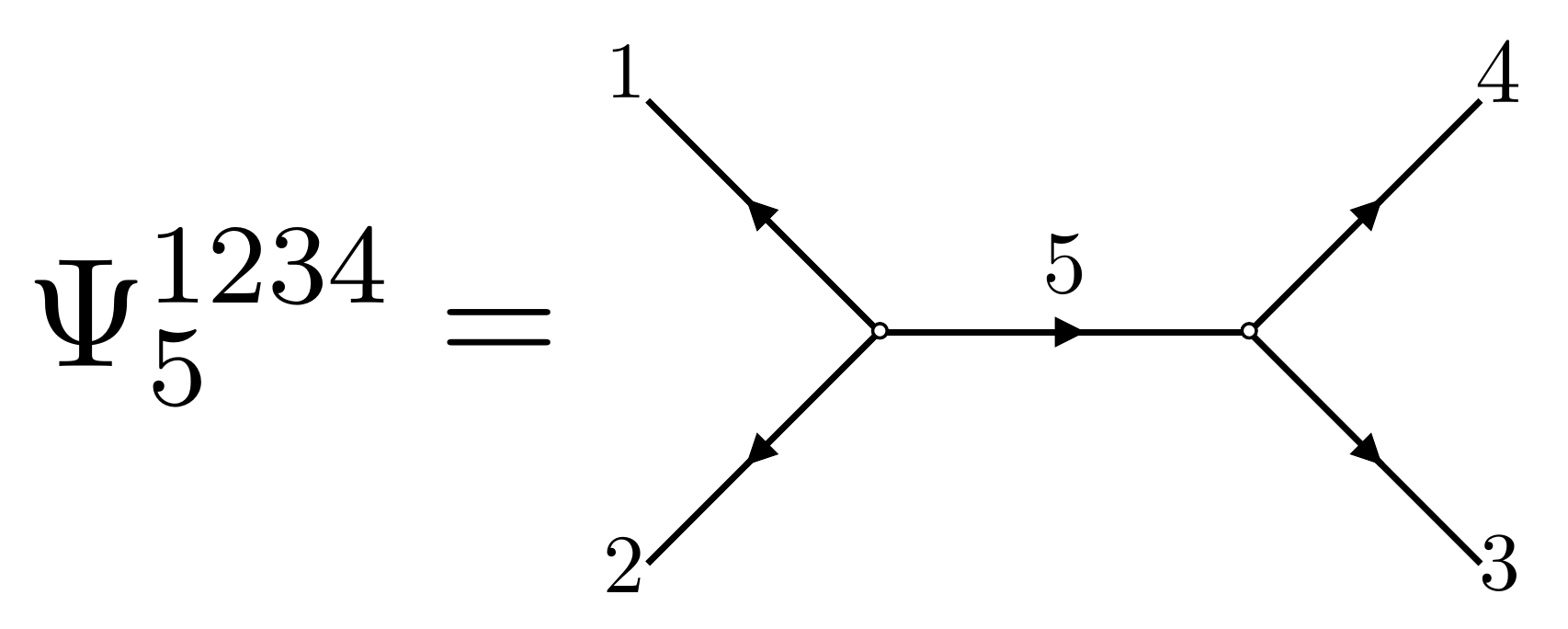}
\end{subfigure}%
\caption{Outgoing/ingoing lines correspond to $\O$ and $\widetilde{\O}$ respectively.}
\label{fig:cpw_graph}
\end{figure*}

Another convenient feature of CPWs is they can be expressed as a conformal integral:
\begin{align}
\Psi^{1234}_{5}(x_1,x_2,x_3,x_4)=\int d^{d}x_{5}\<\O_1(x_1)\O_2(x_2)\O_5(x_5)\>\<\widetilde{\O}_5(x_5)\O_3(x_3)\O_4(x_4)\>,
\label{eq:CPW-as-3pts-glued}
\end{align}
where $\<\O_1(x_1)\O_2(x_2)\O_5(x_5)\>$ is the unique conformally invariant three-point function, or three-point structure, when $\O_{1,2}$ are scalars. We will adopt the convention that these structures do not include the OPE coefficients in their definition.\footnote{We also adopt the convention that inside two and three-point functions $\widetilde{\O}$ denotes an abstract operator with dimension $\widetilde{\Delta}_{\O}$ and does not include any factors from the shadow transform.} More generally, we will refer to integrals involving an operator paired with its shadow as a ``conformal integral".\footnote{In Euclidean signature the integration is over all of space, but in Lorentzian signature we can consider more general pairings of operators and domains of integration \cite{Czech:2016xec,Kravchuk:2018htv}.} It will be convenient to adopt the graphical notation given in figure \ref{fig:cpw_graph} such that a cubic vertex corresponds to a three point function, the CPW is a four-point tree-graph, and the positions of internal operators are integrated over.

Taking the inner product of $s$- and $t$-channel CPWs defines the 6j symbol of the conformal group $SO(d+1,1)$:
\begin{align}
\left(\Psi^{\tilde{1}\tilde{2}\tilde{3}\tilde{4}}_{\widetilde{\Delta}_{5},J_{5}},\Psi^{3214}_{\Delta_{6},J_{6}}\right)=\sixj{1}{2}{3}{4}{5}{6}. \label{eqn:6jdefOverlap}
\end{align}

The 6j symbol tells us how to expand a single CPW in one channel as a spectral integral in another channel,\foot{Note that the 6j symbol is known explicitly in closed form only in $d=1,2,4$ \cite{Hogervorst:2017sfd, Liu:2018jhs}. See also \c{Sleight:2018ryu,Gopakumar:2018xqi}.}
\e{6jcpw}{\Psi^{3214}_{\D_6,J_6}(x_i) = \sum_{J_5=0}^{\infty}\int\limits_{d\o 2}^{{d\o 2}+i\infty} {d\D_5\o 2\pi i}{1\o n_5}\sixj{1}{2}{3}{4}{5}{6}\Psi^{1234}_{\D_5,J_5}(x_i).}
In our graphical notation, a 6j symbol allows us to convert a vertical tree into a sum of horizontal trees and vice-versa, 
\begin{center}
\includegraphics[scale=.3]{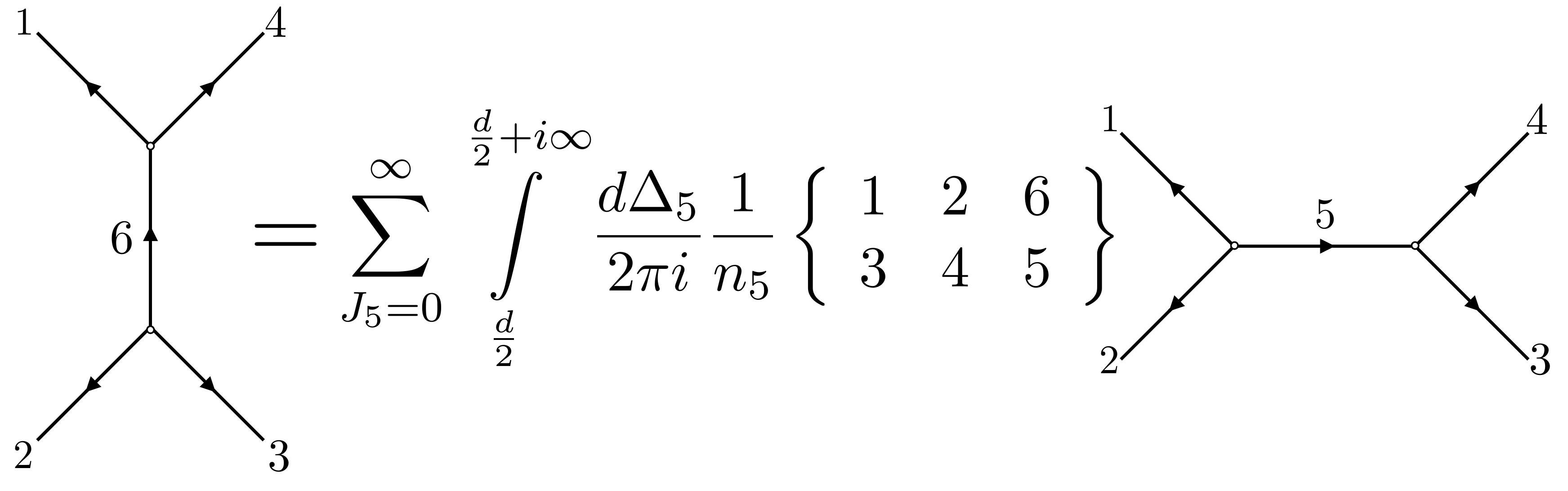}
\label{fig:6jCrossing}
\end{center}

By working with operators on the principal series, we will leverage this equality when studying Witten diagrams. 

Finally, the last conformal integral we need is the Euclidean bubble integral:
\begin{align}
\int\limits_{\partial \rm AdS}d^dx_1 d^dx_2 \<\O_{\frac{d}{2}+i\nu,J}(x)\O_{1}(x_1)\O_{2}(x_2)\>\<\widetilde{\O}_{2}(x_2)\widetilde{\O}_{1}(x_1)\O'_{\frac{d}{2}-i\nu',J'}(x')\>
=
 B_{\O}\delta(\nu-\nu')\delta_{J,J'} \delta(x-x'),
\label{eq:Conf-Bubble}
\end{align}
with the restriction $\nu,\nu'>0$. This identity is shown diagrammatically as:
%
\begin{center}
\includegraphics[scale=.25]{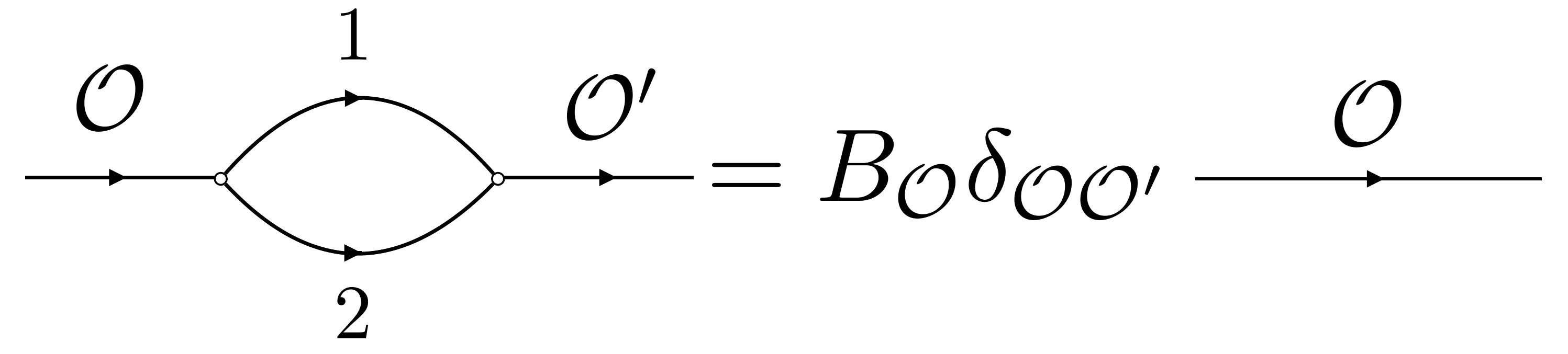}
\end{center}
\noindent The definition of the bubble factor is given in \eqr{eq:CFTbubDef}-\eqr{eq:rhodim} and for internal scalars it is given by \eqr{J0bubble}. We will use this integral repeatedly to compute loop-level Witten diagrams. In particular, we will often use the following form of the bubble identity, which comes from expressing CPWs in terms of three-point functions as in \eqref{eq:CPW-as-3pts-glued}:
\begin{equation}
\int d^dx_3 d^dx_4 \Psi^{1234}_{\frac{d}{2}+i\nu,J}(x_i) \Psi^{\tilde{4} \tilde{3} 5 6}_{\frac{d}{2}+i\nu',J'}(x_i)
=
B_{\O}\delta_{J,J'} \delta(\nu-\nu') \Psi^{1256}_{\frac{d}{2}+i\nu,J}(x_i).
\label{eq:CPWs-bubbled}
\end{equation}

\ssec{dDisc}
Now we can introduce the double discontinuity which is used in the Lorentzian inversion formula \cite{Caron-Huot:2017vep} to reconstruct the OPE data, up to low-spin ambiguities.
 The $t$-channel dDisc$_{t}$ is defined as\foot{To fully reconstruct the $s$-channel OPE data using the inversion formula, one needs the $u$-channel double discontinuity; for simplicity, we will focus on the $s$ and $t$-channel dDiscs and everything will carry over to the $u$-channel. Note that, in general, the Lorentzian inversion formula produces the correct OPE data for operators with $J>J_0$, where $J_0$ controls the Regge growth of the correlator.} 
\begin{align}
\dDisc_{t} ( \mathcal{G}(z, \bar{z})) &\equiv \cos(\alpha)  \mathcal{G}(z,\bar{z})
-\frac{1}{2} \left( e^{i\alpha}\mathcal{G}^{\circlearrowleft}(z,\bar{z})+e^{-i\alpha}\mathcal{G}^{\circlearrowright}(z,\bar{z}) \right),
\label{dDiscdef}
\\
\alpha&=\frac{\pi}{2}(\Delta_2+\Delta_3-\Delta_1-\Delta_4),
\end{align}
where $\mathcal{G}^{\circlearrowleft}(z,\bar{z})$ is found by rotating $\bar{z}$ counterclockwise around the branch cut at $\bar{z}=1$ and clockwise for $\mathcal{G}^{\circlearrowright}(z,\bar{z})$, holding $z$ fixed. dDisc$_s$ is likewise defined by instead continuing around $z=0$ and keeping $\bar{z}$ fixed. 

Written as a discrete sum over real operator dimensions, the $t$-channel conformal block decomposition of $ \mathcal{G}(z,\bar{z})$ is
\begin{align}
\mathcal{G}(z,\bar{z})=\frac{(z\bar{z})^{\frac{\Delta_1+\Delta_2}{2}}}{\left[(1-z)(1-\bar{z})\right]^{\frac{\Delta_2+\Delta_3}{2}}}\sum\limits_{\D, J}p_{\D,J}g_{\D,J}(1-z,1-\bar{z}),
\end{align}
where $p_{\D,J}$ are products of OPE coefficients, 
\e{}{p_{\D,J} \equiv 2^{-J} C_{14\O_{\D,J}}\,C_{23\O_{\D,J}}.}
The action of the dDisc$_{t}$ on this expansion weights each block by $\sin$ factors:
\begin{align}\label{dDisctch}
\dDisc_{t} ( \mathcal{G}(z,\bar{z}))&=2\frac{(z\bar{z})^{\frac{\Delta_1+\Delta_2}{2}}}{\left[(1-z)(1-\bar{z})\right]^{\frac{\Delta_2+\Delta_3}{2}}}\\&\times\sum\limits_{\D,J}\sin\left(\frac{\pi}{2}(\Delta-\Delta_1-\Delta_4)\right)\sin\left(\frac{\pi}{2}(\Delta-\Delta_2-\Delta_3)\right)p_{\D, J}g_{\D, J}(1-z,1-\bar{z}). \nonumber
\end{align}
Thus, exchanges of ``double-trace'' composites $[\O_1\O_4]_{n,\ell}$ and $[\O_2\O_3]_{n,\ell}$ with vanishing anomalous dimension, as in Mean Field Theory, are annihilated by dDisc$_{t}$.
\sec{Review and Warmup: AdS Trees}
\label{sec:tree-level}
We begin by reviewing, in detail, the computation of tree-level Witten diagrams and their conformal block decomposition. This serves as a useful warmup for the one-loop case, which we will compute by gluing trees, and helps develop intuition for the role of the dDisc. The main point will be to introduce the {\bf AdS unitarity cut}, which factorizes tree-level exchanges into products of three-point functions, and its manifest relation to dDisc. The computation follows \c{Liu:2018jhs} and employs the technical toolkit of conformal integrals reviewed in Section \ref{sec:AdSUnitarity} (see e.g. \cite{Dobrev:1977qv,Gadde:2017sjg,Kravchuk:2018htv,Liu:2018jhs,Karateev:2018oml}).

We use conventions and normalizations of \c{Liu:2018jhs}. Bulk coordinates are $y_i$ and boundary coordinates are $x_i$. We will use the notation 
\e{}{\int\limits_{\rm AdS} d^{d+1} y \equiv \int d^{d+1} y ~ \sqrt{g(y)}\quad \text{and} \quad  \int\limits_{\partial\rm AdS} d^d x}
for bulk and boundary integration, respectively. The bulk-to-boundary propagator for a scalar operator $\O_\D$ is $K_{\D}(x_i,y_i)$, and the bulk-to-bulk propagator is $G_{\D}(y_i,y_j)$. We suppress overall coupling-dependence of all diagrams. 

\ssec{Contact Diagrams}
A particularly simple four-point Witten diagram is the $\phi^4$ contact diagram drawn below.
\begin{center}
\includegraphics[scale=.28]{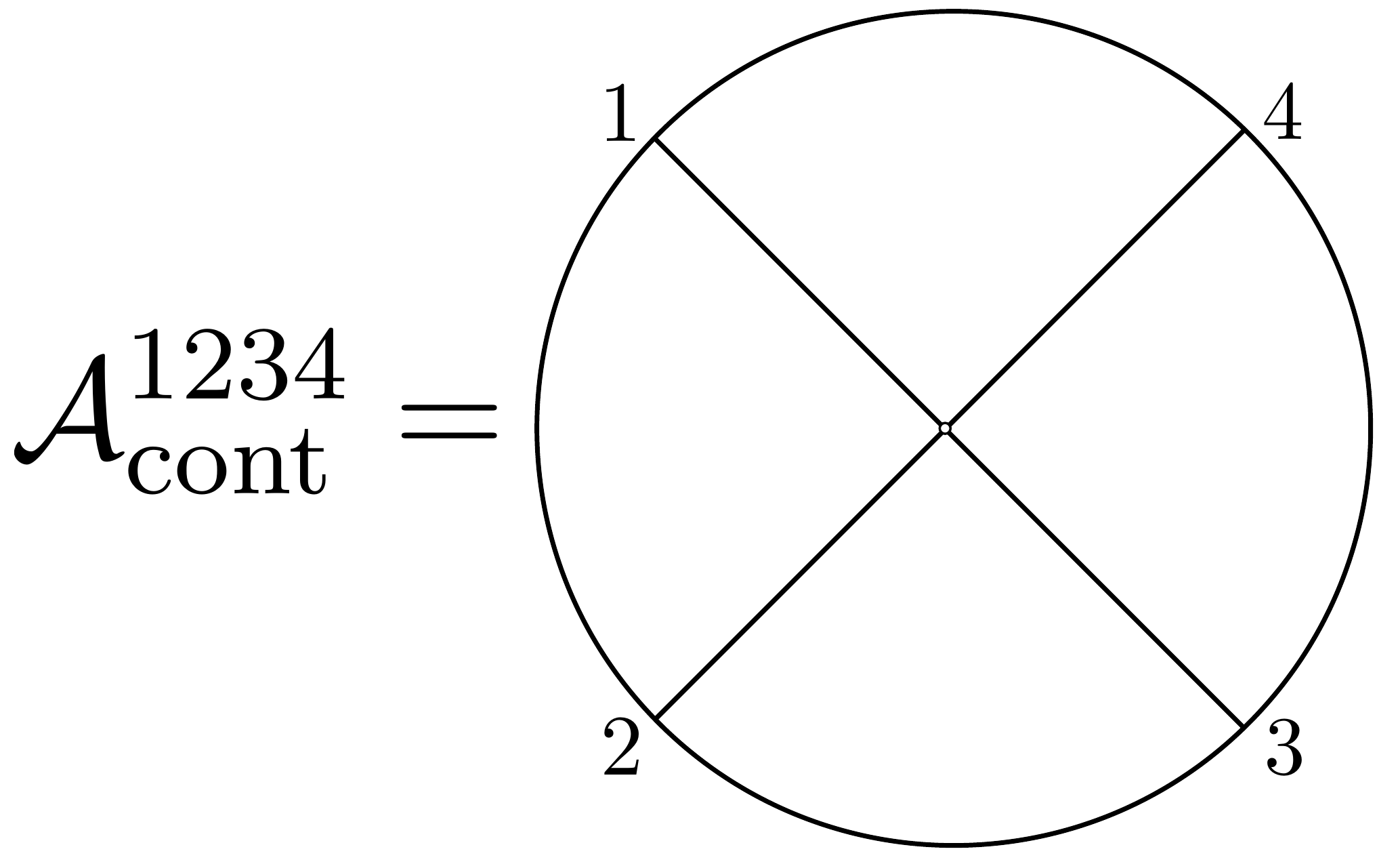}
\end{center}
To find the CPW expansion, we use the scalar AdS harmonic function
\begin{align}
\Omega_{\nu}(y_1,y_2)=\frac{\nu^{2}}{\pi}\int\limits_{\partial \rm AdS}d^d x K_{\frac{d}{2}+i\nu}(x,y_1)K_{\frac{d}{2}-i\nu}(x,y_2), \label{eqn:HarmonicDef}
\end{align}
which is independently invariant under $y_1\leftrightarrow y_2$ and $\nu \rightarrow -\nu$. The harmonic function also satisfies the completeness property:
\begin{align}
\int\limits_{-\infty}^{\infty}d\nu~\Omega_{\nu}(y_1,y_2)=\delta(y_1,y_2). \label{eqn:deltaHarm}
\end{align}
Using this representation of the delta function, the contact diagram can be written as \cite{Zhou:2018sfz}:
\begin{align*}
\hspace{-.3in}\mathcal{A}_{\rm cont}(x_i)=\int\limits_{-\infty}^{\infty}d\nu  \int\limits_{\partial \rm AdS}d^d x_{5} \int\limits_{\rm AdS} d^{d+1}y_1 d^{d+1}y_2  
\frac{\nu^{2}}{\pi} & K_{\Delta_1}(x_1,y_1)K_{\Delta_2}(x_2,y_1)K_{\frac{d}{2}+i\nu}(x_5,y_1)
\\
\times
&
K_{\frac{d}{2}-i\nu}(x_5,y_2) K_{\Delta_3}(x_3,y_2)K_{\Delta_4}(x_4,y_2).
\numberthis
\end{align*}
The AdS integrals have now all been reduced to standard three-point integrals:
\begin{equation}\label{AdSthreept}
\int\limits_{\rm AdS} d^{d+1}y K_{\Delta_{1}}(x_1,y)K_{\Delta_2}(x_2,y)K_{\Delta_3,J_3}(x_3,y)=b_{123}\<\O_{1}(x_1)\O_{2}(x_2)\O_{3}(x_3)\>,
\end{equation}
where the kinematic $b$-factors were derived in \cite{Freedman:1998tz,Costa:2014kfa} and are defined in (\ref{eqn:AdS3pt}). From the definition \eqr{eq:CPW-as-3pts-glued} of the CPW, we have
\begin{align}
\mathcal{A}_{\rm cont}(x_i)=\int\limits_{-\infty}^{\infty}d\nu_5 \frac{\nu_5^{2}}{\pi}b_{12\underline{5}}b_{\tilde{\underline{5}}34}\Psi^{1234}_{\underline{\Delta}_{5},0}(x_1,x_2,x_3,x_4).
\label{eq:cont_CPW_expansion}
\end{align}
Comparing with the CPW expansion (\ref{eq:fourPt-CPW-V1}) and recalling the principal series parameterization $\underline{\D}={d\o2}+i\nu$, we can read off the OPE function for scalar contact diagrams as
\begin{align}
\rho^{1234}_{\rm cont}(\underline\Delta_5,J_5)=-\delta_{J_5,0}(d-2\underline\Delta_5)^{2}b_{12\un 5}b_{\underline{\tilde{5}}34}.
\label{eq:cont_rho}
\end{align}

The scalar contact diagram only has support on $J=0$ operators, and the poles from the $b$-factors in (\ref{eq:cont_CPW_expansion}) are at the double-trace locations\footnote{The $b$-factor has additional families of poles at $\Delta = \pm (\Delta_1-\Delta_2)-2n$ and $d-\Delta_1-\Delta_2-2n$, but they lie in the upper half plane of $\nu$ and will not be picked up.}:
\begin{eqnarray}
\underline{\Delta}_{5}&=&\Delta_{1}+\Delta_{2}+2n, \label{eq:cont_rho_polespt1}
\\
\underline{\Delta}_{5}&=&\Delta_{3}+\Delta_{4}+2n.
\label{eq:cont_rho_polespt2}
\end{eqnarray}
If we split the CPW using (\ref{eq:CPWtoCB}), we determine the conformal block decomposition. This simply appends a $K^{34}_{\tilde{\underline{5}}}$ factor to $\rho^{1234}_{\rm cont}(\underline\Delta_5,J_5)$.
The conformal block decays exponentially for large dimensions, and this allows us to close the $\nu_5$ contour in the lower half-plane. In the process we only pick up the poles (\ref{eq:cont_rho_polespt1}) and (\ref{eq:cont_rho_polespt2}). In the special case that $\Delta_1+\Delta_2=\Delta_3+\Delta_4$ we find a double pole which produces a logarithm after closing the contour. This familiar feature reflects the tree-level anomalous dimension of the double-trace operators.

Now, due to the manifest non-analyticity in spin of the OPE data in any channel,\foot{To derive the $t$- and $u$- channel expansions we use (\ref{eqn:deltaHarm}) to ``split" the contact diagram in different ways.} 
\e{}{\dDisc_{s,t,u}(\mathcal{A}_{\rm cont}(x_i))=0.}
As we explained earlier and will substantiate throughout this paper, the diagrammatic reason for this is the absence of internal lines.

\ssec{Exchange Diagrams}

Next we will review the CPW decomposition of tree-level exchange diagrams. The main goal is to show that dDisc projects onto internal cuts in a natural way. We will also find it useful to recall how the 6j symbol naturally appears in the process of decomposing exchange diagrams in the crossed channel \cite{Liu:2018jhs}. 

To determine the CPW expansion of the $t$-channel exchange graph,
\begin{center}
\includegraphics[scale=.28]{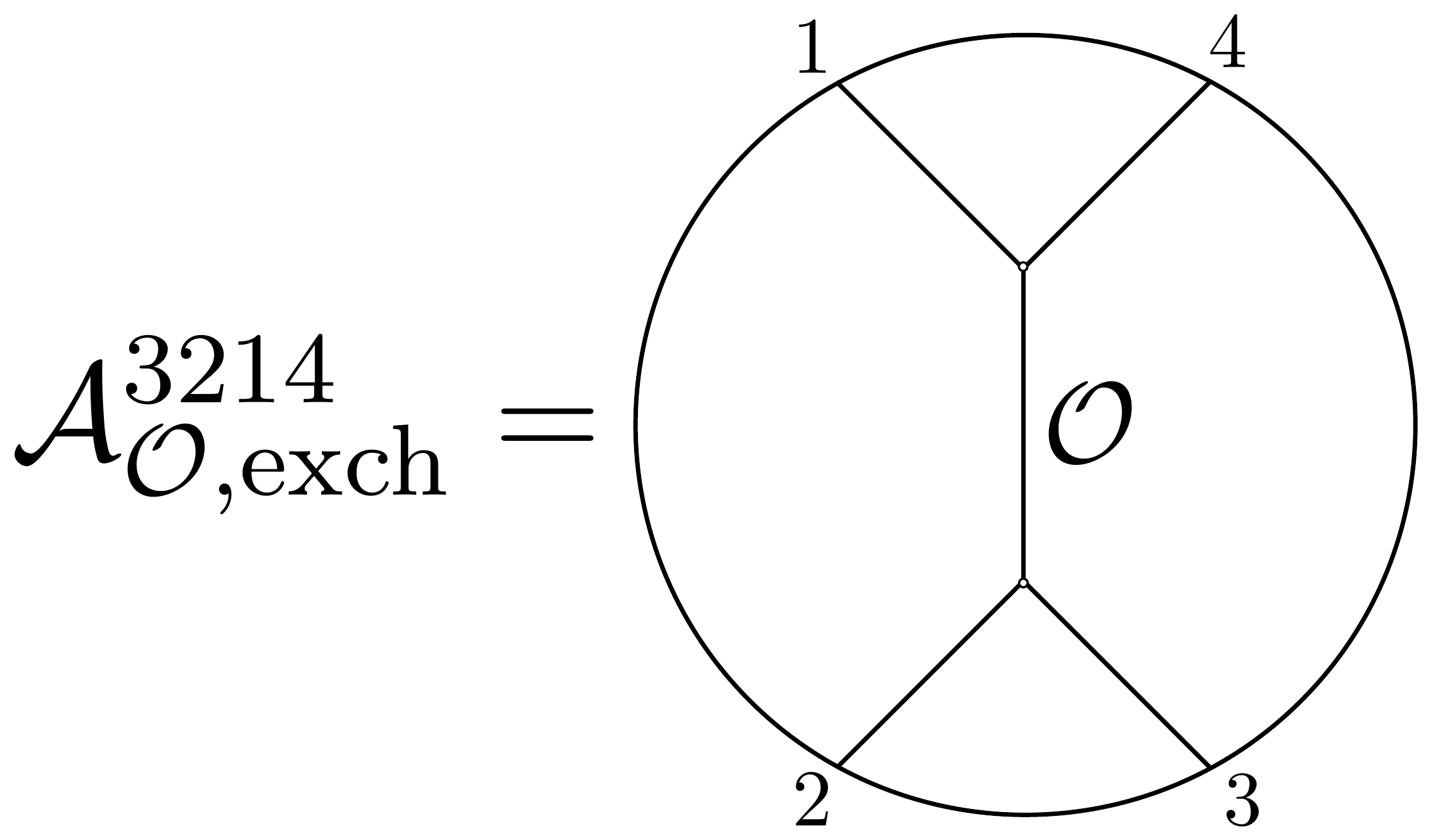}
\end{center}
we will use the split representation for the bulk-to-bulk propagators, see \eqr{Gsplit}. Using \eqr{AdSthreept} on the resulting product of two three-point functions gives a CPW expansion %
\e{eq:Witten_CPW_V1}{
\mathcal{A}^{3214}_{\O,\,\rm exch}(x_i)=\int\limits_{-\infty}^{\infty}d\nu_{\O}P(\nu_{\O},\Delta_{\O})b_{32\underline{\O}}b_{\underline{\widetilde{\O}}14}\Psi^{3214}_{\underline{\O}}(x_i).}
The OPE function is thus 
\begin{align}\label{rhoexch}
\rho^{3214}_{\O,\, \rm exch}(\underline{\Delta}_{\O},J_{\O})=-\delta_{J_{\O},0}\frac{(d-2\underline{\Delta}_{\O})^{2}}{(\underline{\Delta}_{\O}-\Delta_{\O})(\underline{\widetilde{\Delta}}_{\O}-\Delta_{\O})}b_{32\underline{\O}}b_{\underline{\widetilde{\O}}14}.
\end{align}
There are three types of operators contributing to the block decomposition:
\begin{eqnarray}
\underline{\Delta}_{\O}&=&\Delta_\O, \label{eqn:Exch_STPole}
\\
\underline{\Delta}_{\O}&=&\Delta_{1}+\Delta_{4}+\ell+2n,
\\
\underline{\Delta}_{\O}&=&\Delta_{2}+\Delta_{3}+\ell+2n.
\end{eqnarray}
The first pole comes from the measure of the spectral integral; this is the ``single-trace pole''. As in the contact diagram case, the remaining poles come from the $b$-factors, and are at the location of the double-trace operators. 

The tree-level diagram provides a simple but useful example of how cuts simplify a diagram. On the one hand, $\Cut_{\O}[\mathcal{A}^{3214}_{\O,\,\rm exch}]$ is, by definition, the contribution from only the single trace pole (\ref{eqn:Exch_STPole}):
\begin{align}
\Cut_{\O}[\mathcal{A}^{3214}_{\O,\,\rm exch}(x_i)]=(d-2\Delta_{\O})b_{32\O}\,b_{\widetilde{\O}14}K^{14}_{\widetilde\O} g^{3214}_{\O}(x_i).
\end{align}
The cut OPE function manifestly factorizes onto a product of the three-point factors $b_{12\O}b_{\widetilde{\O}34}$. This is the unique, internal line cut and is also proportional to the $\dDisc_{t}$ of the diagram:
\begin{align}
{\dDisc_t(\mathcal{A}^{3214}_{\O,\,\rm exch}(x_i)) 
= 2\sin\left(\frac{\pi}{2}(\Delta_\O-\Delta_2-\Delta_3)\right)\sin\left(\frac{\pi}{2}(\Delta_\O-\Delta_1-\Delta_4)\right) \Cut_{\O}[\mathcal{A}^{3214}_{\O,\,\rm exch}(x_i)].}
\end{align}
That is, the dDisc is putting the virtual $\O$ line on shell. This is a tree-level foreshadowing of the loop relations to come. 

\sssec{Crossed-channel decomposition}

It will be useful for later to gain some experience with 6j symbols. These arise when we decompose this $t$-channel exchange in the $s$-channel. First, insert \eqr{6jcpw} into \eqr{eq:Witten_CPW_V1} to branch the $t$-channel CPWs into the crossed channel:
\begin{align}\label{eq:TreeExchangeSV0}
\mathcal{A}^{3214}_{\O,\,\rm exch}(x_i)=\int\limits_{-\infty}^{\infty}d\nu_{\O}P(\nu_\O,\Delta_\O)b_{32\underline{\O}}b_{\widetilde{\underline{\O}}14}
\sum_{J_{\O'}=0}^{\infty}\int\limits_{\frac{d}{2}}^{\frac{d}{2}+i\infty}\frac{d\Delta_{\O'}}{2\pi i}{\sixj{1}{2}{3}{4}{\O'}{\underline\O}}\frac{1}{n_{\O'}}\Psi^{1234}_{\O'}(x_i).
\end{align}
At this point, we could close the $\nu_\O$ contour to the right so that we are left with a single spectral integral, but generically the 6j symbol has many poles whose physical significance is not clear. It is more convenient to use (\ref{eq:CPWtoCB}) to decompose the 6j symbol as \c{Liu:2018jhs} 
\begin{eqnarray}
\sixj{1}{2}{3}{4}{5}{6}&=&K^{14}_{\tilde{6}}\sixjBlock{1}{2}{3}{4}{5}{6}+K^{23}_{6}\sixjBlock{1}{2}{3}{4}{5}{\tilde{6}}, \label{eqn:6j_Split}
\\
\sixjBlock{1}{2}{3}{4}{5}{6}&=&\left(\Psi^{\tilde{1}\tilde{2}\tilde{3}\tilde{4}}_{\tilde{5}},g^{3214}_{6}\right), \label{eqn:InvBlockSym}
\end{eqnarray}
where in application to \eqr{eq:TreeExchangeSV0} we have $6 \rar \un\O$ and $5\rar \O'$. The symbol (\ref{eqn:InvBlockSym}) represents the inversion of a single $t$-channel block (as opposed to a $t$-channel CPW). This must be formally defined as the ``half'' of the inversion of a $t$-channel CPW proportional to $K^{14}_{\tilde{6}}$, since only CPWs obey well-defined orthogonality and single-valuedness properties. 

To proceed we need the location of the poles of the 6j symbol, which were discussed in \cite{Liu:2018jhs}. One nice feature of the split \eqref{eqn:6j_Split} is that (\ref{eqn:InvBlockSym}) has no poles in $\Delta_{6}$ to the right of the principal series. We can therefore insert (\ref{eqn:6j_Split}) into (\ref{eq:TreeExchangeSV0}) and close the $\Delta_6$ contour to the right for the first term and to the left for second term in (\ref{eqn:6j_Split}). Another important feature of (\ref{eqn:InvBlockSym}) is that it is zero when $\Delta_{6}=\Delta_{1}+\Delta_{4}+\ell_{6}+2n$ and $\Delta_{6}=\Delta_{2}+\Delta_{3}+\ell_{6}+2n$. These are the familiar zeros from the dDisc in the inversion formula and will cancel the $t$-channel double-trace poles from the $b$ factors in (\ref{eq:TreeExchangeSV0}) when we close the $\Delta_6$ contour. 

Therefore, we are left with only the single-trace pole for the $\nu_{\O}$ contour and the $s$-channel CPW decomposition of the $t$-channel exchange diagram is
\begin{align*}\label{exchcross}
\mathcal{A}^{3214}_{\O,\,\rm exch}(x_i)&=\sum\limits_{J_{\O'}=0}^{\infty}\int\limits_{\frac{d}{2}-i\infty}^{\frac{d}{2}+i\infty}\frac{d\Delta_{\O'}}{2\pi i}\left(d-2\Delta_\O\right)b_{32\O}b_{\widetilde{\O}14}K^{14}_{\widetilde{\O}}\sixjBlock{1}{2}{3}{4}{\O'}{\O}\frac{1}{n_{\O'}}K^{34}_{\widetilde{\mathcal{O}}'}\,g^{1234}_{\O'}(x_i).
\numberthis
\end{align*}
We can close the $\Delta_{\O'}$ contour to the right to obtain the conformal block decomposition. We first need that the symbol (\ref{eqn:InvBlockSym}) has poles in $\Delta_{5}$ at the following locations:
\begin{eqnarray}
\Delta_{5}&=&\Delta_{1}+\Delta_2+2n+\ell,
 \\
\Delta_{5}&=&\Delta_{3}+\Delta_4+2n+\ell,
\\
\Delta_{5}&=&\widetilde{\Delta}_{3}+\Delta_4+2n+\ell,
\\
\Delta_{5}&=&\Delta_{3}+\widetilde{\Delta}_4+2n+\ell.
\end{eqnarray}
These latter two sets of poles are not physical and will be cancelled by zeros in $K^{34}_{\widetilde{\O}'}$. In the end, only the first two sequences of poles contribute, as is known from previous AdS computations \cite{Freedman:1998bj,Heemskerk:2009pn,Heemskerk:2010ty,Penedones:2010ue}. The full set of poles for the 6j symbol can be derived from these using tetrahedral symmetry \cite{Liu:2018jhs}, but for this work we can focus on these poles of (\ref{eqn:InvBlockSym}).

\sec{AdS One-Loop Diagrams}
\label{sec:AdS1Loop}
In this section, we turn to the direct bulk computation of scalar one-loop diagrams. Our approach will be as follows. We will apply the identities in Section \ref{sec:AdSUnitarity} to obtain loop diagrams as spectral integrals over CPWs. We will explain how to identify the physical poles that contribute to the spectral integrals.\foot{At intermediate steps, various unphysical operators will appear as poles, but we find these are always have zero residue. This cancellation is not serendipitous but is expected from general arguments: for example, we should not see operators which lie below the unitarity bound, and the appearance of triple-trace operators at one loop would violate the expected $1/N$ counting.} We then focus on the dDisc of the diagrams and show its relation to the internal $\hCut$ operator. The main result is that the spectral integrals will become trivial when computing the dDisc. 

For one-particle irreducible (1PI) diagrams, this algorithm amounts to factorizing the diagram into products of two tree diagrams. In the end, there are no spectral integrals remaining and the $\dDisc$ reduces to a sum over blocks with known coefficients. One-particle reducible diagrams introduce some subtleties whereupon we can also have factorization onto lower-loop and/or lower-point diagrams. In this case there may still be spectral integrals after taking the $\dDisc$. One new result is that some one-particle reducible one-loop diagrams will introduce new operator exchanges not observed at tree-level. 

We will continue to stress the analogy to the unitarity method in flat space scattering amplitudes: the $\dDisc$ localizes amplitudes onto the single-trace poles \eqref{eq:PDef} and factorizes the diagram, just as for the imaginary part of the S-matrix in flat space. The Lorentzian inversion formula then functions as a dispersion relation for the CFT data.

In the next section, we will return to the original unitarity method of \cite{Aharony:2016dwx} and make explicit contact with the results of this section. Both methods will agree exactly.  

\subsection{One-Particle Irreducible Diagrams}
\subsubsection{Bubble}
\label{subsec:AdS1Loop_Bubble}

We start with the $\phi^4$-type $s$-channel bubble diagram with non-derivative vertices. The procedure for calculating this diagram is shown below: we will use the split representation on the two internal lines to reduce the bubble to two contact diagrams sewn together.
\begin{figure}[H]
\begin{center}
\includegraphics[scale=.45]{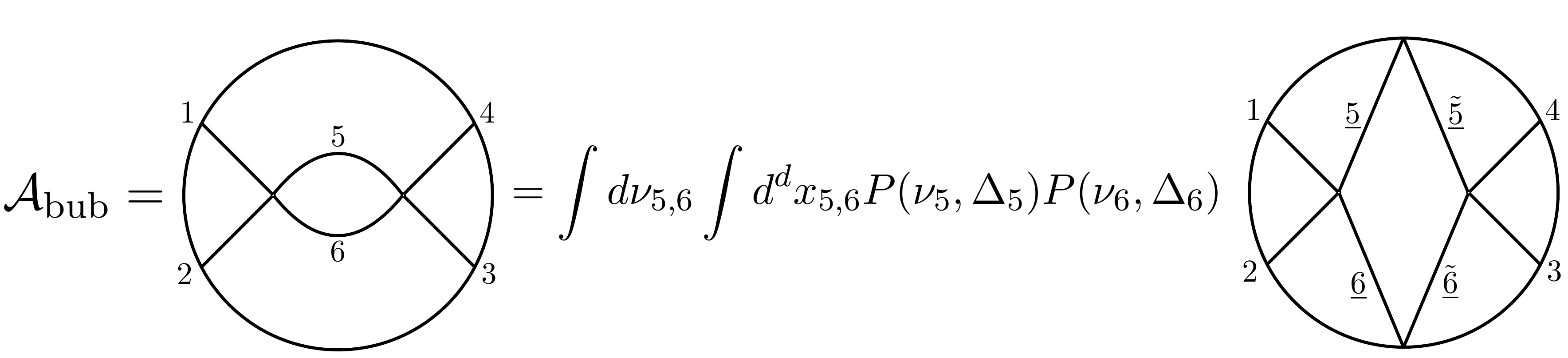}
\end{center}
\label{fig:AdSBubbleToContact}
\end{figure}
\noindent As an equation, this corresponds to
\e{}{\mathcal{A}_{\rm bub}^{1234}(x_i)= \mathcal{A}_{\rm cont}^{12\underline{6}\underline{5}}\otimes
\mathcal{A}_{\rm cont}^{\underline{\tilde{5}}\underline{\tilde{6}}34},}
where we recall the definition of $\otimes$ in \eqr{crossdef}. We can now plug in the CPW decomposition of the constituent contact diagrams. 
The boundary integrals are the CFT bubble integral \eqref{eq:Conf-Bubble} applied to CPWs, so using the identity \eqref{eq:CPWs-bubbled} we have
\begin{equation}
\mathcal{A}_{\rm bub}^{1234}(x_i)=
\int\limits_{-\infty}^{\infty} 
d\nu_5d\nu_6 
P(\nu_5,\Delta_5)P(\nu_6,\Delta_6)
\int\limits_{\frac{d}{2}}^{\frac{d}{2}+i\infty} 
\frac{d\Delta}{2\pi i}
\rho_{\rm cont}^{12\underline{6}\underline{5}}(\Delta,0)\rho_{\rm cont}^{\tilde{\underline{5}}\tilde{\underline{6}}34}(\Delta,0)B_{\Delta,0}\Psi^{1234}_{\Delta,0}(x_i). \label{eq:bubblePt1}
\end{equation}
Equivalently, the $s$-channel OPE function is
\e{}{\rho^{1234}_{\rm bub}(\D,J) = \delta_{J,0}\int\limits_{-\infty}^{\infty} 
d\nu_5d\nu_6 
P(\nu_5,\Delta_5)P(\nu_6,\Delta_6) \rho_{\rm cont}^{12\underline{6}\underline{5}}(\Delta,0)\rho_{\rm cont}^{\tilde{\underline{5}}\tilde{\underline{6}}34}(\Delta,0)B_{\Delta,0}.}
Thus, the conformal block expansion of the bubble is determined by the spectral integrals over $\nu_{5,6}$. Note that only $J=0$ double traces contribute in the $s$-channel because we are gluing two $\phi^4$ vertices together in the $s$-channel.

To understand the pole structure it is useful start with the poles in the $\Delta$ plane. The contact diagram OPE functions $\rho_{\rm cont}^{12\underline{6}\underline{5}}(\Delta,0)$ and $\rho_{\rm cont}^{\tilde{\underline{5}}\tilde{\underline{6}}34}(\Delta,0)$, given in \eqr{eq:cont_rho}, yield the poles
\begin{eqnarray}
\Delta&=&\Delta_1+\Delta_2+2n, \label{eq:bubDTExt1}
\\
\Delta&=&\Delta_3+\Delta_4+2n,  \label{eq:bubDTExt2}
\\
\Delta&=&\underline{\Delta}_{5}+\underline{\Delta}_{6}+2n, \label{eq:bubDTInt1}
\\
\Delta&=&\underline{\widetilde{\Delta}}_{5}+\underline{\widetilde{\Delta}}_{6}+2n.  \label{eq:bubDTInt2}
\end{eqnarray}
Picking up the poles (\ref{eq:bubDTExt1}) or (\ref{eq:bubDTExt2}) corresponds to doing an external vertical line cut while picking up the poles (\ref{eq:bubDTInt1}) or (\ref{eq:bubDTInt2}) corresponds to doing an internal vertical line cut. 

We now show that $\dDisc_{s}$ of this diagram is simpler than the full diagram, and equivalent to the $\hCut_{56}$ operator. As reviewed in Section \ref{sec:AdSUnitarity}, taking the $\dDisc_{s}$ removes the contribution of (\ref{eq:bubDTExt1}) and (\ref{eq:bubDTExt2}).\footnote{To avoid divergences in the principal series integral from the $\sin$ factors we should first close the $\Delta$ contour and then apply the dDisc to the conformal block sum.} That is, $\dDisc_{s}$ projects out the external vertical line cuts. Next, we will show the only other allowed exchanges are $[\O_5\O_6]_{n,\ell}$. We also show their contribution is determined by $\hCut_{56}$, or that the full contribution of this exchange comes from closing the $\nu_{5,6}$ spectral integrals on shell.

To see this, we first close the $\Delta$ contour on the poles in (\ref{eq:bubDTInt1}) which produces the conformal block $g^{1234}_{[\underline{5}\underline{6}]_{n,\ell}}(x_i)$. Turning to the remaining $\nu_{5,6}$ integrals, we are forced to close them in the lower-half plane so that the conformal blocks are exponentially suppressed. In the process, one contribution we pick up comes from the single-trace poles of $P(\nu_5,\Delta_5)$ and $P(\nu_6,\Delta_6)$: this puts the operators $\underline{\O}_5$ and $\underline{\O}_6$ on shell and leaves us with the block $g^{1234}_{[56]_{n,\ell}}(x_i)$. 

Shadow symmetry of the integrand guarantees that if we close the $\Delta$ contour on the poles in (\ref{eq:bubDTInt2}) we find the same result. In this case we must close the $\nu_{5,6}$ integrals in the upper-half-plane for convergence. The presence of these extra poles is why we defined the $\hCut_{56}$ in (\ref{eq:HatCut}) as $\Cut_{56}$ with an extra projection.

To complete the argument that $\dDisc_{s}$ truly localizes on the $[\O_5\O_6]$ family, we now rule out other poles in $\nu_{5,6}$. If we close $\Delta$ on the (\ref{eq:bubDTInt1}) pole we find a pole in $\nu_5$ from $\rho_{\rm cont}^{12\underline{6}\underline{5}}(\Delta,0)$ located at:
\begin{align}
\underline{\Delta}_{6}=\Delta_1+\Delta_2-\underline{\Delta}_{5}+2m+2n,
\end{align}
which would give a contribution $ g^{1234}_{[\underline{5}\underline{6}]_{n,\ell}}(x_i) \rar g^{1234}_{[12]_{n+m,\ell}}(x_i)$. However, dDisc$_s[g^{1234}_{[12]_{n+m,\ell}}(x_i)]=0$. We repeat this exercise for all other poles for $\nu_{5,6}$ and find they do not contribute to $\dDisc_{s}$. 

Thus, taking dDisc$_s$ does isolate the $[\O_5\O_6]_{n,\ell}$ family, and the final result is
\es{}{\dDisc_{s}(\mathcal{A}_{\rm bub}^{1234}(x_i))&=-\left(d-2\Delta_{5}\right)\left(d-2\Delta_{6}\right)\\&\times \sum\limits_{n=0}^{\infty} \hspace{-.2in} \res\limits_{\hspace{.2in}\Delta=\Delta_{[56]_{n,0}}}\left(\rho_{\rm cont}^{1265}(\Delta,0)\rho_{\rm cont}^{\tilde{5}\tilde{6}34}(\Delta,0)
B_{\Delta,0}K^{\Delta_3,\Delta_{4}}_{\widetilde{\Delta},0}\dDisc_{s}(g^{1234}_{\Delta,0}(x_i))\right)~.}
Recalling that $\dDisc_{s}(g^{1234}_{\Delta,0}(x_i))$ simply multiplies $g^{1234}_{\Delta,0}(x_i)$ by a prefactor (cf. \eqr{dDisctch}) which is the same for all $[\O_5\O_6]_{n,0}$ blocks, we have proven \eqr{ddcutintro} for the bubble:
\e{}{{\dDisc_s(\mathcal{A}_{\rm bub}^{1234}) = 2\sin\left({\pi\o2}(\D_5+\D_6-\D_1-\D_2)\right)\sin\left({\pi\o2}(\D_5+\D_6-\D_3-\D_4)\right)\hCut_{56}[\mathcal{A}_{\rm bub}^{1234}] }.}

The simplification of the $\dDisc$ of the diagram is analogous to simplifications from unitarity cuts of flat space scattering amplitudes. In the latter, internal cuts of a one-loop diagram -- i.e. calculating its imaginary piece -- are simple: they put the internal legs on shell and factorize the amplitude onto tree-level pieces. Likewise, as shown above for the AdS bubble, this is exactly what dDisc has done. For the AdS bubble, just as for S-matrices, it is more difficult to determine the real part of the amplitude, which involves doing external line cuts.\foot{Suppose we were to consider an external line cut, e.g. closing on the poles (\ref{eq:bubDTExt1}) or (\ref{eq:bubDTExt2}). In this case we have to perform the $\nu_{5,6}$ integrals to determine the correct OPE coefficient, but now an infinite number of poles contribute.}

\sssec*{\it Identical operators}
It may also be useful to understand this phenomenon of dDisc simplification when all operators are taken to be identical, i.e. $\O_1=\O_2=...=\O_6=\O$. In this case we first close $\underline{\Delta}_{5,6}$ on the single trace located at $\Delta_{\O}$. We now have a triple pole at $\Delta=2\Delta_{\O}$: from (\ref{eq:bubblePt1}), a double pole comes from $\rho^{12\underline{6}\underline{5}}_{\rm cont}(\Delta,J)$ and a single pole comes from $\rho^{\underline{\tilde{5}}\underline{\tilde{6}}34}_{\rm cont}(\Delta,J)$. After closing the $\Delta$ contour this triple pole gives us a $\ln^{2}(z\bar{z})$ divergence, which is proportional to the squared anomalous dimension of the $[\O\O]_{n,\ell}$ states \cite{Aharony:2016dwx}:
\begin{align}
\mathcal{A}^{1234}_{\rm bub}(x_i)\big|_{\ln^2(z\bar{z})}=\frac{1}{2}T_{s}(x_i)\sum\limits_{n,\ell} p^{(0)}_{n,\ell}(\gamma_{n,\ell}^{(1)})^{2}g_{n,\ell}(z,\bar{z}).
\end{align}
The $\dDisc_{s}$ projects onto this term because $\dDisc_s(\ln(z\zb)g_{n,\ell}(z,\zb))=\dDisc_s(g_{n,\ell}(z,\zb))=0$. So at the level of the spectral representation (\ref{eq:bubblePt1}) only the triple pole can contribute to $\dDisc_{s}$. Because this triple pole only appears when $\underline{\Delta}_{5,6}$ are put on shell, this is another way to see the simplifying effect of dDisc.

\subsubsection{Triangle}
Next we will turn to the triangle diagram in a $\phi^{3}+\phi^{4}$ theory, drawn below.
\begin{center}
\includegraphics[scale=.28]{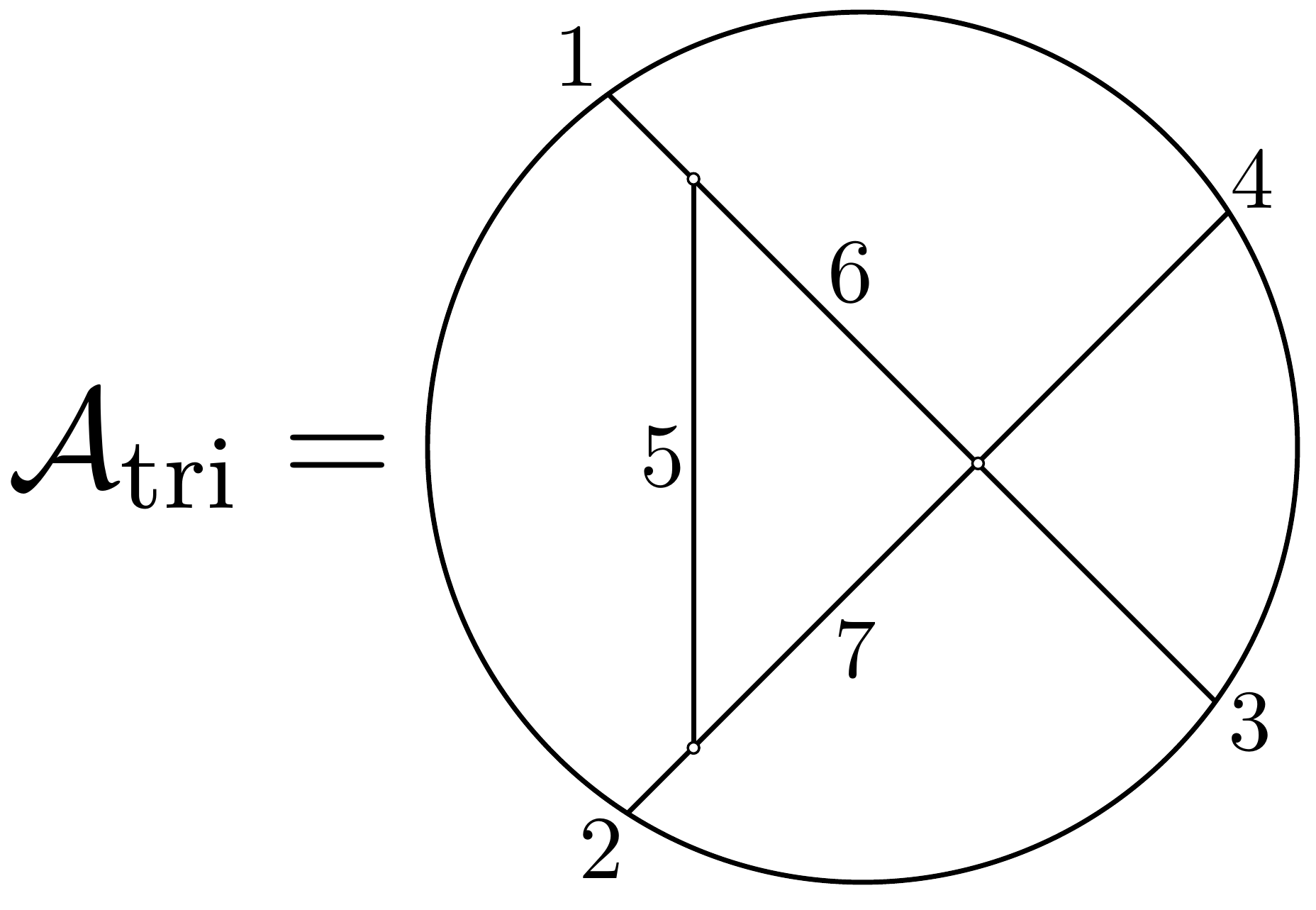}
\label{AdStri}
\end{center}
This will introduce the 6j symbol into the calculations, but otherwise will proceed similarly to the bubble diagram. 

When we use the split representation for lines 6 and 7, the triangle diagram becomes a $t$-channel tree exchange diagram glued in the $s$-channel to a contact diagram.
\e{Atriint}{\mathcal{A}_{\rm tri}^{1234}(x_i)= \mathcal{A}^{\underline{7}21\underline{6}}_{5, \rm exch}\,\otimes\,\mathcal{A}_{\rm cont}^{\tilde{\underline{6}}\tilde{\underline{7}}34}.}
Our end goal is to obtain the block decomposition in a single channel. Since we are gluing the two diagrams together along $x_{6,7}$, it is simplest to calculate the $s$-channel $\O_1\O_2\ \rightarrow \ \O_3\O_4$ decomposition, as this will allow us to use the CFT bubble identity. 

Recall from the discussion surrounding \eqr{eq:TreeExchangeSV0} that using crossing on the tree-level exchange diagram\foot{To recap the steps: use crossing to go from the $\O_6\O_1\ \rightarrow \ \O_2\O_7$ channel to the $\O_1\O_2 \ \rightarrow\  \O_6\O_7$ channel, extend the contour so the CPW is replaced by the block, and then close the $\Delta_{\O'}$ contour to pick up the $\Delta_{5}$ pole.} comprising the left ``half'' of the triangle diagram -- the $\mathcal{A}^{\underline{7}21\underline{6}}_{5, \rm exch}(x_{6},x_1,x_2,x_7)$ factor in \eqr{Atriint} -- introduces a 6j symbol. This leads to \eqr{exchcross} with the relabeling $\mathcal{A}^{3214}_{\O, \rm exch} \rar \mathcal{A}^{\underline{7}21\underline{6}}_{\O_5, \rm exch}$. 
%
%
Plugging that into the triangle diagram and using the bubble identity the same way we did in Section \ref{subsec:AdS1Loop_Bubble}, we have
\begin{align}\label{trifull}
\mathcal{A}^{1234}_{\rm tri}(x_i)&=-\int\limits_{-\infty}^{\infty}d\nu_6d\nu_7\int\limits_{\frac{d}{2}-i\infty}^{\frac{d}{2}+i\infty}\frac{d\Delta_\O}{2\pi i} P(\nu_6,\Delta_6)P(\nu_7,\Delta_7)
\\
 & \quad \res_{\Delta_{\O'}=\Delta_{5}}\rho^{\underline{6}12\underline{7}}_{5, \rm exch}(\Delta_{\O'},0)
\rho^{\tilde{\underline{6}} \tilde{\underline{7}}34}_{\rm cont}(\Delta_\O,0)
\nonumber 
K^{1\underline{6}}_{\tilde{5}}
\sixjBlock{1}{2}{\underline{7}}{\underline{6}}{\O}{5}\frac{B_{\O}}{n_{\O}}K^{34}_{\widetilde{\O}}\,g^{1234}_{\O}(x_i).
\end{align}
Again, because of the $\phi^4$ vertex in the triangle diagram, the blocks only have support for $J=0$. 

We now analyze the pole structure. In the $\Delta$ plane we have poles at the following locations:
\begin{eqnarray}
\Delta&=&\Delta_1+\Delta_2+2n, \label{eq:triPolesExt1}
\\
\Delta&=&\Delta_3+\Delta_4+2n, \label{eq:triPolesExt2}
\\
\Delta&=&\underline{\Delta}_{6}+\underline{\Delta}_7+2n, \label{eq:triPolesInt1}
\\
\Delta&=&\widetilde{\underline{\Delta}}_{6}+\underline{\Delta}_7+2n, \label{eq:triPolesInt2}
\\
\Delta&=&\underline{\Delta}_{6}+\widetilde{\underline{\Delta}}_{7}+2n, \label{eq:triPolesInt3}
\\
\Delta&=&\widetilde{\underline{\Delta}}_{6}+\widetilde{\underline{\Delta}}_{7}+2n. \label{eq:triPolesInt4}
\end{eqnarray}
The contact diagram OPE function gives us the poles (\ref{eq:triPolesExt2}) and (\ref{eq:triPolesInt4}), while the inversion of the exchange graph gives us the poles (\ref{eq:triPolesExt1}), (\ref{eq:triPolesInt1}), (\ref{eq:triPolesInt2}), and (\ref{eq:triPolesInt3}).

As with the bubble diagram, evaluating the triangle in full involves performing infinite sums over poles in $\nu_6, \nu_7$ to obtain the full OPE function, and so is somewhat involved. Once again, however, the $s$-channel dDisc is simpler. Taking $\dDisc_{s}$ removes the external double-trace poles (\ref{eq:triPolesExt1}) and (\ref{eq:triPolesExt2}), leaving us with the states corresponding to vertical internal line cuts. By shadow symmetry, all poles (\ref{eq:triPolesInt1})-(\ref{eq:triPolesInt4}) give the same contribution so we can focus on (\ref{eq:triPolesInt1}).\foot{See the analogous discussion for the bubble. For example, if we pick up the (\ref{eq:triPolesInt1}) pole we must close the $\nu_6, \nu_7$ in the lower half plane, for (\ref{eq:triPolesInt2}) we must close $\nu_6$ in the upper half plane and $\nu_7$ in the lower half plane, and so on.} As with the bubble, if we pick up this pole, close $\nu_{6,7}$ in the lower-half plane, and focus on the term with a non-zero $\dDisc_{s}$, then we only pick up the single-trace poles from the $P(\nu,\Delta)$ factors. Putting everything together,
\begin{align}
\dDisc_{s}(\mathcal{A}^{1234}_{\rm tri}(x_i))&= \left(d-2\Delta_{6}\right)\left(d-2\Delta_{7}\right)K^{16}_{5} \res_{\Delta_{\O'}=\Delta_{5}}\rho^{6127}_{5, \rm exch}(\Delta_{\O'},0) \\&\times\sum\limits_{n=0}^{\infty}\hspace{-0.2 in}\res\limits_{\hspace{.2in}\Delta_\O=\Delta_{[67]_{n,0}}}\rho^{\tilde{6}\tilde{7}34}_{\rm cont}(\Delta_\O,0)\sixjBlock{1}{2}{7}{6}{\O}{5}{\frac{B_{\O}}{n_{\O}}}K^{34}_{\widetilde{\O}}\dDisc_{s}(g^{1234}_{\O}(x_i)).\nonumber
\end{align}
This expression involves various factors, but the punchline is familiar. Only the sum over the $[\O_6\O_7]_{n,0}$ family is picked out by dDisc$_s$, and this is proportional to taking a vertical internal line cut:
\e{}{{\dDisc_s(\mathcal{A}^{1234}_{\rm tri}) = 2\sin\left({\pi\o2}(\D_6+\D_7-\D_1-\D_2)\right)\sin\left({\pi\o2}(\D_6+\D_7-\D_3-\D_4)\right)\hCut_{67}[\mathcal{A}^{1234}_{\rm tri}] }}
Unlike the full diagram \eqr{trifull}, we do not have any spectral integrals here: $\dDisc_{s}$ has put the internal states on shell. 


\sssec{Box}
\label{sec:AdSBox}

The AdS box diagram has many interesting features which the other one-loop diagrams do not. 
\begin{itemize}
\item It is the only one-loop diagram which is non-zero\foot{It can also be shown explicitly using the lightcone bootstrap \cite{Fitzpatrick:2012yx,Komargodski:2012ek} that the $\qDisc$ of the box is non-zero, see \cite{dsdi}, Appendix C. In the lightcone limit $z\ll1-\bar{z}\ll1$ the $\qDisc$ of the AdS box reduces to the $\qDisc$ of the ``large spin box diagram'' in \cite{dsdi}, so the calculations in this limit are equivalent.} when we take a simultaneous $\dDisc_s$ and $\dDisc_{t}$, also known as the $\qDisc$ \cite{Caron-Huot:2017vep}:
\es{}{\qDisc(\mathcal{A}_{\rm box}(x_i)) &\equiv \dDisc_{s} \dDisc_{t}(\mathcal{A}_{\rm box}(x_i)) \neq 0}
whereas\footnote{The $s$-channel bubble and triangle diagrams have zero dDisc$_t$. If this were not the case, the inversion formula would imply that their $s$-channel expansion contains operators of unbounded spin, contradicting our direct computation. The unitarity cuts for the analogous scattering amplitudes are also zero for the same underlying reason: these diagrams have no internal $t$-channel cuts.
}
\e{}{\qDisc(\mathcal{A}_{\rm tri}(x_i)) =
\qDisc(\mathcal{A}_{\rm bub}(x_i)) =0.}
\item In Mellin space, the box diagram has simultaneous poles in the two independent Mellin variables $s$ and $t$. This is equivalent to having a nonzero $\qDisc$. 

\item Ignoring the operator labels, the graph is more symmetric and has an equally natural decomposition in both the $s$ and $t$-channel. 

\item The OPE decomposition involves double-trace operators of unbounded spin in both the $s$ and $t$-channels. In general, the Lorentzian inversion formula implies that a nonzero $\qDisc$ means that the OPE data is analytic in spin in both channels. Diagrammatically, this is because the box has both internal horizontal and vertical cuts.
 
\end{itemize}

\noindent The box diagram we will study is
\begin{center}
\includegraphics[scale=.28]{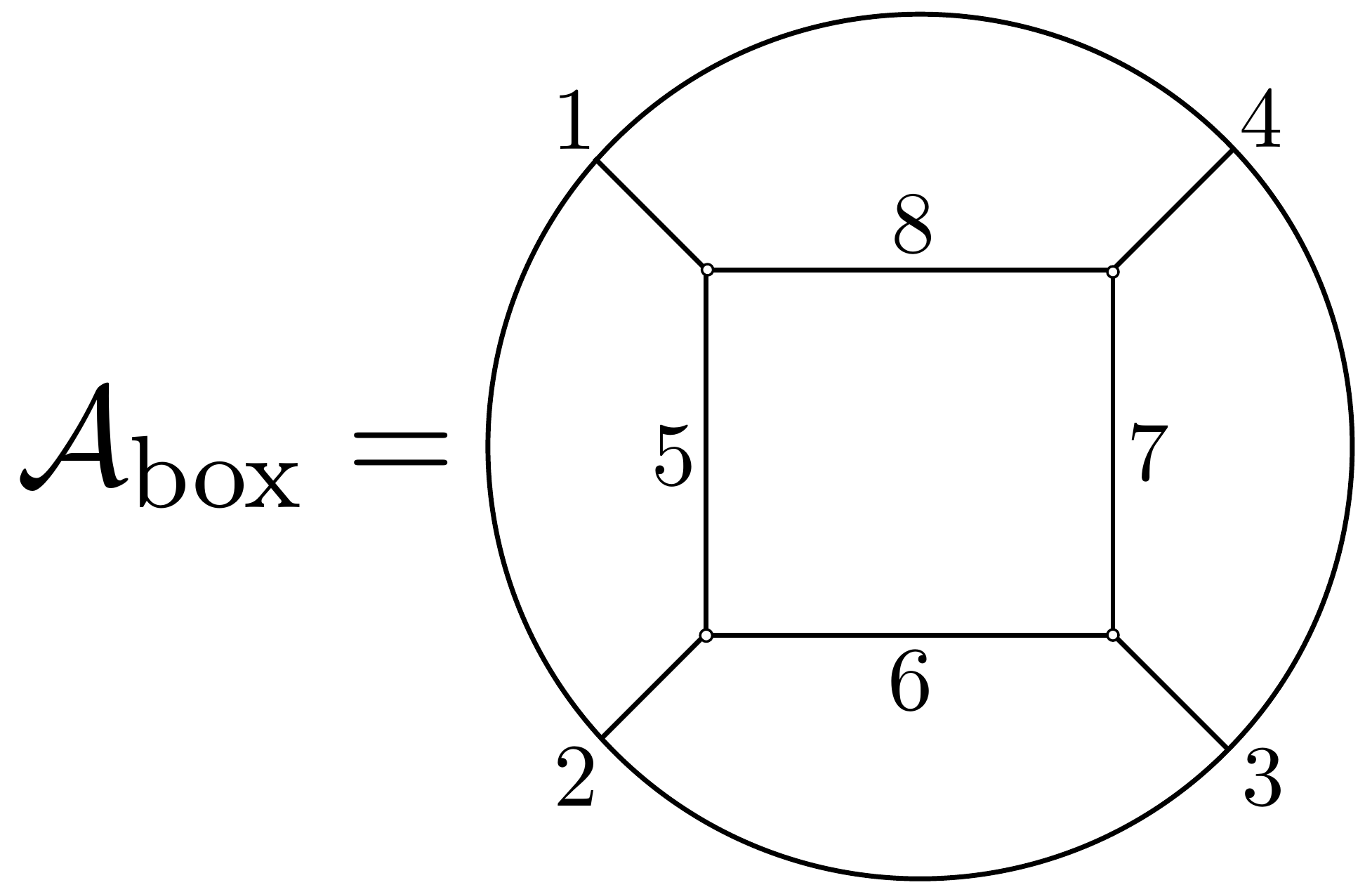}
\end{center}
We will start with the $s$-channel $\O_1\O_2 \ \rightarrow \ \O_3\O_4$ decomposition without loss of generality. We use our bag of tricks -- the split representation, crossing, and the CFT bubble identity -- to obtain the box OPE function. Using the split representation for the $6$ and $8$ legs, the box can be written as two trees sewn together,
\e{AdSBoxGluing}{\mathcal{A}_{\bx}(x_i)= \mathcal{A}^{\underline{6}21\underline{\tilde{8}}}_{5, \rm exch}\otimes \mathcal{A}_{7,\rm exch}^{3\underline{\tilde{6}}\underline{8}4}.}
%
Using crossing on each exchange diagram (cf. \eqr{eq:TreeExchangeSV0}) we find $s$-channel CPWs glued together. Next we use the CFT bubble identity \eqr{eq:CPWs-bubbled} to do the boundary integrals and obtain the $s$-channel decomposition written in terms of the OPE data of the two constituent trees
\begin{align}
\mathcal{A}_{\bx}(x_i)=&
(d-2\Delta_5)(d-2\Delta_7)
\left(\int\limits_{-\infty}^{\infty}\prod\limits_{i=6,8}
d\nu_i P(\nu_i,\Delta_i)\right)
\res_{\Delta_{\O'}=\Delta_{\tilde{5}}} \rho_{\tilde{5}, \rm exch}^{\tilde{8}126}(\Delta_{\O'},J) 
\res_{\Delta_{\O'}=\Delta_{7}}  \rho_{7, \rm exch}^{48\tilde{6}3}(\Delta_{\O'},J)
\nonumber 
\\ 
&\sum_{J_{\O}=0}^{\infty}\int\limits_{\frac{d}{2}-i\infty}^{\frac{d}{2}+i\infty}\frac{d\Delta_{\O}}{2\pi i} K^{2\underline{6}}_{\tilde{5}}K^{4\underline{8}}_{\tilde{7}}\sixjBlock{2}{1}{\tilde{\underline{8}}}{\underline{6}}{\O}{5}\sixjBlock{\underline{8}}{\tilde{\underline{6}}}{3}{4}{\O}{7}\frac{B_{\O}}{n_{\O}^{2}}K^{34}_{\widetilde{\O}}g^{1234}_{\O}(x_i).
\label{eq:BoxEqualsTreeSquared}
\end{align}
By way of orientation, note the similarity to the analogous expression \eqr{trifull} for the triangle, only we have had to use crossing on two trees, not just one, and hence contact data is now replaced by exchange data. 

Now to the main point, that the block decomposition of the box makes the three classes of cuts manifest: the external line cuts for $\Delta_{\O}=\Delta_1+\Delta_2+2n+J$ and $\Delta_{\O}=\Delta_3+\Delta_4+2n+J$, as well as the internal line cuts for $\Delta=\Delta_6+\Delta_8+2n+J$. In this case, the poles in $\Delta$ all come from inverting individual blocks \eqref{eqn:InvBlockSym} and therefore $J$ is nonzero. Once again, taking the $\dDisc_s$ places all spectral integrals on shell. We quote the final result of applying our algorithm, which is by now hopefully familiar:
\begin{align}
\dDisc_{s}(\mathcal{A}_{\bx}(x_i))=\sum\limits_{n,\ell}\hspace{-.2in}&\res\limits_{\hspace{.2in}\Delta_\O=\Delta_{[68]_{n,\ell}}}\prod\limits_{i=5}^{8}(d-2\Delta_i)b_{15\tilde{8}}b_{2\tilde{5}6}b_{3\tilde{6}7}b_{4\tilde{7}8}
\nonumber \\ &K^{26}_{\tilde{5}}K^{48}_{\tilde{7}}\sixjBlock{2}{1}{\tilde{8}}{6}{\O}{5}\sixjBlock{8}{\tilde{6}}{3}{4}{\O}{7}\frac{B_{\O}}{n_{\O}^{2}}K^{34}_{\widetilde{\O}}\,\dDisc_{s}(g^{1234}_{\O}(x_i)).
\label{eq:dDisc_box}
\end{align} 
As anticipated, dDisc$_s$ isolates the family specified by the internal vertical line cut that puts the 6 and 8 lines on shell:
\e{}{{\dDisc_s(\mathcal{A}_{\rm box}) = 2\sin\left({\pi\o2}(\D_6+\D_8-\D_1-\D_2)\right)\sin\left({\pi\o2}(\D_6+\D_8-\D_3-\D_4)\right)\hCut_{68}[\mathcal{A}_{\rm box}] }.}
Analogous expressions can be read off in the $t$-channel by relabeling. Note that conformal blocks $g^{1234}_{\O}(x_i)$ for all spins appear explicitly in both channels. 

In Appendix \ref{boxmellin} we make some comments about the box amplitude in Mellin space.

\sssec*{\it qDisc and the pentagon identity for 6j symbols}
We can also study the quadruple discontinuity \cite{Caron-Huot:2017vep} -- the ``qDisc'' -- which is the composition of two discontinuities in different channels. Commutativity of dDisc$_s$ and dDisc$_t$ implies a non-trivial relation between infinite sums of conformal blocks. Taking dDisc$_t$ of \eqr{eq:dDisc_box} and equating it with the reverse order, we note that all of the AdS kinematic factors in the above expression cancel out to yield 
\begin{align}
&\dDisc_{t}\bigg[\sum\limits_{n,\ell}\hspace{-.2in}\res\limits_{\hspace{.2in}\Delta_{O}=\Delta_{[68]_{n,\ell}}}K^{26}_{\tilde{5}}K^{48}_{\tilde{7}}K^{34}_{\widetilde{\O}}\sixjBlock{1}{2}{6}{\tilde{8}}{\O}{5}\sixjBlock{8}{\tilde{6}}{3}{4}{\O}{7}
\frac{B_{\O}}{n_{\O}^{2}}\dDisc_{s}(g^{1234}_{\O}(x_i))\bigg]
\nonumber\\
=~
&\dDisc_{s}\bigg[\sum\limits_{n',\ell'}\hspace{-.2in}\res\limits_{\hspace{.2in}\Delta_{\O'}=\Delta_{[57]_{n',\ell'}}}K^{37}_{\tilde{6}}K^{15}_{\tilde{8}}K^{14}_{\widetilde{\O}'}\sixjBlock{3}{2}{\tilde{5}}{7}{\O'}{6}\sixjBlock{5}{\tilde{7}}{4}{1}{\O'}{8}
\frac{B_{\O'}}{n_{\O'}^{2}}\dDisc_{t}(g^{3214}_{\O'}(x_i))\bigg]. \label{eq:qdiscboxV2}
\end{align}
In fact, this identity, and more generally crossing symmetry for the AdS box, follows from the pentagon identity of the conformal group $SO(d+1,1)$ \cite{Gadde:2017sjg}, given in \eqr{eq:CFTPentV2}. We explain this in more detail in Appendix \ref{app:Pentagon}.


\subsection{One-Particle Reducible Diagrams}
\label{sec:non1PI}
So far we have studied 1PI diagrams, but the full set of one-loop diagrams includes one-particle {\it reducible} diagrams. The upshot here is: {\it vertex corrections and corrections to bulk-to-bulk propagators induce new double-trace poles not present at tree-level, while loop corrections to bulk-to-boundary propagators do not.}

The presence of new poles in some diagrams is, perhaps, surprising. One may expect that the one-particle reducible diagrams, like propagator and vertex corrections, are simply proportional to tree-level quantities. One hint that this is wrong is that there can be regions of bulk integration where the diagrams degenerate to the 1PI diagrams studied in the previous section (e.g. when a bulk-to-bulk propagator shrinks to zero size), and these {\it do} have double-trace poles not present at tree-level.\footnote{Another way to see how these diagrams can degenerate to 1PI diagrams is to let an exchanged field have an asymptotically large mass and replace the propagator a contact interaction.} The question becomes whether these degenerations do contribute to the final result. We will confirm this on general grounds below. 

\subsubsection{Mass Corrections in $\f^3$ theory}
In AdS diagrams there are two types of propagator corrections, corresponding to mass renormalization of bulk-to-boundary or bulk-to-bulk propagators. 

\begin{figure}[t]
\begin{center}
\includegraphics[scale=.35]{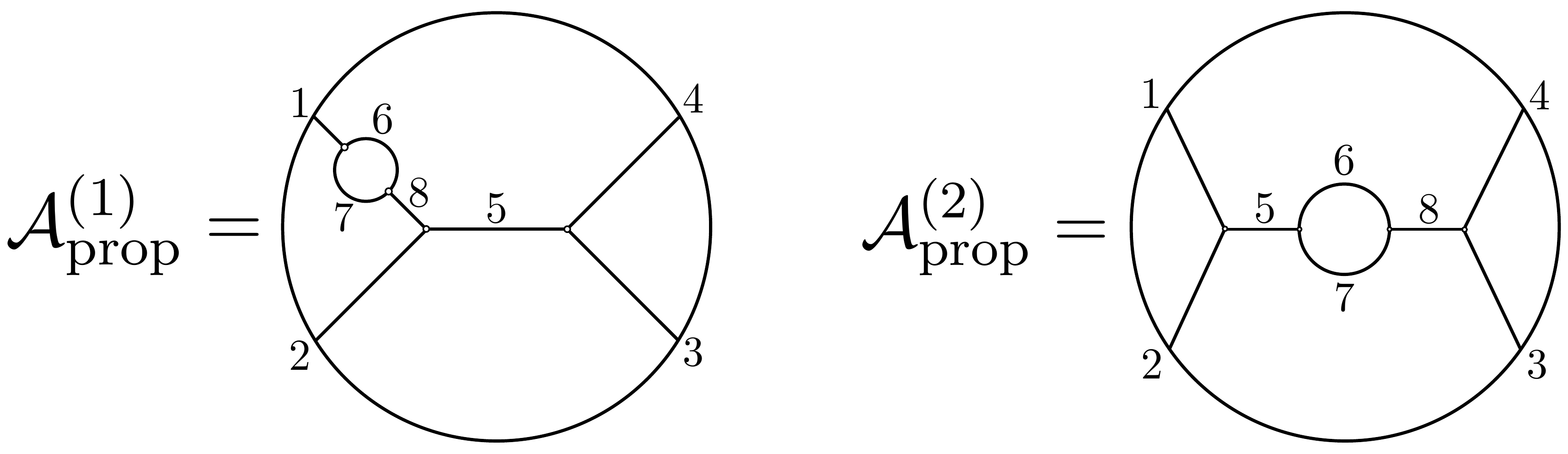}
\end{center}
\caption{One-loop corrections $\f^{3}$ theory. The Witten diagrams on the left and right correspond to renormalization of the bulk-to-boundary and bulk-to-bulk propagator respectively.}
\label{fig:AdS_Bubble_BdyBulk}
\end{figure}

Consider first a one-loop correction to the $\f^3$ tree diagram with a loop on a bulk-to-boundary propagator, shown as $\cA_{\rm prop}^{(1)}$ in the first diagram of figure \ref{fig:AdS_Bubble_BdyBulk}. 
We will demonstrate that this diagram is proportional to the tree-level exchange diagram. To prove this, it is sufficient to study the loop-corrected bulk-to-boundary propagator alone, which we call $K_{\D_1}^\1$. To calculate it, we will use the split representation for all three bulk-to-bulk propagators. We then use the bubble integral \eqref{eq:Conf-Bubble} to obtain a result proportional to a single bulk-to-boundary propagator:
\begin{align}
K^\1_{\Delta_1}(x_1,y_1)&=\int\limits_{\rm AdS} d^{d+1}y_{2}d^{d+1}y_{3}K_{\Delta_{1}}(x_1,y_2)G_{\Delta_6}(y_2,y_3)G_{\Delta_7}(y_2,y_3)G_{\Delta_{8}}(y_3,y_1)
\nonumber \\&\propto\left(\int\limits_{-\infty}^{\infty}d\nu_6d\nu_7P\left(\nu_6,\Delta_6\right)P(\nu_7,\Delta_7)P\left(\frac{i}{2}(d-2\Delta_1),\Delta_8\right)b_{1\underline{6}\underline{7}}b_{\tilde{\underline{6}}\tilde{\underline{7}}\tilde{8}}B_{\O_1}\right)K_{\Delta_1}(x_1,y_1).  \label{eqn:1lpBulkBdy1}
\end{align}
We have used that conformal symmetry implies $\int_{\partial \rm AdS} d^d x' \braket{\mathcal{O}(x) \mathcal{O} (x') } K_{\widetilde{\Delta}}(x',y) \propto K_{\Delta}(x,y)$ \cite{Costa:2014kfa}.\footnote{We thank Ant\'onio Antunes for discussions on this point.} The result is proportional to $K_{\Delta_1}(x_1,y_1)$: (\ref{eqn:1lpBulkBdy1}) involves two spectral integrals, but they only produce an overall proportionality constant. Therefore, this diagram simply renormalizes the tree-level exchange, i.e. there are no cuts associated to $\O_{6}$, $\O_{7}$, or $\O_{8}$. Note that if we set $\Delta_8 = \Delta_1$, there is a divergence from the third $P$ factor. In this case the divergence needs to be regularized to extract one-loop corrections to $\Delta_{1}$ \cite{Giombi:2017hpr}.

We next study the exchange diagram where the bulk-to-bulk propagator is renormalized, $\cA^{(2)}_{\text{prop} }$. By splitting the two bulk-to-bulk propagators in the bubble we can reduce this diagram to the gluing of two $s$-channel tree-level exchange diagrams,
\e{}{\cA^{(2)}_{\text{prop}  }= \cA_{5,\, \rm exch}^{12\underline{6}\underline{7}}\,\otimes \,\cA_{8,\, \rm exch}^{\underline{\t 6}\underline{\t 7}34}.}
Using the CFT bubble identity, we find the OPE function takes a simple form 
\begin{align}
\rho^{1234,\,(2)}_{\text{prop}}(\underline{\D}_5,J_5) = 2\int\limits_{-\infty}^{\infty} d\nu_5d\nu_6d\nu_7&P(\nu_5,\Delta_5)P(\nu_6,\Delta_6)P(\nu_7,\Delta_7)P(\nu_5,\Delta_8)
b_{12\underline{5}}b_{\tilde{\underline{5}}\underline{6}\underline{7}}b_{\tilde{\underline{6}}\tilde{\underline{7}}\underline{5}}b_{\tilde{\underline{5}}34}K^{34}_{\underline{\t5}}B_{\underline{\D}_{5},0}.
\end{align}
In contrast with the previous case, the loop correction to the propagator here introduces new poles in $\nu_5$, and therefore new physical states: taking $\O_5=\O_8$ so that the loop renormalizes a bulk-to-bulk propagator, the diagram has new $[\O_6\O_7]_{n,\ell}$ poles in addition to the $[\O_1\O_2]_{n,\ell}$, $[\O_3\O_4]_{n,\ell}$ and $\O_5$ poles that are present for the tree-level exchange. We will discuss one-loop single-trace poles in more detail in the next subsection.

\subsubsection{Vertex Correction in $\f^3+\f^4$ theory}
\label{sec:Vertex}

Now we turn to vertex corrections. We focus on a bubble vertex correction in $\f^3+\f^4$ theory for concreteness, but the conclusion will be the same for the vertex correction arising in $\f^3$ theory. We label the diagram as
\begin{figure}[h!]
\begin{center}
\includegraphics[scale=.3]{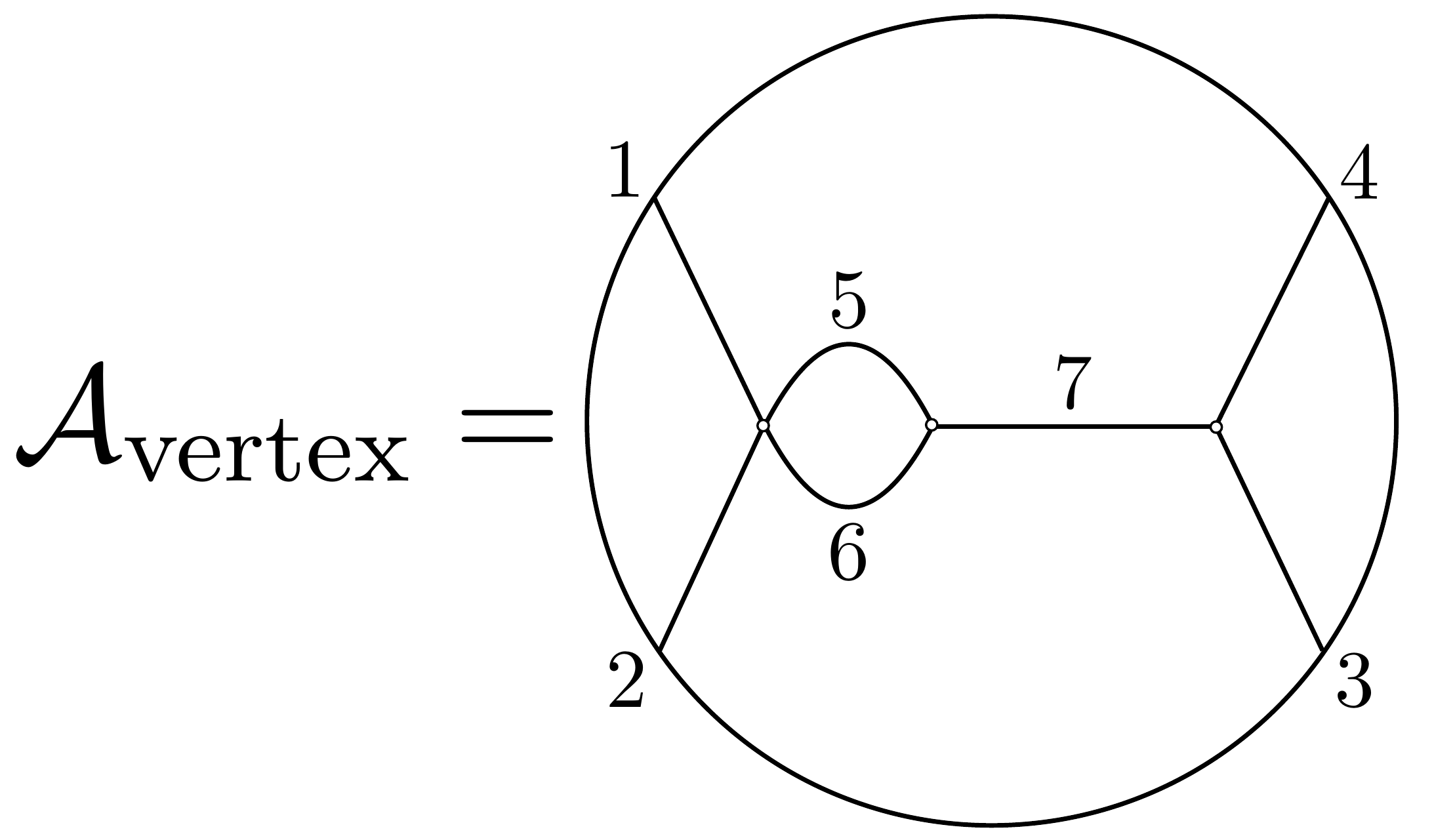}
\end{center}
\label{fig:Witten_Vertex_Correction}
\end{figure}

%
\noindent This diagram is the off-shell gluing
\e{vertexglue}{\cA_{\rm vertex} = \cA_{\rm cont}^{12\underline{5}\underline{6}}\,\otimes \,\cA_{7 ,\,\rm exch}^{\underline{\t 5}\underline{\t 6}34}.}
Using the split representation for the $5$ and $6$ legs yields the $s$-channel OPE function 
\e{}{\rho_{\rm vertex}^{1234}(\D,J) = \delta_{J,0}\int\limits_{-\infty}^{\infty}d\nu_{5}d\nu_{6}P(\nu_5,\Delta_5)P(\nu_6,\Delta_6)\rho^{12\underline{6}\underline{5}}_{\rm cont}(\Delta,0)\rho^{\tilde{\underline{5}}\tilde{\underline{6}}34}_{\rm exch,7}(\Delta,0)B_{\Delta,0}.}
From our previous computations of $\rho^{12\underline{6}\underline{5}}_{\rm cont}(\Delta,0)$ and $\rho^{\tilde{\underline{5}}\tilde{\underline{6}}34}_{\rm exch,7}(\Delta,0)$ in \eqr{eq:cont_rho} and \eqr{rhoexch}, respectively, we have, in addition to the usual external double trace poles, three families of poles in $\D$ corresponding to internal line cuts:
\begin{eqnarray}
\Delta&=&\Delta_7, \label{eq:vertexCPWPole1}
\\
\Delta&=&\underline{\Delta}_{5}+\underline{\Delta}_{6}+2n, \label{eq:vertexCPWPole2}
\\
\Delta&=&\underline{\widetilde{\Delta}}_{5}+\underline{\widetilde{\Delta}}_{6}+2n. \label{eq:vertexCPWPole3}
\end{eqnarray}
In the language of cutting diagrams, all vertical line cuts in this one-particle reducible graph are allowed. 

Using the split representation for the $5$ and $6$ propagators has made it manifest that the double-trace operators $[\O_{5}\O_{6}]_{n,0}$ will appear in the CPW decomposition.\foot{The presence of these poles, and explicit computation of their residues for specific values of $d$ and $\D_i$, was found in \c{EllisPriv} by brute force calculation using the methods of \c{Yuan:2018qva}. We thank Ellis Ye-Yuan for early conversations that alerted us to these poles.} These are analogous to the $[\O_6\O_7]_{n,\ell}$ poles appearing in the mass renormalization diagram of the previous subsection. The contribution of the $[\O_{5}\O_{6}]_{n,0}$ operators is found by closing the $\Delta$ contour on \eqref{eq:vertexCPWPole2} and localizing the spectral integrals on their single-trace poles, with a similar procedure for the \eqref{eq:vertexCPWPole3} poles.

What about the $\O_7$ pole? It is helpful for interpretation to study this diagram by first splitting the $\O_7$ line: in this approach we can phrase the result in terms of the one-loop correction to the cubic vertex, lining up nicely with expectations from standard QFT. Starting from \eqr{vertexglue}, if we split the $\O_7$ rung we find
\begin{align}
\mathcal{A}_{\rm vertex}(x_i)&=\int\limits_{\partial \rm AdS}d^d x_{7}\int\limits_{-\infty}^{\infty}d\nu_{7}P(\nu_7,\Delta_7)\mathcal{A}^{12\underline{7};\,56}_\1(x_1,x_2,x_7)\mathcal{A}^{\tilde{\underline{7}}34}_{\rm tree}(x_7,x_3,x_4),
\end{align}
where the subdiagrams are
\begin{align}
\mathcal{A}^{127;\,56}_{\1}(x_1,x_2,x_7)&\equiv\int\limits_{\rm AdS}d^{d+1}y_{1}d^{d+1}y_{2}K_{\Delta_1}(x_1,y_1)K_{\Delta_2}(x_2,y_1)G_{\Delta_5}(y_1,y_2)G_{\Delta_{6}}(y_1,y_2)K_{\Delta_7}(x_7,y_2),
\\
\mathcal{A}^{734}_{\rm tree}(x_7,x_3,x_4)&\equiv\int\limits_{\rm AdS}d^{d+1}y_1 K_{\Delta_7}(x_7,y_1)K_{\Delta_3}(x_3,y_1)K_{\Delta_4}(x_4,y_1).
\end{align}
Clearly, $\mathcal{A}^{127;\,56}_{\1}(x_i)$ contains the one-loop correction to the OPE coefficient $C_{127}$ due to a mixed cubic and quartic diagram, while $\mathcal{A}^{734}_{\rm tree}(x_i)$ is the tree-level contribution to $C_{734}$. By conformal symmetry, these are both proportional to CFT three-point structures,
\begin{align}
\mathcal{A}^{127;\,56}_{\1}(x_1,x_2,x_7)&=C_{\1}^{127;\,56}\<\O_1\O_2\O_7\>,\label{3pt1loop}
\\
\mathcal{A}^{734}_{\rm tree}(x_7,x_3,x_4)&=b_{734}\<\O_7\O_3\O_4\>,
\end{align}
where $C_{\1}^{127;\,56}$ is defined by conformally pairing \eqr{3pt1loop} with its shadow \c{Liu:2018jhs},
\e{}{C_{\1}^{127;\,56}\equiv{\left(\mathcal{A}^{127;\,56}_{\1}(x_1,x_2,x_7),\la \widetilde\O_1\widetilde\O_2\widetilde\O_7\ra\right)_{E}\o \left(\la \O_1\O_2\O_7\ra, \la \widetilde\O_1\widetilde\O_2\widetilde\O_7\ra\right)_{E}}.}
The definition of the three-point pairing \cite{Kravchuk:2018htv,Karateev:2018oml} is given in \eqref{eq:ThreePointEuclidean}.
This gives the same CPW decomposition as before, but written in a different way:
\begin{align}
\mathcal{A}_{\rm vertex}(x_i)&=\int\limits_{\partial \rm AdS}d^d x_{7}\int\limits_{-\infty}^{\infty}d\nu_{7}P(\nu_7,\Delta_7)\,C_{\1}^{127;\,56}\,b_{\tilde{\underline{7}}34}\Psi^{1234}_{\underline{\Delta}_7,0}(x_i).
\label{eq:vertex7Split}
\end{align}
The diagram has the form of the tree-level exchange graph for $\O_{7}$ exchange, only with one of the kinematic, tree-level, three-point $b$ factors replaced by the {\it one-loop} factor $C_{\1}$, i.e. the one-loop correction to the cubic coupling. Moreover, the function $C_{\1}$ contains the poles at $\Delta_7=\Delta_5+\Delta_6+2n$ we found earlier: by splitting the loop in $\mathcal{A}^{127;\,56}_{\1}(x_1,x_2,x_7)$, the spectral representation of $C_{\1}^{127;\,56}$ is 
\begin{align}
C_{\1}^{127;\,56} &= \int\limits_{-\infty}^{\infty} d\nu_{5}d\nu_{6}\int\limits_{\partial \rm AdS}d^d x_5d^dx_6 \, b_{\tilde{\underline{5}}\tilde{\underline{6}}7}P(\nu_5,\Delta_5)P(\nu_6,\Delta_6)\mathcal{A}^{12\underline{6}\underline{5}}_{\rm cont}(x_1,x_2,x_6,x_5)
\frac{\<\widetilde{\underline{\O}}_{5}\widetilde{\underline{\O}}_{6}\O_7\>}{\<\O_{1}\O_{2}\O_7\>}
\nonumber \\
&=\int\limits_{-\infty}^{\infty} d\nu_{5}d\nu_{6}\,b_{\tilde{\underline{5}}\tilde{\underline{6}}7}P(\nu_5,\Delta_5)P(\nu_6,\Delta_6)\rho^{12\underline{6}\underline{5}}_{\rm cont}(\Delta_7,0)B_{\Delta_7,0},
\end{align}
where in the last line we plugged in the CPW decomposition for the contact diagram and did the position integrals. This function has poles at $\Delta_7=\Delta_5+\Delta_6+2n$ as claimed above. In Appendix \ref{vertcorr}, we comment on how the $\O_7$ exchange can be reabsorbed by one-loop counterterms and its relation to the CFT unitarity method.


\sec{Matching Bulk and Boundary Unitarity Methods}
\label{sec:UnitaritiesUnified}

In the previous sections we developed AdS unitarity methods directly in the bulk. We are now in position to make contact with the original, holographic unitarity method of \cite{Aharony:2016dwx}. We begin by recalling properties of the OPE data associated to bulk amplitudes, and rephrase the approach of \cite{Aharony:2016dwx} in terms of the dDisc and the Lorentzian inversion formula. We then show that there is a transparent and natural relation between gluing bulk diagrams, on the one hand, and multiplying OPE data in the conformal block decomposition, on the other. In this section, in an effort to make clear which manipulations are to be thought of as bulk and which as boundary, we refer to CFT correlators as $\cG$ and AdS amplitudes as $\cA$, despite their equivalence. For simplicity we continue to refer to both as tree-level, one-loop, etc.

We remind the reader that the contents of this section are summarized in figure \ref{fig:Gluing_Dictionary} for the example of AdS $\phi^3+\phi^4$ theory. 

\ssec{Review of CFT Unitarity Method}

The method of \cite{Aharony:2016dwx} answers the following question: given the planar OPE data of a large $N$ CFT, how do we compute leading non-planar corrections? In bulk terms: given the tree-level OPE data, how do we compute one-loop corrections to that data? We briefly review this now. 

For simplicity, we start by considering identical, external scalars $\O$ dual to a self-interacting scalar field in AdS, as in \cite{Aharony:2016dwx}. Recall that tree-level, four-point AdS amplitudes generate anomalous dimensions, $\g_{n,\ell}$, for double-trace operators $[\O\O]_{n,\ell}$. These admit a $1/N$ expansion,
\e{}{\g_{n,\ell} = {\g^{(1)}_{n,\ell}\o N^{2}} + {\g^{(2)}_{n,\ell}\o N^4}+\ldots}
Modulo non-analyticities in spin $\ell$, the one-loop term $\g^{(2)}_{n,\ell}$ must be fixed by the tree-level term $\g^{(1)}_{n,\ell}$. This is because the loop-level amplitude is fixed once the Feynman rules -- equivalently, the on-shell tree-level amplitudes -- are specified, modulo finite one-loop counterterm ambiguities which can only introduce non-analytic OPE data \cite{Heemskerk:2009pn}. 

It is straightforward to see how $\g^{(2)}_{n,\ell}$ is fixed in terms of $\g^{(1)}_{n,\ell}$. Let us use the language of Lorentzian inversion. Recalling that
\e{}{\dDisc_t(g_{\D,\ell}(1-z,1-\zb)) = 2\sin^2\left({\pi\o2}(\D-\ell-2\D_\O)\right)g_{\D,\ell}(1-z,1-\zb)}
and
\e{}{\dDisc_t(\ln(1-\zb)) = 0}
one concludes that for the CFT correlator $\la \O\O\O\O\ra$ at one loop, ${\rm dDisc}_t(\cG_{\rm 1-loop})$ is {\it quadratic} in the tree-level anomalous dimension, and {\it independent} of the one-loop anomalous dimension. These imply   
\e{cftunit}{{\rm dDisc}_t(\cG_{\rm 1-loop}) \supset {\pi^2\o 2}\sum_{n,\ell} p_{n,\ell}^{(0)} (\g^{(1)}_{n,\ell})^2 g_{n,\ell}(1-z,1-\zb),}
where we denote $g_{n,\ell}(1-z,1-\zb)$ as the $t$-channel conformal block for $[\O\O]_{n,\ell}$ exchange. In this sense, tree-level data is a ``source" for the one-loop data (and beyond). Through the Lorentzian inversion formula, one can mechanically extract $\g^{(2)}_{n,\ell}$ (see e.g. \cite{Aharony:2016dwx,Alday:2017vkk,Alday:2017xua,Aprile:2017bgs,Aprile:2017xsp}).

For non-identical operators $\la\O_1\O_2\O_3\O_4\ra$, the concept is the same, only some details differ. Generically, the $\sin$ factors in \eqr{dDisctch} do not degenerate, and the $1/N^{2}$ factors come from the OPE coefficients of external operators with some new double traces first appearing at one-loop. In particular, $\dDisc_t(\cG_{\rm 1-loop})$ takes the form \eqr{dDisctch} with $\O_{\D,J} =  [\O_5\O_6]_{n,\ell}$, where $\O_5$ and $\O_6$ are some other primaries in the CFT:
\begin{align}\label{dDisctchni}
\dDisc_{t} ( \mathcal{G}(z,\bar{z}))&=2\sin\left(\frac{\pi}{2}(\Delta_5+\D_6-\Delta_1-\Delta_4)\right)\sin\left(\frac{\pi}{2}(\Delta_5+\D_6-\Delta_2-\Delta_3)\right)\\&\times\frac{(z\bar{z})^{\frac{\Delta_1+\Delta_2}{2}}}{\left[(1-z)(1-\bar{z})\right]^{\frac{\Delta_2+\Delta_3}{2}}}\sum\limits_{n,\ell}p_{[56]_{n,\ell}}g_{[56]_{n,\ell}}(1-z,1-\bar{z}), \nonumber
\end{align}
where
\e{dDisctchni2}{p_{[56]_{n,\ell}} = 2^{-\ell}C_{12[56]_{n,\ell}}C_{43[56]_{n,\ell}}.}
Each of these OPE coefficients is of order $1/N^{2}$ because the external operators are not the double-trace constituents. The product is then of order $1/N^{4}$, playing the role of the product $(\g^{(1)}_{n,\ell})^2$ for the identical operator case.\footnote{There is also an intermediate case: if for example $[\O_5\O_6]_{n,\ell}= [\O_1\O_2]_{n,\ell}$, then $p_{[56]_{n,\ell}}$ is replaced by a product of an anomalous dimension $\g^{(1)}_{n,\ell}$ with a tree-level OPE coefficient $C_{34[56]_{n,\ell}}$.}

\ssec{Cutting, Gluing and Contraction in CFT}
We can now explain the map between bulk and boundary unitarity methods. The main point is that equation \eqr{cftunit} is the OPE version of gluing bulk trees at one loop. In other words,

\begin{quotation}
{\it \noindent There is a direct correspondence between which tree-level OPE data one plugs into the sums \eqr{cftunit} and \eqr{dDisctchni}, and which bulk diagrams one glues together.}
\end{quotation}

\noindent For simplicity in making contact with \cite{Aharony:2016dwx}, first consider identical external scalars $\O$. Consider an arbitrary bulk theory containing the scalar field dual to $\O$. The theory may include spinning operators, such as the graviton, to which it couples. The sum of all $s$- $t$- and $u$-channel tree diagrams, when branched into a single channel, gives the total tree-level anomalous dimension $\g_{n,\ell}^{(1)}$ for $[\O\O]_{n,\ell}$ operators. Let us write it as
\e{}{\g_{n,\ell}^{(1)} = \g_{n,\ell}^{(1),\mathbf{s}}+\g_{n,\ell}^{(1),\mathbf{t}}+\g_{n,\ell}^{(1),\mathbf{u}},}
where $\g_{n,\ell}^{(1),\mathbf{x}}$ denotes the contribution from diagrams in channel $\mathbf{x}$. Inserting this into the right-hand side of \eqr{cftunit}, one can expand ${\rm dDisc}_t(\cG_{\rm 1-loop})$ into products of individual channel contributions. The map is that each term in the product reproduces the piece of ${\rm dDisc}_{t}(\cA_{\rm 1-loop})$ obtained by gluing the bulk diagrams in the respective channels! 

That is, define $\g^{\mathbf{x}| \mathbf{y}}$ as the product of anomalous dimensions due to tree-level exchange in the $\mathbf{x}$ and $\mathbf{y}$ channels,
\e{}{\g^{\mathbf{x}| \mathbf{y}}_{n,\ell}\equiv \g_{n,\ell}^{(1),\mathbf{x}}\, \g_{n,\ell}^{(1),\mathbf{y}} .}
In this notation, the dDisc in the $t$-channel (say) of the {\it full, crossing-symmetric} correlator $\cG_\1$ includes a piece
\e{ddxy}{{\dDisc_t( \cG_{\rm 1-loop}) \supset \dDisc_t( \cG_{\rm 1-loop}^{\mathbf{x}| \mathbf{y}}) = {\pi^2\o 2}\sum_{n,\ell} a_{n,\ell}^{(0)} \g^{\mathbf{x}| \mathbf{y}}_{n,\ell}g_{n,\ell}(1-z,1-\zb)}~,}
with one for each pair of channels $(\mathbf{x,y})$. If we denote the off-shell conformal gluing \eqr{crossdef} of tree diagrams in the $\mathbf{x}$ and $\mathbf{y}$ channels as
\e{}{\cA_{\rm 1-loop}^{\mathbf{x}| \mathbf{y}} \equiv \cA_{\rm tree}^{\mathbf{x}}\otimes \cA_{\rm tree}^{\mathbf{y}} ,}
then {\it ${\rm dDisc}_t( \cG_{\rm 1-loop}^{\mathbf{x}| \mathbf{y}})$ is a weighted sum of cuts $\hCut_{t}( \cA_{\rm 1-loop}^{\mathbf{x}| \mathbf{y}})$ over internal lines.} In particular, we have a map between individual diagrams in the bulk and pieces of the tree-level anomalous dimension; this is a refinement of the map emphasized in \c{Aharony:2016dwx} between the full bulk amplitude summed over channels and the OPE data of the full crossing-symmetric correlator.\foot{See footnote 13 of \c{Aharony:2016dwx}.}

For 1PI diagrams, there is a unique internal $t$-channel cut, so these two operators are proportional:
%
\e{}{\dDisc_t( \cG_{\rm 1-loop}^{\mathbf{x}| \mathbf{y}}) \propto  \hCut_{t}( \cA_{\rm 1-loop}^{\mathbf{x}| \mathbf{y}}).}
For non-1PI diagrams, there are multiple internal cuts -- recall the amplitudes $\cA^{(2)}_{\rm prop}$ and $\cA_{\rm vertex}$ in Section \ref{sec:non1PI} -- in which case $\dDisc_t$ is a sum over the different $t$-channel cuts.

The above generalizes immediately to generic, non-identical operators. For 1PI diagrams, 
\e{ddcut}{\dDisc_t(\mathcal{G}_\1^{\mathbf{x}| \mathbf{y}}) = 2\sin\left({\pi\o2}(\tau_t-\D_1-\D_2)\right)\sin\left({\pi\o2}(\tau_t-\D_3-\D_4)\right)\hCut_{t}[\mathcal{A}_\1^{\mathbf{x}| \mathbf{y}}]}
where $\tau_t = \D_t-\ell_t$ is the twist of the $t$-channel double-trace operators first appearing in this correlator at one loop and $\mathbf{x}|\mathbf{y}$ refers to a product of OPE coefficients instead of anomalous dimensions (cf. \eqr{dDisctchni} and \eqr{dDisctchni}). This relation \eqr{ddcut} was explicitly proven in earlier sections for the bubble, triangle and box diagrams. For non-1PI diagrams there is an analogous extension which sums over all internal cuts.

\sssec{Contraction}
In addition to explaining the CFT duals of cutting and gluing bulk diagrams, there is a simple CFT operation that corresponds to {\it contraction}. Whereas in the bulk, contraction means replacing an exchange diagram with a contact diagram, in the CFT one just replaces a factor of $\g^{(1)}_{n,\ell}$ from an exchange diagram with its contact-diagram counterpart:
\e{}{\g^{(1),\,\rm exch}_{n,\ell}~~ \mapsto ~~\g^{(1),\,\rm cont}_{n,\ell} \qquad \leftrightarrow \qquad\text{Bulk contraction.}}
If the exchanged operator has spin-$J$, one uses the $\g^{(1),\,\rm cont}_{n,\ell}$ derived from a $\phi^4$-type interaction dressed with $2J$ derivatives. This follows from general covariance of scalar-scalar-spin-$J$ couplings. 

This is straightforward to prove for scalar exchanges. Recall that contraction can be achieved by taking the mass of the exchanged field to infinity. Indeed, it was shown in \cite{Fitzpatrick:2010zm} that the leading asymptotic of $\g^{(1),~\rm exch_{\Delta}}_{n,\ell}$ at large internal $\D$ becomes $\g^{(1),\,\rm cont}_{n,\ell}$,
\e{}{\lim_{\D\rar\infty} \D^2 \g^{(1),\, \rm exch_{\Delta}}_{n,\ell} = \g^{(1),\,\rm cont}_{n,\ell}}
where we recall that $\D \sim m L_{\rm AdS}$ at large mass. 

\sssec{Comments}
\quad\ The above map generalizes directly to higher loops. The main idea is essentially recursive: the $\dDisc$ allows us to construct $L$-loop diagrams using lower-loop diagrams as input. More precisely, {\it at arbitrary loop order, ${\rm dDisc}_t$ is a weighted sum of internal cuts $\hCut_t$.} The one-loop diagrams computed from tree-level data serve as input at two loops, and so on.\foot{For example, the leading log comes from a term $(\g^{(1)}_{n,\ell})^3$ in dDisc($\cG_{\rm 2-loop}$).} The bulk unitarity method will be demonstrated in the next section for the double-ladder diagram. 

We emphasize that this prescription for how to build up individual loop diagrams from the CFT is not restricted to scalar theories. Anomalous dimensions associated to spinning operator exchanges have pieces that are non-analytic in spin \cite{Costa:2014kfa,Alday:2017gde}, but this is simply fed into the same formulas above. 

For non-1PI diagrams, one can always add local counterterms to remove the part of the amplitude that is simply proportional to the tree, thus defining one-loop renormalized masses and coupling constants. Stated slightly differently, there exists a choice of renormalization scheme in which one-loop diagrams are all 1PI, and thus have only a single internal cut in a given channel. In such a scheme, dDisc and $\hCut$ are proportional for all one-loop diagrams. 

In Appendix \ref{app:Checks_Bulk_Vs_Bdy} we explicitly demonstrate the map between dDisc and $\hCut$ for the box, $\cA_{\rm box}$, and the vertex correction, $\cA_{\rm vertex}$. We have also checked that the boundary unitarity method agrees, in the manner described above, with a bulk computation of \c{Yuan:2018qva, EllisPriv} which uses a different, more direct approach than the bulk unitarity method presented here. More precisely, we have checked this correspondence for the case of the four-point triangle in $\phi^3+\phi^4$ theory in AdS$_5$ with $\D_\phi=2$. On the CFT side, one inserts (suppressing the tree-level ``$(1)$'' superscripts for clarity) $\g^{\mathbf{x}|\mathbf{y}}_{n,\ell} = \g^{\mathbf{s},\phi^3}_{n,\ell}\g^{\phi^4}_{n,\ell}$ into \eqr{ddxy}.

\section{Higher Loops and Points}
\label{sec:AdSHigherPoint}
Let us apply the recursive approach to higher-loop amplitudes using the bulk unitarity method. While a systematic study is left for the future, we are now in a good position to demonstrate the higher-loop procedure in a $\phi^3$ theory for the double-ladder diagram. 

In general, higher-point functions naturally enter in the study of four-point diagrams beyond one-loop, as will the crossing properties of higher-point CPWs. We will demonstrate this for the double ladder, which, by the split representation, can be written as the gluing of two five-point tree-level diagrams. Thus we first study the five-point tree and its crossing properties.\footnote{For further work on higher-point tree diagrams and conformal blocks, see \cite{Rosenhaus:2018zqn,Parikh:2019ygo,Jepsen:2019svc,Parikh:2019dvm,Carmi:2019ocp,Fortin:2019zkm}}

Let us emphasize: while some properties are special to the double-ladder and five-point tree, there is nothing fundamentally unique about these diagrams in the application of our unitarity methods. 

\subsection{Warmup: Five-Point Tree}
\label{sec:5pttree}
The five-point tree we will study is is drawn below.
\begin{figure}[h!]
\begin{center}
\includegraphics[scale=.28]{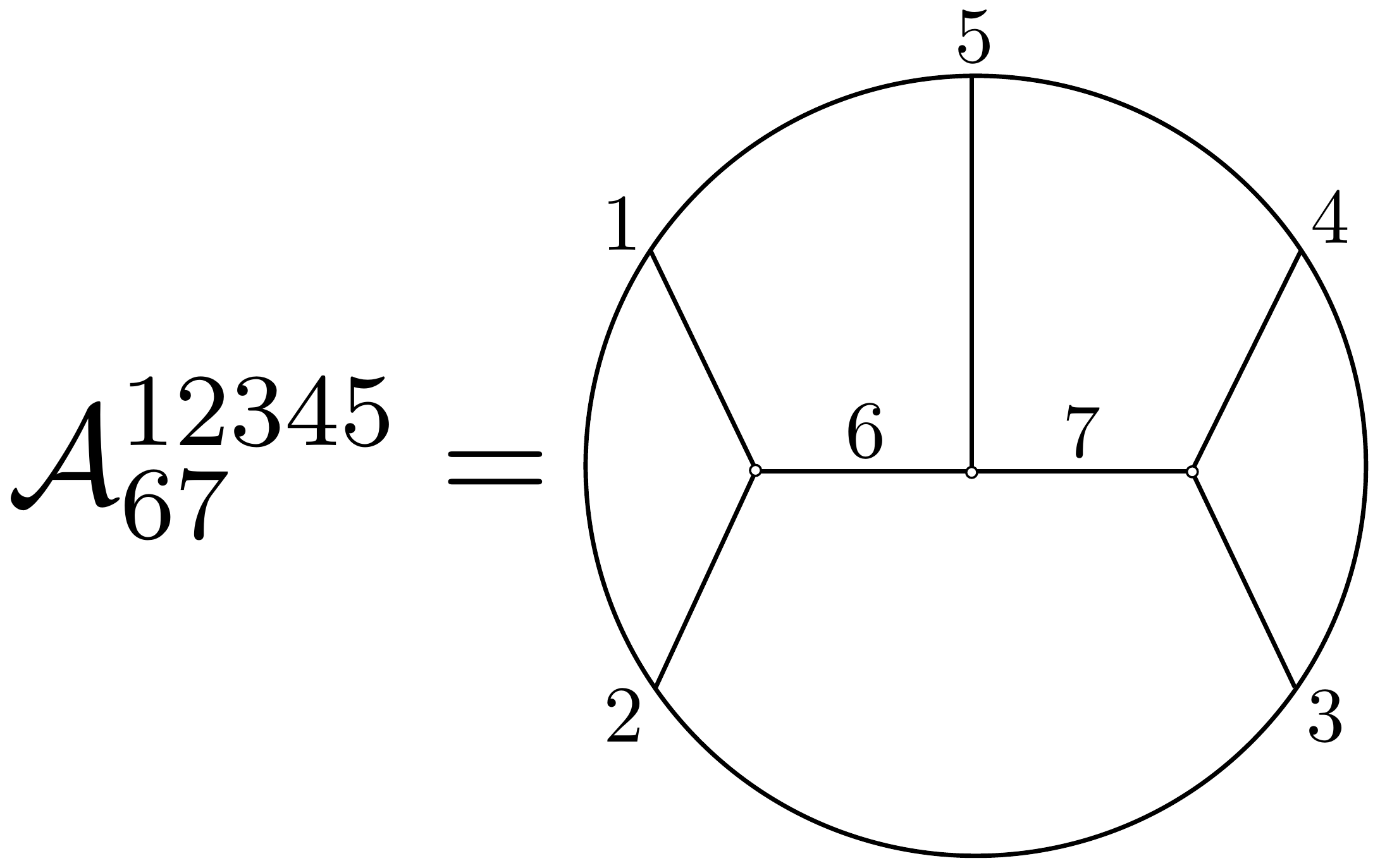}
\end{center}
\label{fig:five_Pt_Tree}
\end{figure}

\noindent To proceed, we define the five-point CPW expansion as:
\begin{align}
\mathcal{A}^{12345}_{67}(x_i)&=\sum\limits_{J_6,J_{7}=0}^{\infty}  
\int\limits_{\frac{d}{2}}^{\frac{d}{2}+i\infty}\frac{d\Delta_6 ~ d\Delta_7}{(2\pi i)^2}
\rho^{12345}(\Delta_6,J_6,\Delta_7,J_7)\Psi^{12345}_{67}(x_i),
\\
\Psi^{12345}_{67}(x_i) &\equiv \int d^{d}x_{6}d^{d}x_{7}\<\O_1\O_2\O_6\>\<\widetilde{\O}_{6}\O_{5}\widetilde{\O}_{7}\>\<\O_7\O_3\O_4\>.
\end{align}
As the external operators here are scalars, all the internal operators will be symmetric traceless tensors. It will be helpful for understanding crossing to note that the five-point CPW can be obtained by gluing a three-point function to a four-point CPW:
\begin{equation}
\Psi^{12345}_{67}(x_i) =
\int d^d x_7 \Psi^{125\tilde{7}}_6(x_1,x_2,x_5,x_7) \braket{\mathcal{O}_7 \mathcal{O}_3 \mathcal{O}_4}
=
\int d^d x_6 \braket{\mathcal{O}_1 \mathcal{O}_2 \mathcal{O}_6} \Psi^{\tilde{6}534}_{\tilde{7}}(x_6,x_5,x_3,x_4) .
\label{eq:fivepts_from_fourpts}
\end{equation}
We can also define the five-point conformal block as the solution to the Casimir equations with the following behavior when we take the limit $x_1\rightarrow x_2$ and then take $x_3\rightarrow x_4$
\begin{align}
g^{12345}_{67}(x_i)\approx (-2)^{J_5}(-2)^{J_6}|x_{12}|^{\Delta_{6}-\Delta_1-\Delta_2}|x_{34}|^{\Delta_7-\Delta_3-\Delta_4}\<\O_{6}(x_1)\O_{5}(x_5)\O_{7}(x_3)\>.
\end{align}
With this definition we have
\begin{align}
\Psi^{12345}_{67}(x_i)=K^{5\tilde{7}}_{\tilde{6}}K^{56}_{\tilde{7}}g^{12345}_{67}(x_i)+K^{12}_{6}K^{5\tilde{6}}_{\tilde{7}}g^{12345}_{\tilde{6}7}(x_i)+K^{34}_{7}K^{5\tilde{7}}_{6}g^{12345}_{6\tilde{7}}(x_i)+K^{34}_{7}K^{12}_{6}g^{12345}_{\tilde{6}\tilde{7}}(x_i).
\end{align}
This relation is proven in Appendix \ref{app:5ptCPW}. The relation between the five-point OPE function and the OPE coefficients is:
\begin{align}
\left(-\frac{1}{2}\right)^{J_{6}}\left(-\frac{1}{2}\right)^{J_{7}}C_{126}C_{657}C_{734}=\res\limits_{\Delta_{6'}=\Delta_6}\res\limits_{\Delta_{7'}=\Delta_7}\rho^{12345}(\Delta_{6'},J_{6},\Delta_{7'},J_{7})K^{5\tilde{7}'}_{\tilde{6}'}K^{56'}_{\tilde{7}'}.
\end{align}

We will not study crossing for the five-point CPW in general, only the crossing relations required to understand the double-ladder. We will start by decomposing the five-point exchange diagram, given above, in terms of an OPE channel with the same graphical structure:
\begin{align}
\mathcal{A}^{12345}_{67}(x_i)=\int\limits_{-\infty}^{\infty}d\nu_{6}d\nu_{7}P(\nu_6,\Delta_6)P(\nu_7,\Delta_7)b_{12\underline 6}b_{\underline{\tilde{6}}5\underline{\tilde{7}}}b_{\underline734}\Psi^{12345}_{\underline6\underline7}(x_i). \label{eq:fiveptExchangeWitten}
\end{align}
Closing the contours, we find poles in $\nu_6$ at
\begin{eqnarray}
\underline{\Delta}_{6}&=&\Delta_6, \label{eq:5pt6PolST}
\\
\underline{\Delta}_{6}&=&\Delta_{1}+\Delta_{2}+2n, \label{eq:5pt6PolDT1}
\\ 
\underline{\Delta}_{6}&=&\Delta_{5}+\Delta_{7}+2n, \label{eq:5pt6PolDT2}
\end{eqnarray}
and similarly for the $\nu_7$ poles we have
\begin{eqnarray}
\underline{\Delta}_{7}&=&\Delta_7, \label{eq:5pt7PolST}
\\
\underline{\Delta}_{7}&=&\Delta_{3}+\Delta_{4}+2n, \label{eq:5pt7PolDT1}
\\
\underline{\Delta}_{7}&=&\Delta_{5}+\Delta_{6}+2n. \label{eq:5pt7PolDT2}
\end{eqnarray}
From these double-trace poles we can now construct triple-trace operators. For example, if we pick up the pole \eqref{eq:5pt6PolDT2} this sets $\O_6=[\O_5\O_7]_{n_1,\ell_1}$, for some $n_1,\ell_1$. Then we can pick up the pole \eqref{eq:5pt7PolDT1} which sets $\O_7=[\O_3\O_4]_{n_2,\ell_2}$. Therefore, $\O_6$ has now become a triple-trace operator $\O_6 =[\O_3\O_4\O_5]_{\Delta,\ell}$.\footnote{Triple-trace operators with spin are degenerate but to be compact we label them by their dimension and spin and drop any extra index.} Likewise, by picking up the poles \eqref{eq:5pt6PolDT1} and \eqref{eq:5pt7PolDT2} and we can also from the triple-trace operator $\O_7=[\O_1\O_2\O_5]_{\Delta,\ell}$.

This aligns with expectations from Mean Field Theory and large $N$ counting. In the canonical large $N$ normalization, the five-point tree has scaling $\mathcal{A}^{12345}_{67} \sim N^{-3}$. To leading order in $1/N$, $\<\O_1\O_2\O_5[\O_1\O_2\O_5]\> \sim N^0$ (this is the Mean Field Theory result) while $\<\O_3\O_4[\O_1\O_2\O_5]\> \sim N^{-3}$. These results are also consistent with those of \cite{Parikh:2019ygo,Jepsen:2019svc} which used geodesic Witten diagrams \cite{Hijano:2015zsa}.

Now we turn to the crossed channel decomposition. Expanding the five-point diagram in a dual channel requires the 9j symbol for the conformal group, which is currently not known in closed form. Nevertheless, we can make progress using the 6j symbol on internal parts of the five-point partial wave, which amounts to expressing the 9j symbol as a spectral integral of products of 6j symbols. Crossing can be applied to each pair of glued three-point functions in succession, as these are partial waves \eqref{eq:fivepts_from_fourpts}. In preparation for the AdS double-ladder, we will apply crossing twice to obtain
\begin{align}
\mathcal{A}^{12345}_{67}(x_i)&=\int\limits_{-\infty}^{\infty}d\nu_{6}d\nu_{7}\sum\limits_{J_{8,9}=0}^{\infty}\int\limits_{\frac{d}{2}}^{\frac{d}{2}+i\infty}\frac{d\Delta_{8}d\Delta_9}{(2\pi i)^{2}}P(\nu_6,\Delta_6)P(\nu_7,\Delta_7)b_{12\underline{6}}b_{\tilde{\underline{6}}5\tilde{\underline{7}}}b_{\underline{7}34}
\nonumber \\ & \qquad\frac{1}{n_8n_9}\sixj{2}{\tilde{\underline{7}}}{5}{1}{8}{\tilde{\underline{6}}}\sixj{2}{3}{4}{8}{9}{\underline{7}}\frac{K^{15}_{\tilde{8}}}{K^{4\tilde{9}}_{\tilde{8}}}\Psi^{15234}_{89}(x_i). \label{eq:Crossing5pt}
\end{align}
In this channel we again find that triple-traces are exchanged, this time manifest as poles in the 6j symbol. For instance, we have triple-trace operators when $\O_8=[\O_1\O_5]$ and $\O_9=[\O_4\O_8]=[\O_1\O_4\O_5]$ or when $\O_9=[\O_2\O_3]$ and $\O_8=[\O_4\O_9]=[\O_2\O_3\O_4]$.\footnote{The pole at $\mathcal{O}_8=[\mathcal{O}_2 \mathcal{O}_{\tilde{7}}]$ is canceled by a zero in the second 6j symbol, and thus does not lead to a triple-trace exchange.} We now have sufficient information about the five-point Witten diagram to address the double-ladder.
\subsection{Double-Ladder}
\label{sec:twoloopladder}
Now we study the double-ladder diagram:
\begin{figure}[h!]
\begin{center}
\includegraphics[scale=.28]{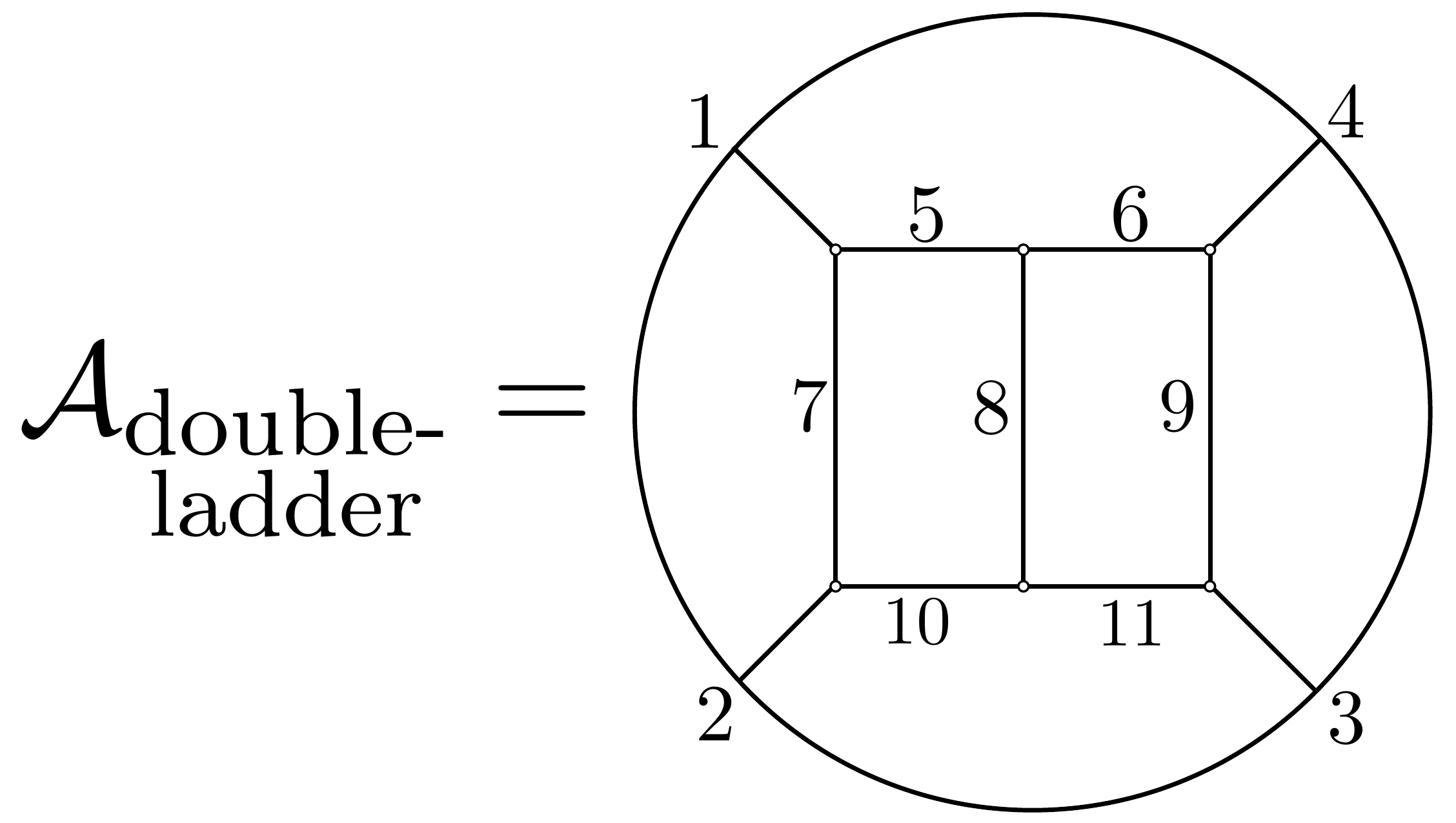}
\end{center}
\label{fig:two_loop_ladder}
\end{figure}
\\ \noindent The double-ladder has several non-trivial cuts: two vertical cuts for the $[\O_5\O_{10}]$ and $[\O_6\O_{11}]$ families, a horizontal cut for the triple-trace family $[\O_7\O_8\O_9]$, and two diagonal cuts for $[\O_5\O_8\O_{11}]$ and $[\O_6\O_{8}\O_{10}]$. Part of the goal of this section is to show no other cuts are possible for this diagram. We also want to understand how to compute the dDisc of this diagram.

There are several natural, but distinct, ways to decompose the double-ladder:

\begin{itemize}
\item If we split the 5, 6, 10 and 11 legs, it is a gluing of three vertical four-point exchanges:
\e{doubleladthreetreesplit}{\mathcal{A}_{\substack{\text{double-}\\ \rm ladder}}(x_i) = \mathcal{A}^{\underline{10},2,1,\underline{5}}_{7,\, \rm exch} \,\otimes\, \mathcal{A}^{\underline{11},\un{\widetilde{10}},\un{\widetilde 5},\underline{6}}_{8,\, \rm exch} \,\otimes\, \mathcal{A}^{3,\un{\widetilde{11}},\un{\widetilde{6}},4}_{9,\, \rm exch}.}
\item If we split the 7, 8 and 9 legs, it is a gluing of two horizontal five-point trees:
\e{eq:2box_FivePtSplit}{\mathcal{A}_{\substack{\text{double-}\\ \rm ladder}} (x_i)= \mathcal{A}^{\underline{7}14\underline{8}\underline{9}}_{6,5} \,\otimes\,\mathcal{A}^{\underline{\tilde{7}}23\underline{\tilde{9}}\underline{\tilde{8}}}_{10,11}.}
\item If we split the 5 and 10 legs (say), it is a gluing of a vertical four-point exchange and a four-point box:
\e{doubleladboxsplit}{\mathcal{A}_{\substack{\text{double-}\\ \rm ladder}}(x_i) = \mathcal{A}^{\underline{10},2,1,\underline{5}}_{7,\, \rm exch}\, \otimes \mathcal{A}^{\tilde{\underline{5}},\widetilde{\underline{10}},3,4}_{\rm box}.}
\item If we split the 5, 8, and 11 legs it is a gluing of two five-point trees:
\e{eq:dladsplit5pt2}{\mathcal{A}_{\substack{\text{double-}\\ \rm ladder}} (x_i)=\mathcal{A}^{\underline 5,1,\underline{11},\underline 8,2}_{7,10}\otimes \mathcal{A}^{\underline{\tilde{5}},\underline{\tilde{8}},\underline{\widetilde{11}},3,4}_{6,9}}
\end{itemize}
We will discuss each of these representations below. 

It will be convenient to apply the split representation on all rungs and reduce the AdS double-ladder to a spectral integral of the corresponding two-loop conformal integral as our starting point, 
\begin{align}
\mathcal{A}_{\substack{\text{double-}\\ \rm ladder}}(x_i)=\int\limits_{-\infty}^{\infty}&\prod\limits_{i=5}^{11}d\nu_i
P(\nu_i,\Delta_i)
\int\limits_{\partial \rm AdS} dx_{5,...,11}
b_{\underline{5},1,\tilde{\underline{7}}}b_{\underline{7},2,\underline{10}}
b_{\widetilde{\underline{10}},\underline{11},\underline{8}}b_{\tilde{\underline{8}},\tilde{\underline{5}},\underline{6}}
b_{\tilde{\underline{6}},4,\tilde{\underline{9}}}b_{\underline{9},3,\widetilde{\underline{11}}}
\nonumber
\\ 
&\<\O_5 \O_1\widetilde{\O}_{7}\>
\<\O_7 \O_{2} \O_{10}\>
\<\widetilde{\O}_{10}\O_{11}\O_{8}\>
\<\widetilde{\O}_{8}\widetilde{\O}_{5}\O_{6}\> 
\<\widetilde{\O}_{6} \O_{4}\widetilde{\O}_{9}\>
\<\O_{9} \O_{3}\widetilde{\O}_{11}\>.
\label{eqn:twoLoopCFTBoxInt}
\end{align}

\sssec{$s$-channel}
We first obtain the $s$-channel decomposition, $\O_1\O_2 \rar \O_3\O_4$. We begin by performing the $x_{7,8,9}$ integrals in (\ref{eqn:twoLoopCFTBoxInt}). This implements the cut given by \eqr{doubleladthreetreesplit}. Using the 6j symbol to obtain the crossed-channel decomposition of the three constituent exchange diagrams, and then using the CFT bubble identity \eqref{eq:Conf-Bubble} to do the remaining position-space integrals, we have
\begin{align}
\hspace{-.5in}\mathcal{A}_{\substack{\text{double-}\\ \rm ladder}}(x_i)&=\int\limits_{-\infty}^{\infty}\prod\limits_{i=5}^{11}d\nu_i P(\nu_i,\Delta_i)\sum\limits_{J_{\chi_1}=0}^{\infty}\int\limits_{\frac{d}{2}}^{\frac{d}{2}+i\infty}\frac{d\Delta_{\chi_{1}}}{2\pi i}
b_{\underline{5},1,\tilde{\underline{7}}}b_{\underline{7},2,\underline{10}}
b_{\widetilde{\underline{10}},\underline{11},\underline{8}}b_{\tilde{\underline{8}},\tilde{\underline{5}},\underline{6}}
b_{\tilde{\underline{6}},4,\tilde{\underline{9}}}b_{\underline{9},3,\widetilde{\underline{11}}}
\frac{B_{\chi_{1}}^{2}}{n_{\chi_{1}}^{3}}
\nonumber \\
&\hspace{.4in}\sixj{1}{2}{\underline{10}}{\underline{5}}{\chi_{1}}{\underline{7}}\sixj{\tilde{\underline{5}}}{\widetilde{\underline{10}}}{\underline{11}}{\underline{6}}{\chi_{1}}{\underline{8}}\sixj{\tilde{\underline{6}}}{\widetilde{\underline{11}}}{3}{4}{\chi_{1}}{\underline{9}}\Psi^{1234}_{\chi_{1}}(x_i). \label{eq:twoloopbox_sch}
\end{align}
We can then use the split of the 6j symbol (\ref{eqn:6j_Split}), to close the $\nu_6,\nu_7,\nu_8$ contours, and in doing so we pick up only the single-trace poles. The block decomposition becomes
\begin{align}
\mathcal{A}_{\substack{\text{double-}\\ \rm ladder}}(x_i)&=\prod\limits_{j=7}^{9}(d-2\Delta_{j})\int\limits_{-\infty}^{\infty}\prod\limits_{i=5,6,10,11}d\nu_i P(\nu_i,\Delta_i)\sum\limits_{J_{\chi_1}=0}^{\infty}\int\limits_{\frac{d}{2}}^{\frac{d}{2}+i\infty}\frac{d\Delta_{\chi_{1}}}{2\pi i}
b_{\underline{5},1,\tilde{7}}b_{7,2,\underline{10}}
b_{\widetilde{\underline{10}},\underline{11},8}b_{\tilde{8},\tilde{\underline{5}},\underline{6}}
b_{\tilde{\underline{6}},4,\tilde{9}}b_{9,3,\widetilde{\underline{11}}}
\frac{B_{\chi_{1}}^{2}}{n_{\chi_{1}}^{3}}
\nonumber \\
&\hspace{.4in}
K^{1 \underline{5}}_{\tilde{7}}
K^{\tilde{\underline{5}} \underline{6}}_{\tilde{8}}
K^{\tilde{\underline{6}}4}_{\tilde{9}}
\sixjBlock{1}{2}{\underline{10}}{\underline{5}}{\chi_{1}}{7}\sixjBlock{\tilde{\underline{5}}}{\widetilde{\underline{10}}}{\underline{11}}{\underline{6}}{\chi_{1}}{8}\sixjBlock{\tilde{\underline{6}}}{\widetilde{\underline{11}}}{3}{4}{\chi_{1}}{9} K^{34}_{\widetilde{\chi}_1} g^{1234}_{\chi_{1}}(x_i). \label{eq:twoloopbox_schV2}
\end{align}
The poles for physical operators are now manifest in the same way as at tree-level and one-loop. Each factor from inverting a tree-level exchange diagram furnishes double-trace poles. In addition to the external double-trace poles we also have the poles 
\begin{eqnarray}
\Delta_{\chi_1}&=&\underline{\Delta}_5+\underline{\Delta}_{10}+2n, \label{eq:2loopbox_intCut1}
\\
\Delta_{\chi_1}&=&\underline{\Delta}_{6}+\underline{\Delta}_{11}+2n, \label{eq:2loopbox_intCut2}
\end{eqnarray}
corresponding to the two internal vertical cuts. We see that the same process we applied at one loop cleanly delivers the physical states in this two-loop example. We leave implicit the other double-trace poles related by shadow symmetry.

The $\dDisc_{s}$ removes the external double-trace poles, and if we close the $\Delta_{\chi_1}$ contour on the $\underline{\Delta}_{5}+\underline{\Delta}_{10}+2n$ poles, the $\nu_{5,10}$ spectral integrals localize on their single-trace poles. Unlike at one loop, however, this does not localize all spectral integrals: the $\nu_{6,11}$ spectral integrals remain. These integrals cannot be done in a simple form that we are aware of, and when we close these contours, we pick up an infinite sequence of poles. This result dovetails with the statement in flat space scattering amplitudes that a unitarity cut of a two-loop diagram factorizes it into a tree and one-loop diagram. In our case, we can see this structure more plainly by splitting the $\O_{5,10}$ lines in the double-ladder at the outset: the double-ladder becomes a tree exchange diagram and box diagram glued together, cf. \eqr{doubleladboxsplit}. Therefore isolating the $[\O_5\O_{10}]_{n,\ell}$ contribution requires an external line cut for the box sub-graph. 

However, the representation \eqref{eq:twoloopbox_schV2} does not make manifest that the $s$-channel expansion also contains triple-trace operators corresponding to a ``diagonal'' cut. To see these operators we can consider the expression for the double-ladder given in (\ref{eq:dladsplit5pt2}). Before discussing the bulk approach, we will first explain how to see these operators via the approach of \cite{Aharony:2016dwx}, as reviewed in Section \ref{sec:UnitaritiesUnified}. Our starting point is that the AdS theory gives rise to the tree-level five-point Witten diagrams $\mathcal{A}^{5,1,11,8,2}_{7,10}$ and $\mathcal{A}^{5,8,11,3,4}_{6,9}$. Restoring the coupling dependence, each Witten diagram scales like $N^{-3}$. Therefore, when we solve crossing symmetry we find $\<\O_1\O_2 [\O_5\O_8\O_{11}]\>\sim \<\O_3\O_4[\O_5\O_8\O_{11}]\>\sim N^{-3}$, and hence this triple-trace operator must contribute to $\<\O_1\O_2\O_3\O_4\>$ at order $N^{-6}$, i.e. two-loop order. 

We now recover this fact directly in the bulk approach. By applying crossing for the first five-point sub-diagram in \eqref{eq:dladsplit5pt2} we find:
\begin{align}\label{5treecross}
\mathcal{A}^{1,5,11,8,2}_{7,10}(x_i)&=\int\limits_{-\infty}^{\infty}d\nu_{7}d\nu_{10}P(\nu_7,\Delta_7)P(\nu_{10},\Delta_{10})b_{1,5,\underline{7}}b_{\underline{\tilde{7}},2,\underline{10}}b_{\underline{\widetilde{10}},8,11}
\\ &\hspace{.4in}\sum\limits_{J_{\chi_1}}\int\limits_{\frac{d}{2}}^{\frac{d}{2}+i\infty}\frac{d\Delta_{\chi_1}}{2\pi i}\sixj{5}{\underline{10}}{2}{1}{\tilde{\chi}_1}{\tilde{\underline{7}}}\frac{1}{n_{\chi_1}}\Psi^{2,1,11,8,5}_{\chi_1,\underline{\widetilde{10}}}(x_i)\nonumber
\end{align}
We see that the triple-trace operator in the five-point CPW emerges via a combination of the 6j symbol and a $b$ factor, e.g. we should close the $\Delta_{\chi_1}$ contour\footnote{We first should use symmetries of the integrand under a shadow transform to extend the contour along the entire axis, but this will not affect the poles we are considering.} on the $[\O_5\underline{\O}_{10}]$ pole and then close the $\nu_{10}$ contour on a double-trace pole in $b_{\widetilde{\underline{10}},8,11}$, along with other poles related by shadow symmetry. For the second, five-point tree in \eqref{eq:twoloopbox_schV2} we will apply crossing twice:
\begin{align}
\mathcal{A}^{\tilde{5},\tilde{8},\widetilde{11},3,4}_{6,9}(x_i)=&\int\limits_{-\infty}^{\infty}d\nu_6 d\nu_9 P(\nu_6,\Delta_6)P(\nu_9,\Delta_9)b_{\tilde{5},\tilde{8},\underline{6}}b_{\tilde{\underline{6}},4,\tilde{\underline{9}}}b_{\underline{9},\widetilde{11},3}
\nonumber \\ &\sum\limits_{J_{\chi_2},\,J_{\chi_3}}\int\limits_{\frac{d}{2}}^{\frac{d}{2}+i\infty}\frac{d\Delta_{\chi_2}d\Delta_{\chi_3}}{(2\pi i)^{2}}\sixj{\tilde{\underline{6}}}{\widetilde{11}}{3}{4}{\chi_2}{\underline{9}}\sixj{\tilde{8}}{\widetilde{11}}{\chi_2}{\tilde{5}}{\chi_3}{\tilde{\underline{6}}}\frac{1}{n_{\chi_2}n_{\chi_3}}\Psi^{\tilde{8},\widetilde{11},3,4,\tilde{5}}_{\chi_{3},\tilde{\chi}_{2}}(x_i)
\end{align}
We actually only need to apply crossing once to see the contribution of the $[\widetilde{\O}_{5}\widetilde{\O}_{8}\widetilde{\O}_{11}]$ in the CPW decomposition of this diagram, exactly mirroring what we did for the first sub-diagram in \eqr{5treecross},\foot{In this case we would close the $\chi_2$ contour on the $[\underline{\widetilde{\O}}_{6}\widetilde{\O}_{11}]$ double-trace pole and then close the $\nu_6$ contour on the $[\widetilde{\O}_{5}\widetilde{\O}_{8}]$ double-trace pole in the first $b$-factor.} but we need to use crossing twice in order to glue the two sub-diagrams together and produce the CPW expansion for the double-ladder. Performing this gluing and using the CFT bubble integral we find:
\begin{align}
\mathcal{A}_{\substack{\text{double-}\\ \rm ladder}}(x_i)=&\int\limits_{-\infty}^{\infty}d\nu_{5,...,11}\prod\limits_{i=5}^{11}P(\nu_i,\Delta_i)b_{1,\underline{5},\underline{7}}b_{\underline{\tilde{7}},2,\underline{10}}b_{\underline{\widetilde{10}},\underline{8},\underline{11}}b_{\tilde{\underline{5}},\tilde{\underline{8}},\underline{6}}b_{\tilde{\underline{6}},4,\tilde{\underline{9}}}b_{\underline{9},\widetilde{\underline{11}},3}
\nonumber \\ &\sum\limits_{J_{\chi_1}}\int\limits_{\frac{d}{2}}^{\frac{d}{2}+i\infty}\frac{d\Delta_{\chi_1}}{(2\pi i)}\sixj{\underline{5}}{\underline{10}}{2}{1}{\tilde{\chi}_1}{\tilde{\underline{7}}}\sixj{\tilde{\underline{6}}}{\widetilde{\underline{11}}}{3}{4}{\chi_1}{\underline{9}}\sixj{\tilde{\underline{8}}}{\widetilde{\underline{11}}}{\chi_1}{\tilde{\underline{5}}}{\underline{10}}{\tilde{\underline{6}}}\frac{1}{n_{\chi_1}^{2}n_{\underline{10}}}
\nonumber \\ & \hspace{1.65cm} (-1)^{J_{\chi_1}}B_{\underline{10}}B_{\chi_1}\Psi^{1234}_{\chi_1}(x_i).
\end{align}
Finally, we see that the triple-trace operator $[\O_5\O_8\O_{11}]$ emerges in the double-ladder as it does in the individual, five-point exchange diagrams.
%
%
\sssec{$t$-channel}

Next we obtain the $t$-channel decomposition, $\O_3\O_2 \rar \O_1\O_4$. The $t$-channel expansion is most naturally thought of in terms of five-point functions, using the cut \eqr{eq:2box_FivePtSplit}. This cut makes it clear how five-point tree data feeds into the horizontal cut of the double-ladder. To proceed, we first substitute the partial wave decompositions of the five-point ladder (\ref{eq:fiveptExchangeWitten}) into \eqr{eq:2box_FivePtSplit}. We then use four-point crossing twice on each diagram, as in (\ref{eq:Crossing5pt}), such that the gluing of the two five-point partial waves reduces to two bubble integrals. In this way, we obtain the $t$-channel partial wave expansion,
\begin{align}
\hspace{-.5in}\mathcal{A}_{\substack{\text{double-}\\ \rm ladder}}(x_i)=\int\limits_{-\infty}^{\infty}\prod\limits_{i=5}^{11}&d\nu_i P(\nu_i,\Delta_i)\sum\limits_{J_{\chi_2},J_{\chi_3}=0}^{\infty}\int\limits_{\frac{d}{2}}^{\frac{d}{2}+i\infty}\frac{d\Delta_{\chi_{2}}}{2\pi i}\frac{d\Delta_{\chi_{3}}}{2\pi i}
b_{\underline{5},1,\tilde{\underline{7}}}b_{\underline{7},2,\underline{10}}
b_{\widetilde{\underline{10}},\underline{11},\underline{8}}b_{\tilde{\underline{8}},\tilde{\underline{5}},\underline{6}}
b_{\tilde{\underline{6}},4,\tilde{\underline{9}}}b_{\underline{9},3,\widetilde{\underline{11}}} 
\nonumber \\
&K^{\underline{6},\tilde{\underline{8}}}_{\tilde{\underline{5}}}K^{\underline{8},\underline{11}}_{\widetilde{\underline{10}}}K^{4,\tilde{\underline{9}}}_{\tilde{\underline{6}}}K^{3,\underline{9}}_{\widetilde{\underline{11}}}\sixjBlock{\tilde{\underline{8}}}{\tilde{\underline{7}}}{1}{\underline{6}}{\chi_{2}}{\underline{5}}\sixjBlock{\underline{11}}{2}{\underline{7}}{\underline{8}}{\chi_{2}}{\underline{10}}
\nonumber \\ &\sixjBlock{\tilde{\underline{9}}}{\widetilde{\chi}_{2}}{1}{4}{\chi_{3}}{\underline{6}}\sixjBlock{3}{2}{\chi_{2}}{\underline{9}}{\chi_{3}}{\underline{11}}\frac{B_{\chi_{2}}B_{\chi_{3}}}{n_{\chi_{2}}^{2}n_{\chi_{3}}^{2}}\Psi^{3214}_{\chi_{3}}(x_i). \label{eq:twoloopbox_tchPt1}
\end{align}
We also see that this diagram produces more non-trivial spectral integrals and it takes more work to see that spurious poles ultimately do not contribute. For example, as a function of $\Delta_{\chi_{3}}$, our expression has poles at $\Delta_{\chi_3}=\Delta_{\chi_2} +\underline{\Delta}_{9}+2n+\ell$ while as a function of $\Delta_{\chi_{2}}$ there are double-trace poles
\begin{eqnarray}
\Delta_{\chi_2}&=&\underline{\Delta}_7+\underline{\Delta}_8+2n+\ell, \label{eq:twoLoop_tch_1}
\\ \Delta_{\chi_2}&=&\Delta_2+\underline{\Delta}_{11}+2n+\ell, \label{eq:twoLoop_tch_2}
\\ \Delta_{\chi_2}&=&\Delta_1+\underline{\Delta}_6+ 2n+\ell, \label{eq:twoLoop_tch_3}
\end{eqnarray}
along with poles related by shadow symmetry. The first set of poles (\ref{eq:twoLoop_tch_1}) will give physical operators while (\ref{eq:twoLoop_tch_2}) and (\ref{eq:twoLoop_tch_3}) are artifacts of how we chose to perform the crossing transformations. When we actually evaluate the residues we find the two families (\ref{eq:twoLoop_tch_2}) and (\ref{eq:twoLoop_tch_3}) have vanishing residue and the only physical, triple-trace operators which appear are $[\O_7\O_8\O_9]$, as anticipated. We see how the procedure carried out here can easily identify multi-trace states beyond one loop.

\section{Discussion}
\label{sec:Conclusion}

There are many open questions and avenues to consider. Obviously, a systematic understanding at higher loops and points would help to flesh out the nuts and bolts of the AdS/CFT unitarity method. There have been a few studies of higher-point correlators, e.g. \cite{Rosenhaus:2018zqn,Parikh:2019ygo,Jepsen:2019svc,Goncalves:2019znr} but they have received much less focus than four-point functions. However, as we study higher-loop, four-point correlators in AdS/CFT they inevitably appear when cutting certain Witten diagrams. We therefore expect understanding the OPE decompositions of higher-point correlators will help in the study of higher loops. 

While the dDisc is especially useful at one loop, its power as a tool in the unitarity method is not as clear at higher loops. In particular, one would like a clean way to isolate individual diagrammatic cuts, as opposed to sums over such cuts, e.g. by higher-discontinuity operators. On the other hand, dDisc is a central object in the non-perturbative Lorentzian inversion formula, and will still be simpler to compute than a full higher-loop amplitude. It would be nice to discover whatever relations there may be between these different types of objects. 

It would also be interesting to understand in more detail how consistency conditions in AdS/CFT at higher loops, such as causality, unitarity, and crossing symmetry, constrain the space of theories. There has been great success in constraining graviton \cite{Camanho:2014apa, Maldacena:2015iua,Hartman:2015lfa,Afkhami-Jeddi:2016ntf,Afkhami-Jeddi:2017rmx,Afkhami-Jeddi:2018own,KPZ2017,Costa:2017twz,Meltzer:2017rtf,Kologlu:2019bco} and general higher derivative interactions \cite{Fitzpatrick:2010zm,Hartman:2015lfa,Alday:2014tsa,Alday:2016htq} at leading order in $1/N$, but the study of these consistency conditions at higher loops remains unexplored. Individual loop amplitudes are also sensitive to the existence of compact, extra dimensions \cite{Alday:2017vkk,Alday:2019qrf}. Therefore, we may expect loop amplitudes correlators to contain rich new physics not visible at tree-level which place stronger constraints on the space of AdS theories.

So far we have also only discussed the scattering of particles with bounded spin, e.g. scalars and gravitons. However, if we want to understand string theory in AdS from the boundary CFT, we also need to include particles with unbounded spin. These particles organize onto Regge trajectories and are key for resolving possible causality violations due to higher derivative interactions. These trajectories have been studied at tree-level using conformal Regge theory \cite{Cornalba:2007fs,Cornalba:2009ax,Costa:2012cb,Li:2017lmh} and by generalizing to higher loops we can also hope to better understand string scattering in AdS \cite{Brower:2006ea,Brower:2007xg,Shenker:2014cwa,Meltzer:2019pyl}.

\section*{Acknowledgements}
We thank Ofer Aharony, Soner Albayrak, Ant\'onio Antunes, Dean Carmi, Clifford Cheung, Liam Fitzpatrick, Amirhossein Tajdini, Murat Kologlu, Per Kraus, Ying Hsuan-Lin, Julio Parra-Martinez, Gim Seng Ng, David Poland, and David Simmons-Duffin for discussions. This research was supported in part by Perimeter Institute for Theoretical Physics. Research at Perimeter Institute is supported by the Government of Canada through the Department of Innovation, Science and Economic Development and by the Province of Ontario through the Ministry of Research and Innovation. The research of DM is supported by the Walter Burke Institute for Theoretical Physics and the Sherman Fairchild Foundation. EP is supported by Simons Foundation grant 488657. AS is grateful to the Walter Burke Institute for Theoretical Physics for hospitality during the course of this work. This material is based upon work supported by the U.S. Department of Energy, Office of Science, Office of High Energy Physics, under Award Number DE-SC0011632. This work was performed at the Aspen Center for Physics, which is supported by National Science Foundation grant PHY-1607611.

\appendix
\section{Conventions}
\label{app:Details}
\subsection{Two and Three Point Correlators}
First let us specify our conventions for two and three-point functions. The standard convention in the CFT literature is to let all operators have unit norm. In this paper we also define the three-point functions to be kinematic structures without OPE coefficients. For scalar operators $\f_i$ and spinning operators $\O_{\Delta,J}$ the two- and three-point functions are\footnote{Note that our normalization differs from \cite{Kravchuk:2018htv} by a factor of $(-2)^J$ }
\begin{align}
\<\O_{\Delta,J}(x_1,z_1)\O_{\Delta,J}(x_2,z_2)\>&=\frac{\left(z_1\cdot I(x_{12}) \cdot z_2\right)^{J}}{x_{12}^{2\Delta_{\O}}},
\\
\<\f_1(x_1)\f_2(x_2)\O_{\Delta_3,J_3}(x_3,z_3)\>&=\frac{\left(X_{3}\cdot z_{3}\right)^{J_3}}{x_{12}^{\Delta_1+\Delta_2-\Delta_3+J_3}x_{13}^{\Delta_1+\Delta_3-J_3-\Delta_2}x_{23}^{\Delta_2+\Delta_3-J_3-\Delta_1}},
\end{align}
where we defined $\O_{\Delta,J}(x,z)=\O^{\mu_1...\mu_J}_{\Delta,J}z_{\mu_1}...z_{\mu_{J}}$ and:
\begin{align}
I_{\mu\nu}(x)=\delta_{\mu\nu}-2\frac{x_{\mu}x_{\nu}}{x^{2}},
\\
X_{3}^{\mu}=\frac{x^{\mu}_{13}}{x_{13}^{2}}-\frac{x^{\mu}_{23}}{x_{23}^{2}}.
\end{align}
With this normalization the physical three-point functions are:
\begin{align}
\<\f_1(x_1)\f_2(x_2)\O_{\Delta_3,J_3}(x_3,z_3)\>_{phys}=C_{123}\<\f_1(x_1)\f_2(x_2)\O_{\Delta_3,J_3}(x_3,z_3)\>.
\end{align}
However, the AdS conventions we follow \cite{Costa:2014kfa,Liu:2018jhs} define a different normalization,
\begin{align}
\<\O_{\Delta,J}(x_1,z_1)\O_{\Delta,J}(x_2,z_2)\>&=\mathcal{C}_{\Delta,J}\frac{\left(z_1\cdot I(x_{12}) \cdot z_2\right)^{J}}{x_{12}^{2\Delta_{\O}}},
\\
\mathcal{C}_{\Delta,J}&=\frac{\Gamma(\Delta)}{2\pi^{\frac{d}{2}}\Gamma(\Delta+1-d/2)}.
\end{align}
This normalization feeds into the integral of three bulk-to-boundary propagators,
\e{}{\int\limits_{\rm AdS} d^{d+1}y K_{\Delta_{1}}(x_1,y)K_{\Delta_2}(x_2,y)K_{\Delta_3,J_3}(x_3,y) = b_{123}\<\O_{1}(x_1)\O_{2}(x_2)\O_{3}(x_3)\>~,}
where
\e{eqn:AdS3pt}{b_{123}\equiv\mathcal{C}_{\Delta_1}\mathcal{C}_{\Delta_{2}}\mathcal{C}_{\Delta_{3},J_{3}}\frac{\pi^{d/2}\Gamma\left(\frac{\Delta_1+\Delta_2+\Delta_{3}+J_{3}-d}{2}\right)\Gamma\left(\frac{\Delta_1+\Delta_{2}-\Delta_{3}+J_{3}}{2}\right)\Gamma\left(\frac{\Delta_2+\Delta_{3}-\Delta_{1}+J_{3}}{2}\right)\Gamma\left(\frac{\Delta_1+\Delta_{3}-\Delta_{2}+J_{3}}{2}\right)}{2^{1-J_{3}}\Gamma(\Delta_1)\Gamma(\Delta_{2})\Gamma(\Delta_{3}+J_{3})}.}

In the body of the paper we have used the conventions of \cite{Costa:2014kfa,Liu:2018jhs} for the propagators and then expanded the full answer in terms of conformal partial waves. However, to determine the OPE coefficients in a theory with unit normalized operators we should rescale our propagators as $G_{\Delta}\rightarrow \mathcal{C}^{-1}_\D G_{\Delta}$ and $K_{\Delta}\rightarrow \mathcal{C}_\D^{-1}K_{\Delta}$. We will take this into account in Appendix \ref{app:checkOPE} when comparing the CFT and AdS unitarity methods for evaluating the box diagram.

\subsection{Conformal Integral Factors}
In this section we will summarize factors which appear when evaluating conformal integrals.

First recall the shadow transform and coefficients are defined by,
\begin{align}
&\int d^{d}x_{3'}\<\widetilde{\O}_{3}(x_3)\widetilde{\O}_{3}(x_{3'})\>\<\O_1(x_1)\O_2(x_2)\O_3(x_{3'})\>=S^{\Delta_1,\Delta_2}_{\Delta_3,J_3}\<\O_1(x_1)\O_2(x_2)\widetilde{\O}_3(x_3)\>,
\\
&S^{\Delta_1,\Delta_2}_{\Delta_3,J_3}=\frac{\pi^{\frac{d}{2}}\Gamma(\Delta_3-\frac{d}{2})\Gamma(\Delta_3+J_3-1)\Gamma\left(\frac{\widetilde{\Delta}_{3}+\Delta_1-\Delta_2+J_3}{2}\right)\Gamma\left(\frac{\widetilde{\Delta}_{3}+\Delta_2-\Delta_1+J_3}{2}\right)}{\Gamma(\Delta_3-1)\Gamma(\widetilde{\Delta}_{3}+J)\Gamma\left(\frac{\Delta_3+\Delta_1-\Delta_2+J_{3}}{2}\right)\Gamma\left(\frac{\Delta_3+\Delta_2-\Delta_1+J_{3}}{2}\right)}.
\end{align}
Then the $K$-factors that relate blocks to partial waves are proportional to shadow coefficients according to $K^{\Delta_1,\Delta_2}_{\Delta_3,J_3}=\left(-\frac{1}{2}\right)^{J_3}S^{\Delta_1,\Delta_2}_{\Delta_3,J_3}$.

Furthermore, the normalization of the conformal partial wave with respect to the conformally invariant inner product is:
\begin{align}
n_{\Delta,J}=\frac{K^{\Delta_3 \Delta_4}_{\widetilde{\Delta},J}K^{\widetilde{\Delta}_{3},\widetilde{\Delta}_{4}}_{\Delta,J}\text{vol}(S^{d-2})}{\text{vol}(\text{SO}(d-1))}\frac{(2J+d-2)\pi\Gamma(J+1)\Gamma(J+d-2)}{2^{d-2}\Gamma^2(J+\frac{d}{2})}.
\label{eq:CPW_Norm}
\end{align}
Here the volume factors are defined by:
\begin{align}
\text{vol}(S^{d-1})&=\frac{2\pi^{\frac{d}{2}}}{\Gamma(\frac{d}{2})}, \qquad
\frac{\text{vol}(\text{SO}(d))}{\text{vol}(\text{SO}(d-1))}=\text{vol}(S^{d-1}).
\end{align}
We also repeatedly used the CFT bubble integral for gluing three point functions\footnote{We use a different ordering for the three point function so this gives the factor of $(-1)^{J}$.} \cite{Kravchuk:2018htv}:
\begin{small}
\begin{align*}
\hspace{-.4in}\int d^{d}x_{1}d^{d}x_{2}\<\O_{1}\O_{2}\O^{a}(x)\>
\<\widetilde{\O}_{2}\widetilde{\O}_{1}\widetilde{\O}'_{b}(x')\>&=\frac{(-1)^{J_{\O}}\left(\<\O_{1}\O_{2}\O\>,\<\widetilde{\O}_{1}\widetilde{\O}_{2}\widetilde{\O}\>\right)_{E}}{\mu(\Delta,J)}\delta^{a}_{b}\delta(x-x')2\pi\delta(\nu-\nu')\delta_{JJ'}
\\ &\equiv B_{\O}\delta^{a}_{b}\delta(x-x') \delta(\nu-\nu'), \label{eq:CFTbubDef}
\numberthis
\end{align*}
\end{small}
where $a$ and $b$ are shorthand for the tensor indices. The three point pairing is defined by
\bea
\left(\<\O_{1}\O_{2}\O\>,\<\widetilde{\O}_{1}\widetilde{\O}_{2}\widetilde{\O}\>\right)_{E}=\frac{1}{2^{d}\text{vol}(\text{SO}(d-1))}\<\O_{1}(0)\O_{2}(e)\O_{3}(\infty)\>\<\widetilde{\O}_{1}(0)\widetilde{\O}_{2}(e)\widetilde{\O}_{3}(\infty)\>, \label{eq:ThreePointEuclidean}
\eea
where indices are implicitly contracted. A simple case, the scalar-scalar-spin-$J$ three-point function, is \cite{Kravchuk:2018htv}
\begin{align}
&\left(\<\f_{1}\f_{2}\O_{3,J}\>,\<\widetilde{\f}_{1}\widetilde{\f}_{2}\widetilde{\O}_{3,J}\>\right)_{E}=\frac{\widehat{C}_{J}(1)}{2^{d}\text{vol}(\text{SO}(d-1))},
\\
&\widehat{C}_{J}(\eta)=\frac{\Gamma(\frac{d-2}{2})\Gamma(J+d-2)}{2^{J}\Gamma(J+\frac{d-2}{2})\Gamma(d-2)}{}_{2}F_{1}\left(-J,J+d-2,\frac{d-1}{2},\frac{1-\eta}{2}\right).
\end{align}
Finally $\mu(\Delta,J)$ is the Plancherel measure for the symmetric traceless operators 
\begin{align}
\mu(\Delta,J)&=\frac{ \text{dim}~\rho_J~ }{2^{d}\text{vol}(\text{SO}(d))}\frac{\Gamma(\Delta-1)\Gamma(d-\Delta-1)(\Delta+J-1)(d-\Delta+J-1)}{\pi^{d}\Gamma(\Delta-\frac{d}{2})\Gamma(\frac{d}{2}-\Delta)},
\\
\text{dim}~\rho_J &= \frac{\Gamma(J+d-2)(2J+d-2)}{\Gamma(J+1)\Gamma(d-1)}. \label{eq:rhodim}
\end{align}
We will often use the bubble factor for scalars,
\e{J0bubble}{B_{\Delta,0} = {1\o 2^d \text{vol}(\text{SO}(d-1))\mu(\D_\O,0)} =  \frac{2\pi^{\frac{3d}{2}}}{\Gamma(\frac{d}{2})}{\Gamma(\D_\O-{d\o2})\Gamma({d\o2}-\D_\O)\o \Gamma(\D_\O)\Gamma(d-\D_\O)}}

\subsection{Five-Point Partial Wave}
\label{app:5ptCPW}
Recall we have defined the five-point partial waves in terms of the conformal integral:
\begin{align}
\Psi^{12345}_{6,7}(x_i)=\int d^{d}x_{6}d^{d}x_{7}\<\O_1\O_2\O_6\>\<\widetilde{\O}_{6}\O_5\widetilde{\O}_{7}\>\<\O_{7}\O_{3}\O_{4}\>, \label{eq:App5ptCPW}
\end{align}
and we want to expand it in terms of conformal blocks. We define the conformal blocks to be solutions of the conformal Casimir equation such that in the limit $x_1\rightarrow x_2$ and $x_3\rightarrow x_4$:
\begin{align}
g^{12345}_{6,7}(x_1,x_2,x_3,x_4,x_5)\approx (-2)^{J_5}(-2)^{J_6}|x_{12}|^{\Delta_{6}-\Delta_1-\Delta_2}|x_{34}|^{\Delta_7-\Delta_3-\Delta_4}\<\O_{6}(x_1)\O_{5}(x_5)\O_{7}(x_3)\>.
\end{align}
The five-point partial wave is also a solution to the conformal Casimir equation, so we must have:
\begin{align}
\Psi^{12345}_{6,7}(x_i)=R_{1}g^{12345}_{6,7}(x_i)+R_{2}g^{12345}_{\tilde{6},7}(x_i)+R_{3}g^{12345}_{6,\tilde{7}}(x_i)+R_{4}g^{12345}_{\tilde{6},\tilde{7}}(x_i), \label{eq:app5CPWtoCB}
\end{align}
where we used that the conformal Casimirs are invariant under $\Delta\rightarrow d-\Delta$\cite{DO1,DO2,DO3}.

To determine the coefficients $R_i$ we will follow the procedure of \cite{ssw} and evaluate the five-point integral in various limits. In the limit $x_1\rightarrow x_2$ the three-point function behaves as:
\begin{align}
\<\O_1(x_1)\O_2(x_2)\O_3^{\mu_1...\mu_J}(x_3)\>\approx|x_{12}|^{\Delta_{3}-J_3-\Delta_1-\Delta_2}x_{12}^{\nu_1}...x_{12}^{\nu_{J_3}}\<\O_{3,\nu_1...\nu_{J_3}}(x_1)\O^{\mu_1...\mu_{J_3}}(x_3)\>.
\end{align}

If we take the limit $x_1\rightarrow x_2$ and $x_3\rightarrow x_4$ under the integrand in (\ref{eq:App5ptCPW}) and compare with (\ref{eq:app5CPWtoCB}) we find:
\begin{align}
R_{1}=K^{5\tilde{7}}_{\tilde{6}}K^{56}_{\tilde{7}}.
\end{align}

To get the other $R_{i}$ we can use symmetry properties of the partial wave:
\begin{align}
\Psi^{12345}_{67}(x_i)&=\int d^{d}x_{6}d^{d}x_{7}\<\O_1\O_2\O_6\>\<\widetilde{\O}_{6}\O_5\widetilde{\O}_{7}\>\<\O_{7}\O_{3}\O_{4}\>
\\ &=\int d^{d}x_{6}d^{d}x_{6'}d^{d}x_{7}\<\O_1\O_2\O_6\>\<\widetilde{\O}_{6}\widetilde{\O}_{6'}\>\<\O_{6}\O_5\widetilde{\O}_{7}\>\<\O_{7}\O_{3}\O_{4}\>\frac{1}{S^{5\tilde{7}}_{6}}
\\ &=\int d^{d}x_{6}d^{d}x_{7}\<\O_1\O_2\widetilde{\O}_6\>\<\O_{6}\O_5\widetilde{\O}_{7}\>\<\O_{7}\O_{3}\O_{4}\>\frac{K^{12}_{6}}{K^{5\tilde{7}}_{6}}
\\ &=\frac{K^{12}_{6}}{K^{5\tilde{7}}_{6}}\Psi^{12345}_{\tilde{6}7}(x_i).
\end{align}

This identity implies:
\begin{align}
\Psi^{12345}_{67}(x_i)&=R_2 g^{12345}_{\tilde{6}7}(x_i)+...=K^{12}_{6}K^{5\tilde{6}}_{\tilde{7}}g^{12345}_{\tilde{6}7}(x_i)+...
\\ &\Rightarrow R_{2}=K^{12}_{6}K^{5\tilde{6}}_{\tilde{7}}.
\end{align}
To get the other coefficients we need the identities:
\begin{align}
\Psi^{12345}_{67}(x_i)&=\frac{K^{34}_{7}}{K^{5\tilde{6}}_{7}}\Psi^{12345}_{6\tilde{7}}(x_i)
 =\frac{K^{34}_{7}}{K^{5\tilde{6}}_{7}}\frac{K^{12}_{6}}{K^{57}_{6}}\Psi^{12345}_{\tilde{6}\tilde{7}}(x_i),
\end{align}
which in turn yields:
\begin{align}
R_{3}=K^{34}_{7}K^{5\tilde{7}}_{6},
\\
R_{4}=K^{34}_{7}K^{12}_{6}.
\end{align}
\section{More on the Box Diagram and 6j Symbols}
\label{app:MoreBox}
In this section we will study the AdS box diagram in more detail.

\ssec{6j Properties}
When studying the box diagram it is useful to have some identities for the 6j symbol. As reviewed in \cite{Liu:2018jhs} the 6j symbol is invariant under the tetrahedral group, $S_{4}$. This group is generated by the following transformations:
\begin{align}
\sixj{1}{2}{3}{4}{5}{6}=&\sixj{\tilde{5}}{3}{6}{\widetilde{1}}{4}{2}=\sixj{2}{5}{\widetilde{4}}{6}{1}{\widetilde{3}}=\sixj{2}{1}{4}{3}{5}{\tilde{6}}.
\end{align}

When studying the box diagram and its OPE it is also useful to have a set of identities which replaces an operator with its shadow. For example:
\begin{align}
\sixj{1}{2}{3}{4}{5}{6}&=\int\frac{d^{d}x_1...d^{d}x_6}{vol\left(SO(d+1,1)\right)}\<\widetilde{\O}_{1}\widetilde{\O}_{2}\widetilde{\O}_{5}\>\<\O_5\widetilde{\O}_{3}\widetilde{\O}_4\>\<\O_3\O_2\O_6\>\<\widetilde{\O}_{6}\O_1\O_4\>
\nonumber \\
&=\int\frac{d^{d}x_1...d^{d}x_6d^{d}x_{6'}}{vol\left(SO(d+1,1)\right)}\<\O_{1'}\widetilde{\O}_{2}\widetilde{\O}_{5}\>\<\O_5\widetilde{\O}_{3}\widetilde{\O}_4\>\<\O_3\O_2\O_{6'}\>\<\widetilde{\O}_{1'}\widetilde{\O}_{1}\>\<\widetilde{\O}_{6}\O_1\O_4\>\frac{1}{S^{\tilde{2}\tilde{5}}_{1}}
\nonumber \\
&=\int\frac{d^{d}x_1...d^{d}x_6}{vol\left(SO(d+1,1)\right)}\<\O_{1}\widetilde{\O}_{2}\widetilde{\O}_{5}\>\<\O_5\widetilde{\O}_{3}\widetilde{\O}_4\>\<\O_3\O_2\widetilde{\O}_{6}\>\<\widetilde{\O}_{6}\widetilde{\O}_1\O_4\>\frac{S^{\tilde{6}4}_{1}}{S^{\tilde{2}\tilde{5}}_{1}}
\nonumber \\ 
&=\frac{S^{\tilde{6}4}_{1}}{S^{\tilde{2}\tilde{5}}_{1}}\sixj{\tilde{1}}{2}{3}{4}{5}{6}. \label{eq:6jundoshadow}
\end{align}
Taking both sides and using (\ref{eqn:6j_Split}) also allows to derive similar identities for the inversion of a single block, (\ref{eqn:InvBlockSym}):
\begin{align}
\sixjBlock{1}{2}{3}{4}{5}{6}=\frac{S^{\tilde{1}4}_{\tilde{6}}S^{\tilde{6}4}_{1}}{S^{14}_{\tilde{6}}S^{\tilde{2}\tilde{5}}_{1}}\sixjBlock{\tilde{1}}{2}{3}{4}{5}{6}.
\end{align}
It is straightforward to use tetrahedral symmetry to derive the other identities, so we will not write them down explicitly. 

\subsection{Pentagon Identity and Conformal Diagrams}
\label{app:Pentagon}
\begin{figure}[t]
\begin{center}
\includegraphics[scale=.28]{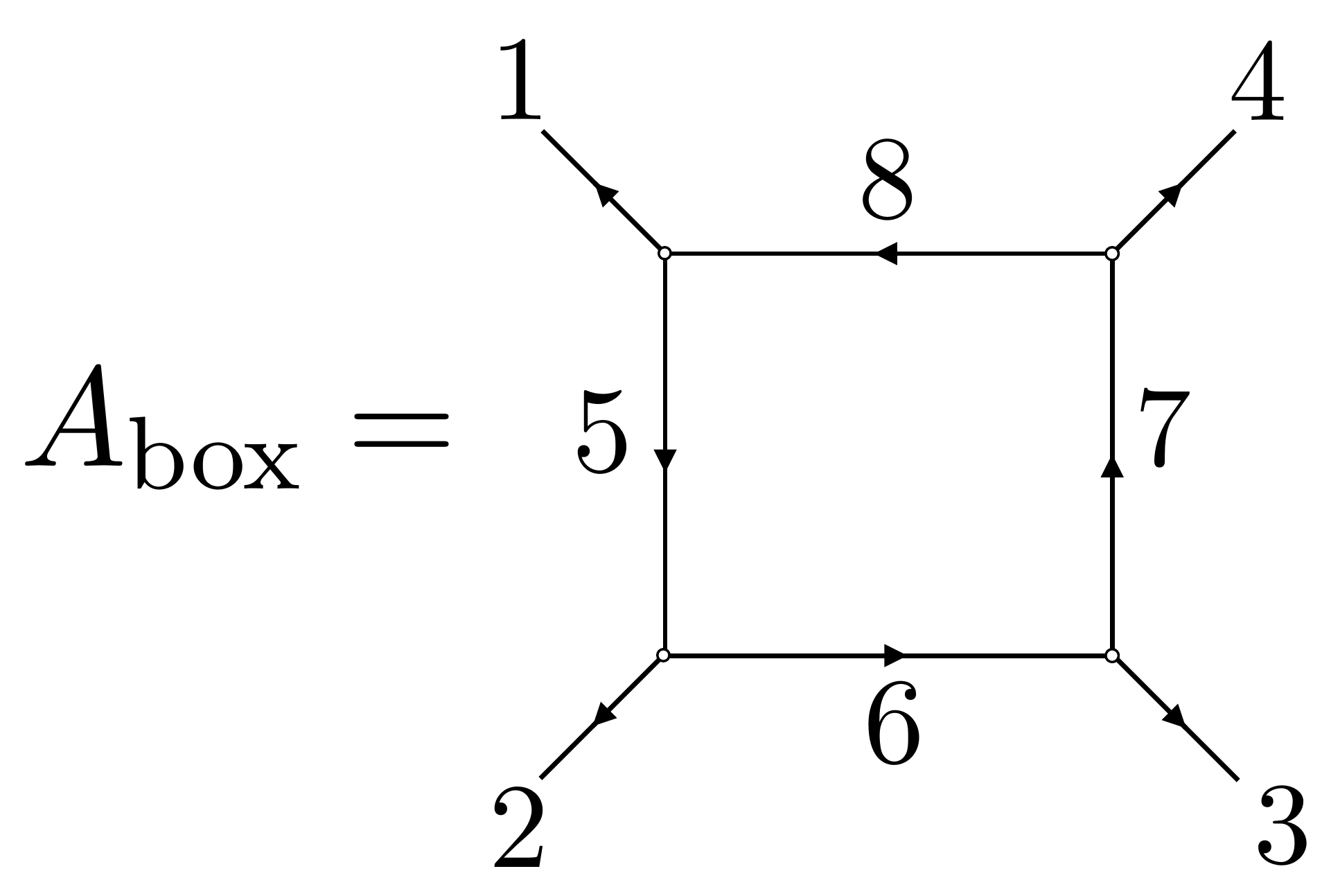}
\caption{The conformal box amplitude is four CFT three-point functions integrated over the position of internal operators. Outgoing and ingoing arrows denote operators and their shadow, respectively.}
\label{fig:CFTboxV2}
\end{center}
\end{figure}

In this section we discuss how identities associated to the AdS box, such as \eqref{eq:qdiscboxV2}, follow from the CFT pentagon identity \cite{Gadde:2017sjg,Liu:2018jhs}. To do this we will use the split representation for all bulk-to-bulk propagators for the AdS box diagram given by \eqref{AdSBoxGluing}. This reduces the AdS box diagram to a spectral integral of a CFT box graph:
\begin{align}
\mathcal{A}_{\bx}(x_i)=&\left(\int\limits_{-\infty}^{\infty}\prod\limits_{i=5}^{8}d\nu_i P(\nu_i,\Delta_i)\right)\int_{\partial \rm AdS}dx_{5,...,8}b_{1\underline{5}\tilde{\underline{8}}}b_{2\tilde{\underline{5}}\underline{6}}b_{3\tilde{\underline{6}}\underline{7}}b_{4\tilde{\underline{7}}\underline{8}}
\nonumber \\ &\<\O_1(x_1)\O_5(x_5)\widetilde{\O}_{8}(x_8)\>\<\O_2(x_2)\widetilde{\O}_{5}(x_5)\O_6(x_6)\>\<\O_3(x_3)\widetilde{\O}_{6}(x_6)\O_7(x_7)\>\<\O_4(x_4)\widetilde{\O}_{7}(x_7)\O_8(x_8)\>
\nonumber \\=&\left(\int\limits_{-\infty}^{\infty}\prod\limits_{i=5}^{8}d\nu_i P(\nu_i,\Delta_i)\right)b_{1\underline{5}\tilde{\underline{8}}}b_{2\tilde{\underline{5}}\underline{6}}b_{3\tilde{\underline{6}}\underline{7}}b_{4\tilde{\underline{7}}\underline{8}}A_{\bx}(x_i), \label{eq:AdSintermsofBox}
\end{align}
where $A_{\bx}$ is the CFT box diagram shown in figure \ref{fig:CFTboxV2} and defined by:
\begin{align}
A_{\bx}(x_i)=\int_{\partial \rm AdS}dx_{5,...,8}&\<\O_1(x_1)\O_5(x_5)\widetilde{\O}_{8}(x_8)\>\<\O_2(x_2)\widetilde{\O}_{5}(x_5)\O_6(x_6)\>
\nonumber \\ &\<\O_3(x_3)\widetilde{\O}_{6}(x_6)\O_7(x_7)\>\<\O_4(x_4)\widetilde{\O}_{7}(x_7)\O_8(x_8)\>.
\end{align}
The CFT diagrams are formed by gluing together three-point structures in a conformally-invariant way \cite{SimmonsDuffin:2012uy, Karateev:2018oml}. 

By using the 6j symbol we can write down the $s$ and $t$-channel expansions of the CFT box\cite{Liu:2018jhs}.
\begin{align}
A_{\bx}(x_i)&=\sum_{J_\O=0}^{\infty}\int\limits_{\frac{d}{2}}^{\frac{d}{2}+i\infty}\frac{d\Delta}{2\pi i} \sixj{1}{2}{6}{\tilde{8}}{\O}{\tilde{5}}\sixj{8}{\tilde{6}}{3}{4}{\O}{7}\frac{B_{\O}}{n_{\O}^{2}}\Psi^{1234}_{\O}(x_1,x_2,x_3,x_4)
\nonumber \\ &=\sum_{J_{\O'}=0}^{\infty}\int\limits_{\frac{d}{2}}^{\frac{d}{2}+i\infty}\frac{d\Delta_{\O'}}{2\pi i} \sixj{3}{2}{\tilde{5}}{7}{\O'}{6}\sixj{\tilde{7}}{5}{1}{4}{\O'}{\tilde{8}}\frac{B_{\O'}}{n_{\O'}^{2}}\Psi^{3214}_{\O'}(x_i). \label{eq:CFTPentV21}
\end{align}
This is the pentagon identity. If we plug in the $s$-channel decomposition of (\ref{eq:CFTPentV21}) into (\ref{eq:AdSintermsofBox}), split the 6j symbol using (\ref{eqn:6j_Split}), and close the $\nu_{5,7}$ contours we of course recover the correct decomposition for the AdS box (\ref{eq:BoxEqualsTreeSquared}). By taking an inner product with a partial wave, we can also write the pentagon identity purely in terms of 6j symbols
\begin{align}
\sum_{J_{\O'}=0}^{\infty}\int\limits_{\frac{d}{2}}^{\frac{d}{2}+i\infty}\frac{d\Delta_{\O'}}{2\pi i} &\sixj{3}{2}{\tilde{5}}{7}{\O'}{6}\sixj{\tilde{7}}{5}{1}{4}{\O'}{\tilde{8}}\sixj{1}{2}{3}{4}{\O}{\O'}\frac{B_{\O'}}{n_{\O'}^{2}} 
\nonumber \\ =&\sixj{1}{2}{6}{\tilde{8}}{\O}{\tilde{5}}\sixj{8}{\tilde{6}}{3}{4}{\O}{7}\frac{B_{\O}}{n_{\O}}. \label{eq:CFTPentV2}
\end{align}
This version of the pentagon identity makes it clear that spectral integrals over a 6j symbol can introduce new poles which do not appear in individual 6j symbols. The integrand in the first line of (\ref{eq:CFTPentV2}) does not have any poles in $\Delta_{\O}$ whose location depends on $\Delta_{6}$ or $\Delta_8$, while in the second line, the poles at $\Delta_{\O}=\Delta_{6}+\widetilde{\Delta}_{8}+2n$ and their shadowed versions are manifest. In the language of the conformal block decomposition, if we invert an infinite number of blocks we can find poles which are not present when inverting a single block\cite{Fitzpatrick:2015qma,dsdi}.

Crossing symmetry of the CFT box also implies crossing symmetry of the AdS box, as the AdS box is simply (\ref{eq:CFTPentV2}) dressed with spectral integrals and $b$-factors. It also follows that the identity \eqref{eq:qdiscboxV2} follows from the pentagon identity. Finally, the previous discussion of seeing new poles from inverting an infinite number of operators carries over for the AdS box.

\ssec{Comments on the Box Mellin Amplitude}\label{boxmellin}
The holographic unitarity method in \c{Aharony:2016dwx} reconstructed one-loop OPE data from tree-level OPE data. It also reconstructed the bulk one-loop Mellin amplitude, $M_{\rm 1-loop}$, from this OPE data. $M_{\rm 1-loop}$ can be written as a sum over poles at double-trace twists. Attention was restricted mostly to one-loop amplitudes involving at least one quartic vertex (i.e. those with vanishing qDisc), for which the sums $\dDisc_t( \cG_{\rm 1-loop})$ in \eqr{cftunit} truncate in spin $\ell$. 

It is helpful to understand how the ansatz for $M_{\rm 1-loop}$ in \c{Aharony:2016dwx}, given in (61) there, obscures the features of the box. On general grounds, a nonzero qDisc requires simultaneous poles in {\it both} Mellin variables $s$ and $t$ channels at double-trace twists; these terms are the Mellin transforms of $\log^2u \log^2 v$ terms that give rise to the nonzero qDisc. However, that ansatz has only simple poles in either $s$ or $t$ if we assume that the residues are analytic in the other variable. Indeed, they are not. This can be confirmed easily in the case of massless $\phi^3$ theory in AdS$_3$, for which the leading residue of $M_{\rm 1-loop}$ in a $t$-channel expansion was given in equations (70) and (72) of \c{Aharony:2016dwx}:
\e{}{M_{\rm 1-loop}(s,t) \propto {R_0(s)\o t-4}+(\text{higher poles})}
where the residue is
\e{}{R_0(s) ={25\o 12}+\sum_{\ell=2,4,\ldots}^\infty {3+2\ell\o (1+\ell)(2+\ell)}\,{}_3F_2(-\ell,3+\ell,{s\o2};2,2;1)~.}
Recall that the double-trace twist equals four because the dual operator has $\D=2$. Evaluating $R_0(4)$, one in fact finds 
\e{}{{}_3F_2(-\ell,3+\ell,2;2,2;1)={}_2F_1(-\ell,3+\ell;2;1)=1~\text{for } \ell\in2\mathbb{Z}~.}
Therefore the sum diverges logarithmically.

It would be worthwhile to compute the complete $M_{\rm 1-loop}$ for the scalar box, which will likely simplify for specific conformal and spacetime dimensions. One-loop amplitudes of KK modes of type IIB supergravity in AdS$_5\times S^5$ have only simultaneous poles \c{Alday:2018kkw, Alday:2019nin}. A natural interpretation of this is as an AdS$_5\times S^5$ avatar of the no-triangle property of one-loop ten-dimensional supergravity in Minkowski space. Knowing $M_{\rm 1-loop}$ for the pure scalar box diagram could help substantiate that hypothesis, by shedding light on the circumstances under which individual poles are absent. We leave that for the future. 

\sec{Explicit Checks of Bulk vs. Boundary Unitarity Methods}
\label{app:Checks_Bulk_Vs_Bdy}

Here we will explicitly check the match between bulk and boundary approaches to unitarity methods for two of the amplitudes considered in this paper: the box diagram $\cA_{\rm box}$ from Section \ref{sec:AdSBox}, and the vertex correction $\cA_{\rm vertex}$ from Section \ref{sec:Vertex}. 

\ssec{Box}
\label{app:checkOPE}

Our strategy is to study the contribution of the entire $[\O_6\O_8]_{n,\ell}$ family in the $\O_1\O_2 \ \rightarrow \ \O_3\O_4$ channel. Since this determines dDisc$_s$, it will be sufficient to show these contributions to the CFT correlator match the direct calculation of the AdS box.

To do this we will solve crossing for the two exchange sub-diagrams, $\mathcal{A}^{6218}_{5, \rm exch}$ and $\mathcal{A}^{3684}_{7,\rm exch}$, which make up the AdS box, $\mathcal{A}_{\bx}(x_i)= \mathcal{A}^{\underline{6}21\underline{\tilde{8}}}_{5, \rm exch}\otimes \mathcal{A}_{7,\rm exch}^{3\underline{\tilde{6}}\underline{8}4}$. By standard large $N$ scaling arguments,
\begin{align}
C_{12[68]_{n,\ell}}\sim C_{34[68]_{n,\ell}}\sim N^{-2}.
\end{align}
Solving crossing for each tree gives the OPE coefficients:
\begin{align}
C_{12[68]_{n,\ell}}C_{68[68]_{n,\ell}}=-\res_{\Delta_{O}=\Delta_{[68]_{n,\ell}}}\left(-\frac{1}{2}\right)^{-\ell}\left(\prod_{i=1,2,5,6,8}\mathcal{C}^{-1}_{\Delta_i,0}\right)(d-2\Delta_{5})\frac{b_{625}b_{\tilde{5}18}K^{18}_{\tilde{5}}K^{68}_{\widetilde{\O}}}{n_{\O}}\sixjBlock{1}{2}{6}{8}{\O}{5},
\\
C_{68[68]_{n,\ell}}C_{34[68]_{n,\ell}}=-\res_{\Delta_{\O}=\Delta_{[68]_{n,\ell}}}\left(-\frac{1}{2}\right)^{-\ell}\left(\prod_{i=3,4,6,7,8}\mathcal{C}^{-1}_{\Delta_i,0}\right)(d-2\Delta_{7})\frac{b_{367}b_{\tilde{7}84}K^{48}_{\tilde{7}}K^{34}_{\widetilde{\O}}}{n_{\O}}\sixjBlock{8}{6}{3}{4}{\O}{7},
\end{align}
where we have suppressed the bulk couplings and the factors of $\mathcal{C}_{\Delta,J}$ come from unit normalizing the operators. The OPE coefficients $C_{68[68]_{n,\ell}}$ are, to leading order in $1/N$, of order $N^0$, and are those of Mean Field Theory. These are known in closed form \cite{Fitzpatrick:2011dm,Karateev:2018oml},
\begin{align}
C_{68[68]_{n,\ell}}^{2}=-\res_{\O=[68]_{n,\ell}}\frac{2^{\ell}K^{\tilde{6}\O}_{\tilde{8}}K^{8\O}_{\tilde{6}}K^{68}_{\widetilde{[68]}_{n,\ell}}}{B_{\O}},
\end{align}
where $B_{\O}$ is the CFT bubble factor defined in (\ref{eq:CFTbubDef}). We then find the following contribution to $\<\O_1\O_2\O_3\O_4\>$:
\begin{align}
\<\O_1\O_2\O_3\O_4\>\supset&\sum\limits_{n,\ell}\left(-\frac{1}{2}\right)^{\ell}C_{12[68]_{n,\ell}}C_{34[68]_{n,\ell}}g_{[68]_{n,\ell}}(z,\bar{z})
\nonumber \\ \supset& -\sum\limits_{n,\ell}\res_{\O=[68]_{n,\ell}}(d-2\Delta_5)(d-2\Delta_7)b_{625}b_{\tilde{5}18}b_{367}b_{\tilde{7}48}\frac{K^{18}_{\tilde{5}}K^{48}_{\tilde{7}}K^{34}_{\widetilde{\O}}}{K^{\tilde{6}\O}_{\tilde{8}}K^{8\O}_{\tilde{6}}n_{\O}^{2}}B_{\O}
\nonumber \\ &\quad\left(\prod\limits_{i=1}^{8}\mathcal{C}^{-1}_{\Delta_i,0}\right)\mathcal{C}^{-1}_{\Delta_6,0}\mathcal{C}^{-1}_{\Delta_8,0}\sixjBlock{1}{2}{6}{8}{\O}{5}\sixjBlock{8}{6}{3}{4}{\O}{7}g^{1234}_{[68]_{n,\ell}}(x_i),\label{eq:finalCFTbox}
\end{align}
where the operators $\O_1,...,\O_8$ are all scalars so we can freely trade $S$ for $K$.

Next let us compare this to the AdS box,
\begin{align} 
\mathcal{A}_{\bx}(x_i)\supset& -\sum\limits_{n,\ell}\res_{\O=[68]_{n,\ell}}\prod_{i=5}^{8}(d-2\Delta_i)b_{1\tilde{5}\tilde{8}}b_{256}b_{3\tilde{6}7}b_{4\tilde{7}8}K^{1\tilde{8}}_{\tilde{5}}K^{48}_{\tilde{7}}K^{34}_{\widetilde{\O}}\frac{B_{\O}}{n_{\O}^{2}}
\nonumber \\ &\left(\prod\limits_{i=1}^{8}\mathcal{C}^{-1}_{\Delta_i,0}\right)\sixjBlock{1}{2}{6}{\tilde{8}}{\O}{5}\sixjBlock{8}{\tilde{6}}{3}{4}{\O}{7}g^{1234}_{[68]_{n,\ell}}(x_i),
\end{align}
where in comparison to (\ref{eq:BoxEqualsTreeSquared}) we used symmetry under $5\rightarrow \tilde{5}$ to make the comparison simpler. Then using identities like (\ref{eq:6jundoshadow}) we can rewrite this as:
\begin{align} 
\mathcal{A}_{\bx}(x_i)\supset -\sum\limits_{n,\ell}&\res_{\O=[68]_{n,\ell}}\prod_{i=5}^{8}(d-2\Delta_i)b_{1\tilde{5}\tilde{8}}b_{256}b_{3\tilde{6}7}b_{4\tilde{7}8}K^{1\tilde{8}}_{\tilde{5}}K^{48}_{\tilde{7}}K^{34}_{\widetilde{\O}}\frac{B_{\O}}{n_{\O}^{2}}
\nonumber \\ &\frac{K^{18}_{\tilde{5}}K^{37}_{\tilde{6}}K^{\tilde{5}1}_{\tilde{8}}}{K^{1\tilde{8}}_{\tilde{5}}K^{8\O}_{\tilde{6}}K^{\O\tilde{6}}_{\tilde{8}}}\left(\prod\limits_{i=1}^{8}\mathcal{C}^{-1}_{\Delta_i,0}\right)\sixjBlock{1}{2}{6}{8}{\O}{5}\sixjBlock{8}{6}{3}{4}{\O}{7}g^{1234}_{[68]_{n,\ell}}(x_i). \label{eq:finalAdSbox}
\end{align}
Plugging in the definitions it is straightforward to check the two expressions, (\ref{eq:finalCFTbox}) and (\ref{eq:finalAdSbox}), agree exactly. 

It should be noted in deriving this relation, although we did use the 6j symbol and the inversion of the block, we did not need their explicit form. Rather all we needed were some general identities for the 6j symbol like (\ref{eq:6jundoshadow}).

\ssec{Vertex Correction}\label{vertcorr}

We can also compare our approach for calculating loop corrections in $\f^{3}+\f^{4}$ theory with the CFT procedure used in \cite{Aharony:2016dwx}. Again, by standard large $N$ scaling arguments,
\e{}{C_{12[56]_{n,0}}\sim C_{34[56]_{n,0}}\sim N^{-2}.}
where $C_{12[56]_{n,0}}$ comes from the contact sub-diagram of $\cA_{\rm vertex}$ while $C_{34[56]_{n,0}}$ comes from the exchange sub-diagram. The contribution of the $[\O_5\O_6]_{n,0}$ family to the $s$-channel double discontinuity of $\<\O_1\O_2\O_3\O_4\>$ is
\begin{align}
\hspace{-.5in}\dDisc_{s}(\<\O_1\O_2\O_3\O_4\>)\supset \sum\limits_{n=0}^{\infty}C_{12[56]_{n,0}}C_{34[56]_{n,0}}\dDisc_{s}(g^{1234}_{[56]_{n,0}}(x_i))
\end{align}
This accounts for the contribution of the $[\O_5\O_6]_{n,0}$ family to the double discontinuity of $\mathcal{A}_{\rm vertex}$ in (\ref{vertexglue}).

From the purely boundary perspective, it appears there is no need to include the exchange of $\O_7$, which we know also contributes to $\mathcal{A}_{\rm vertex}$, see \eqref{eq:vertex7Split}. This raises a question: how should we interpret this boundary approach in terms of Witten diagrams? We claim we can always shift the couplings in the bulk dual such that $\O_{7}$ does not appear at one-loop in $\<\O_1\O_2\O_3\O_4\>$, while still maintaining crossing symmetry. In other words, the boundary approach we consider here is implicitly constructing a linear combination of $\mathcal{A}_{\rm vertex}$ and an exchange diagram.

The proof is as follows: we first note we are free to change the $\<\O_1\O_2\O_7\>$ cubic coupling while still maintaining crossing symmetry. Call the bulk cubic coupling $g_{127}$, which we take to scale as $g_{127}\sim 1/N$. We can consider the shift
\begin{align}
g_{127}\rightarrow g_{127}+\frac{1}{N^{3}}\alpha_{127},
\end{align} 
where $\alpha_{127}$ is order $N^0$. We now have a new exchange diagram for $\<\O_1\O_2\O_3\O_4\>$ which is of order $1/N^{4}$ and proportional to $\alpha_{127}g_{347}$. The $\dDisc_{s}$ of this graph only includes the exchange of $\O_{7}$, so we are free to choose $\alpha_{127}$ such that the single-trace block coming from this diagram cancels that of $\O_7$ in $\mathcal{A}_{\rm vertex}$. This choice leaves us with only double-trace states contributing to $\dDisc_{s}$. Therefore, this unique choice of $\alpha_{127}$ is being picked out by the boundary unitarity method. It would be interesting to understand in general when such shifts are allowed or if consistency conditions like unitarity or causality pose obstructions to shifting cubic couplings.

\bibliographystyle{utphys}
\bibliography{biblio}

\providecommand{\href}[2]{#2}\begingroup\raggedright\begin{thebibliography}{10}

\bibitem{Aharony:2016dwx}
O.~Aharony, L.~F. Alday, A.~Bissi, and E.~Perlmutter, ``{Loops in AdS from
  Conformal Field Theory},''
\href{http://arxiv.org/abs/1612.03891}{{\ttfamily arXiv:1612.03891 [hep-th]}}.

\bibitem{Maldacena:1997re}
J.~M. Maldacena, ``{The Large N limit of superconformal field theories and
  supergravity},'' \href{http://dx.doi.org/10.1023/A:1026654312961,
  10.4310/ATMP.1998.v2.n2.a1}{{\em Int. J. Theor. Phys.} {\bfseries 38} (1999)
  1113--1133}, \href{http://arxiv.org/abs/hep-th/9711200}{{\ttfamily
  arXiv:hep-th/9711200 [hep-th]}}.
[Adv. Theor. Math. Phys.2,231(1998)].

\bibitem{Witten:1998qj}
E.~Witten, ``{Anti-de Sitter space and holography},''
  \href{http://dx.doi.org/10.4310/ATMP.1998.v2.n2.a2}{{\em Adv. Theor. Math.
  Phys.} {\bfseries 2} (1998) 253--291},
\href{http://arxiv.org/abs/hep-th/9802150}{{\ttfamily arXiv:hep-th/9802150
  [hep-th]}}.

\bibitem{Gubser:1998bc}
S.~S. Gubser, I.~R. Klebanov, and A.~M. Polyakov, ``{Gauge theory correlators
  from noncritical string theory},''
  \href{http://dx.doi.org/10.1016/S0370-2693(98)00377-3}{{\em Phys. Lett.}
  {\bfseries B428} (1998) 105--114},
\href{http://arxiv.org/abs/hep-th/9802109}{{\ttfamily arXiv:hep-th/9802109
  [hep-th]}}.

\bibitem{Heemskerk:2009pn}
I.~Heemskerk, J.~Penedones, J.~Polchinski, and J.~Sully, ``{Holography from
  Conformal Field Theory},''
  \href{http://dx.doi.org/10.1088/1126-6708/2009/10/079}{{\em JHEP} {\bfseries
  0910} (2009) 079},
\href{http://arxiv.org/abs/0907.0151}{{\ttfamily arXiv:0907.0151 [hep-th]}}.

\bibitem{Penedones:2010ue}
J.~Penedones, ``{Writing CFT correlation functions as AdS scattering
  amplitudes},'' \href{http://dx.doi.org/10.1007/JHEP03(2011)025}{{\em JHEP}
  {\bfseries 1103} (2011) 025},
\href{http://arxiv.org/abs/1011.1485}{{\ttfamily arXiv:1011.1485 [hep-th]}}.

\bibitem{Fitzpatrick:2011dm}
A.~L. Fitzpatrick and J.~Kaplan, ``{Unitarity and the Holographic S-Matrix},''
  \href{http://dx.doi.org/10.1007/JHEP10(2012)032}{{\em JHEP} {\bfseries 1210}
  (2012) 032},
\href{http://arxiv.org/abs/1112.4845}{{\ttfamily arXiv:1112.4845 [hep-th]}}.

\bibitem{Alday:2017xua}
L.~F. Alday and A.~Bissi, ``{Loop Corrections to Supergravity on $AdS_5 \times
  S^5$},'' \href{http://dx.doi.org/10.1103/PhysRevLett.119.171601}{{\em Phys.
  Rev. Lett.} {\bfseries 119} no.~17, (2017) 171601},
\href{http://arxiv.org/abs/1706.02388}{{\ttfamily arXiv:1706.02388 [hep-th]}}.

\bibitem{Alday:2017vkk}
L.~F. Alday and S.~Caron-Huot, ``{Gravitational S-matrix from CFT dispersion
  relations},'' \href{http://dx.doi.org/10.1007/JHEP12(2018)017}{{\em JHEP}
  {\bfseries 12} (2018) 017},
\href{http://arxiv.org/abs/1711.02031}{{\ttfamily arXiv:1711.02031 [hep-th]}}.

\bibitem{Alday:2018pdi}
L.~F. Alday, A.~Bissi, and E.~Perlmutter, ``{Genus-One String Amplitudes from
  Conformal Field Theory},''
\href{http://arxiv.org/abs/1809.10670}{{\ttfamily arXiv:1809.10670 [hep-th]}}.

\bibitem{Alday:2018kkw}
L.~F. Alday, ``{On Genus-one String Amplitudes on $AdS_5 \times S^5$},''
\href{http://arxiv.org/abs/1812.11783}{{\ttfamily arXiv:1812.11783 [hep-th]}}.

\bibitem{Meltzer:2018tnm}
D.~Meltzer, ``{Higher Spin ANEC and the Space of CFTs},''
  \href{http://dx.doi.org/10.1007/JHEP07(2019)001}{{\em JHEP} {\bfseries 07}
  (2019) 001},
\href{http://arxiv.org/abs/1811.01913}{{\ttfamily arXiv:1811.01913 [hep-th]}}.

\bibitem{Ponomarev:2019ltz}
D.~Ponomarev, E.~Sezgin, and E.~Skvortsov, ``{On one loop corrections in higher
  spin gravity},'' \href{http://dx.doi.org/10.1007/JHEP11(2019)138}{{\em JHEP}
  {\bfseries 11} (2019) 138},
\href{http://arxiv.org/abs/1904.01042}{{\ttfamily arXiv:1904.01042 [hep-th]}}.

\bibitem{Shyani:2019wed}
M.~Shyani, ``{Lorentzian inversion and anomalous dimensions in Mellin space},''
\href{http://arxiv.org/abs/1908.00015}{{\ttfamily arXiv:1908.00015 [hep-th]}}.

\bibitem{Alday:2019qrf}
L.~F. Alday and E.~Perlmutter, ``{Growing Extra Dimensions in AdS/CFT},''
  \href{http://dx.doi.org/10.1007/JHEP08(2019)084}{{\em JHEP} {\bfseries 08}
  (2019) 084},
\href{http://arxiv.org/abs/1906.01477}{{\ttfamily arXiv:1906.01477 [hep-th]}}.

\bibitem{Alday:2019nin}
L.~F. Alday and X.~Zhou, ``{Simplicity of AdS Supergravity at One Loop},''
\href{http://arxiv.org/abs/1912.02663}{{\ttfamily arXiv:1912.02663 [hep-th]}}.

\bibitem{Meltzer:2019pyl}
D.~Meltzer, ``{AdS/CFT Unitarity at Higher Loops: High-Energy String
  Scattering},''
\href{http://arxiv.org/abs/1912.05580}{{\ttfamily arXiv:1912.05580 [hep-th]}}.

\bibitem{Aprile:2017bgs}
F.~Aprile, J.~M. Drummond, P.~Heslop, and H.~Paul, ``{Quantum Gravity from
  Conformal Field Theory},''
  \href{http://dx.doi.org/10.1007/JHEP01(2018)035}{{\em JHEP} {\bfseries 01}
  (2018) 035},
\href{http://arxiv.org/abs/1706.02822}{{\ttfamily arXiv:1706.02822 [hep-th]}}.

\bibitem{Aprile:2017xsp}
F.~Aprile, J.~M. Drummond, P.~Heslop, and H.~Paul, ``{Unmixing Supergravity},''
  \href{http://dx.doi.org/10.1007/JHEP02(2018)133}{{\em JHEP} {\bfseries 02}
  (2018) 133},
\href{http://arxiv.org/abs/1706.08456}{{\ttfamily arXiv:1706.08456 [hep-th]}}.

\bibitem{Aprile:2017qoy}
F.~Aprile, J.~M. Drummond, P.~Heslop, and H.~Paul, ``{Loop corrections for
  Kaluza-Klein AdS amplitudes},''
  \href{http://dx.doi.org/10.1007/JHEP05(2018)056}{{\em JHEP} {\bfseries 05}
  (2018) 056},
\href{http://arxiv.org/abs/1711.03903}{{\ttfamily arXiv:1711.03903 [hep-th]}}.

\bibitem{Giombi:2017hpr}
S.~Giombi, C.~Sleight, and M.~Taronna, ``{Spinning AdS Loop Diagrams: Two Point
  Functions},'' \href{http://dx.doi.org/10.1007/JHEP06(2018)030}{{\em JHEP}
  {\bfseries 06} (2018) 030},
\href{http://arxiv.org/abs/1708.08404}{{\ttfamily arXiv:1708.08404 [hep-th]}}.

\bibitem{Cardona:2017tsw}
C.~Cardona, ``{Mellin-(Schwinger) representation of One-loop Witten diagrams in
  AdS},''
\href{http://arxiv.org/abs/1708.06339}{{\ttfamily arXiv:1708.06339 [hep-th]}}.

\bibitem{Yuan:2017vgp}
E.~Y. Yuan, ``{Loops in the Bulk},''
\href{http://arxiv.org/abs/1710.01361}{{\ttfamily arXiv:1710.01361 [hep-th]}}.

\bibitem{Yuan:2018qva}
E.~Y. Yuan, ``{Simplicity in AdS Perturbative Dynamics},''
\href{http://arxiv.org/abs/1801.07283}{{\ttfamily arXiv:1801.07283 [hep-th]}}.

\bibitem{Bertan:2018afl}
I.~Bertan, I.~Sachs, and E.~D. Skvortsov, ``{Quantum $\phi^4$ Theory in
  AdS${}_4$ and its CFT Dual},''
  \href{http://dx.doi.org/10.1007/JHEP02(2019)099}{{\em JHEP} {\bfseries 02}
  (2019) 099},
\href{http://arxiv.org/abs/1810.00907}{{\ttfamily arXiv:1810.00907 [hep-th]}}.

\bibitem{Bertan:2018khc}
I.~Bertan and I.~Sachs, ``{Loops in Anti?de Sitter Space},''
  \href{http://dx.doi.org/10.1103/PhysRevLett.121.101601}{{\em Phys. Rev.
  Lett.} {\bfseries 121} no.~10, (2018) 101601},
\href{http://arxiv.org/abs/1804.01880}{{\ttfamily arXiv:1804.01880 [hep-th]}}.

\bibitem{Liu:2018jhs}
J.~Liu, E.~Perlmutter, V.~Rosenhaus, and D.~Simmons-Duffin, ``{$d$-dimensional
  SYK, AdS Loops, and $6j$ Symbols},''
\href{http://arxiv.org/abs/1808.00612}{{\ttfamily arXiv:1808.00612 [hep-th]}}.

\bibitem{Carmi:2018qzm}
D.~Carmi, L.~Di~Pietro, and S.~Komatsu, ``{A Study of Quantum Field Theories in
  AdS at Finite Coupling},''
  \href{http://dx.doi.org/10.1007/JHEP01(2019)200}{{\em JHEP} {\bfseries 01}
  (2019) 200},
\href{http://arxiv.org/abs/1810.04185}{{\ttfamily arXiv:1810.04185 [hep-th]}}.

\bibitem{Aprile:2018efk}
F.~Aprile, J.~Drummond, P.~Heslop, and H.~Paul, ``{Double-trace spectrum of
  $N=4$ supersymmetric Yang-Mills theory at strong coupling},''
  \href{http://dx.doi.org/10.1103/PhysRevD.98.126008}{{\em Phys. Rev.}
  {\bfseries D98} no.~12, (2018) 126008},
\href{http://arxiv.org/abs/1802.06889}{{\ttfamily arXiv:1802.06889 [hep-th]}}.

\bibitem{Ghosh:2018bgd}
K.~Ghosh, ``{Polyakov-Mellin Bootstrap for AdS loops},''
\href{http://arxiv.org/abs/1811.00504}{{\ttfamily arXiv:1811.00504 [hep-th]}}.

\bibitem{Mazac:2018ycv}
D.~Mazac and M.~F. Paulos, ``{The analytic functional bootstrap. Part II.
  Natural bases for the crossing equation},''
  \href{http://dx.doi.org/10.1007/JHEP02(2019)163}{{\em JHEP} {\bfseries 02}
  (2019) 163},
\href{http://arxiv.org/abs/1811.10646}{{\ttfamily arXiv:1811.10646 [hep-th]}}.

\bibitem{Beccaria:2019stp}
M.~Beccaria and A.~A. Tseytlin, ``{On boundary correlators in Liouville theory
  on AdS$_{2}$},'' \href{http://dx.doi.org/10.1007/JHEP07(2019)008}{{\em JHEP}
  {\bfseries 07} (2019) 008},
\href{http://arxiv.org/abs/1904.12753}{{\ttfamily arXiv:1904.12753 [hep-th]}}.

\bibitem{Chester:2019pvm}
S.~M. Chester, ``{Genus-2 Holographic Correlator on $AdS_5 \times S^5$ from
  Localization},''
\href{http://arxiv.org/abs/1908.05247}{{\ttfamily arXiv:1908.05247 [hep-th]}}.

\bibitem{Beccaria:2019dju}
M.~Beccaria, H.~Jiang, and A.~A. Tseytlin, ``{Supersymmetric Liouville theory
  in AdS$_{2}$ and AdS/CFT},''
  \href{http://dx.doi.org/10.1007/JHEP11(2019)051}{{\em JHEP} {\bfseries 11}
  (2019) 051},
\href{http://arxiv.org/abs/1909.10255}{{\ttfamily arXiv:1909.10255 [hep-th]}}.

\bibitem{Carmi:2019ocp}
D.~Carmi, ``{Loops in AdS: From the Spectral Representation to Position
  Space},''
\href{http://arxiv.org/abs/1910.14340}{{\ttfamily arXiv:1910.14340 [hep-th]}}.

\bibitem{Aprile:2019rep}
F.~Aprile, J.~Drummond, P.~Heslop, and H.~Paul, ``{One-loop amplitudes in
  $AdS_5\times S^5$ supergravity from $\mathcal{N}=4$ SYM at strong
  coupling},''
\href{http://arxiv.org/abs/1912.01047}{{\ttfamily arXiv:1912.01047 [hep-th]}}.

\bibitem{Drummond:2019hel}
J.~M. Drummond and H.~Paul, ``{One-loop string corrections to AdS amplitudes
  from CFT},''
\href{http://arxiv.org/abs/1912.07632}{{\ttfamily arXiv:1912.07632 [hep-th]}}.

\bibitem{Cornalba:2007zb}
L.~Cornalba, M.~S. Costa, and J.~Penedones, ``{Eikonal approximation in
  AdS/CFT: Resumming the gravitational loop expansion},''
  \href{http://dx.doi.org/10.1088/1126-6708/2007/09/037}{{\em JHEP} {\bfseries
  09} (2007) 037},
\href{http://arxiv.org/abs/0707.0120}{{\ttfamily arXiv:0707.0120 [hep-th]}}.

\bibitem{Fitzpatrick:2011hu}
A.~L. Fitzpatrick and J.~Kaplan, ``{Analyticity and the Holographic
  S-Matrix},'' \href{http://dx.doi.org/10.1007/JHEP10(2012)127}{{\em JHEP}
  {\bfseries 1210} (2012) 127},
\href{http://arxiv.org/abs/1111.6972}{{\ttfamily arXiv:1111.6972 [hep-th]}}.

\bibitem{Caron-Huot:2017vep}
S.~Caron-Huot, ``{Analyticity in Spin in Conformal Theories},''
\href{http://arxiv.org/abs/1703.00278}{{\ttfamily arXiv:1703.00278 [hep-th]}}.

\bibitem{Gadde:2017sjg}
A.~Gadde, ``{In search of conformal theories},''
\href{http://arxiv.org/abs/1702.07362}{{\ttfamily arXiv:1702.07362 [hep-th]}}.

\bibitem{Kravchuk:2018htv}
P.~Kravchuk and D.~Simmons-Duffin, ``{Light-ray operators in conformal field
  theory},''
\href{http://arxiv.org/abs/1805.00098}{{\ttfamily arXiv:1805.00098 [hep-th]}}.

\bibitem{Karateev:2018oml}
D.~Karateev, P.~Kravchuk, and D.~Simmons-Duffin, ``{Harmonic Analysis and Mean
  Field Theory},''
\href{http://arxiv.org/abs/1809.05111}{{\ttfamily arXiv:1809.05111 [hep-th]}}.

\bibitem{SimmonsDuffin:2012uy}
D.~Simmons-Duffin, ``{Projectors, Shadows, and Conformal Blocks},''
  \href{http://dx.doi.org/10.1007/JHEP04(2014)146}{{\em JHEP} {\bfseries 1404}
  (2014) 146},
\href{http://arxiv.org/abs/1204.3894}{{\ttfamily arXiv:1204.3894 [hep-th]}}.

\bibitem{Mack:2009mi}
G.~Mack, ``{D-independent representation of Conformal Field Theories in D
  dimensions via transformation to auxiliary Dual Resonance Models. Scalar
  amplitudes},''
\href{http://arxiv.org/abs/0907.2407}{{\ttfamily arXiv:0907.2407 [hep-th]}}.

\bibitem{Carmi:2019cub}
D.~Carmi and S.~Caron-Huot, ``{A Conformal Dispersion Relation: Correlations
  from Absorption},''
\href{http://arxiv.org/abs/1910.12123}{{\ttfamily arXiv:1910.12123 [hep-th]}}.

\bibitem{Karateev:2017jgd}
D.~Karateev, P.~Kravchuk, and D.~Simmons-Duffin, ``{Weight Shifting Operators
  and Conformal Blocks},''
\href{http://arxiv.org/abs/1706.07813}{{\ttfamily arXiv:1706.07813 [hep-th]}}.

\bibitem{Costa:2018mcg}
M.~S. Costa and T.~Hansen, ``{AdS Weight Shifting Operators},''
  \href{http://dx.doi.org/10.1007/JHEP09(2018)040}{{\em JHEP} {\bfseries 09}
  (2018) 040},
\href{http://arxiv.org/abs/1805.01492}{{\ttfamily arXiv:1805.01492 [hep-th]}}.

\bibitem{Costa:2014kfa}
M.~S. Costa, V.~Goncalves, and J.~Penedones, ``{Spinning AdS Propagators},''
  \href{http://dx.doi.org/10.1007/JHEP09(2014)064}{{\em JHEP} {\bfseries 09}
  (2014) 064},
\href{http://arxiv.org/abs/1404.5625}{{\ttfamily arXiv:1404.5625 [hep-th]}}.

\bibitem{Bekaert:2014cea}
X.~Bekaert, J.~Erdmenger, D.~Ponomarev, and C.~Sleight, ``{Towards holographic
  higher-spin interactions: Four-point functions and higher-spin exchange},''
  \href{http://dx.doi.org/10.1007/JHEP03(2015)170}{{\em JHEP} {\bfseries 03}
  (2015) 170},
\href{http://arxiv.org/abs/1412.0016}{{\ttfamily arXiv:1412.0016 [hep-th]}}.

\bibitem{Bekaert:2015tva}
X.~Bekaert, J.~Erdmenger, D.~Ponomarev, and C.~Sleight, ``{Quartic AdS
  Interactions in Higher-Spin Gravity from Conformal Field Theory},''
  \href{http://dx.doi.org/10.1007/JHEP11(2015)149}{{\em JHEP} {\bfseries 11}
  (2015) 149},
\href{http://arxiv.org/abs/1508.04292}{{\ttfamily arXiv:1508.04292 [hep-th]}}.

\bibitem{Sleight:2017fpc}
C.~Sleight and M.~Taronna, ``{Spinning Witten Diagrams},''
  \href{http://dx.doi.org/10.1007/JHEP06(2017)100}{{\em JHEP} {\bfseries 06}
  (2017) 100},
\href{http://arxiv.org/abs/1702.08619}{{\ttfamily arXiv:1702.08619 [hep-th]}}.

\bibitem{Ponomarev:2019ofr}
D.~Ponomarev, ``{From bulk loops to boundary large-N expansion},''
\href{http://arxiv.org/abs/1908.03974}{{\ttfamily arXiv:1908.03974 [hep-th]}}.

\bibitem{ssw}
D.~Simmons-Duffin, D.~Stanford, and E.~Witten, ``{A spacetime derivation of the
  Lorentzian OPE inversion formula},''
\href{http://arxiv.org/abs/1711.03816}{{\ttfamily arXiv:1711.03816 [hep-th]}}.

\bibitem{Costa:2012cb}
M.~S. Costa, V.~Goncalves, and J.~Penedones, ``{Conformal Regge theory},''
  \href{http://dx.doi.org/10.1007/JHEP12(2012)091}{{\em JHEP} {\bfseries 1212}
  (2012) 091},
\href{http://arxiv.org/abs/1209.4355}{{\ttfamily arXiv:1209.4355 [hep-th]}}.

\bibitem{Czech:2016xec}
B.~Czech, L.~Lamprou, S.~McCandlish, B.~Mosk, and J.~Sully, ``{A Stereoscopic
  Look into the Bulk},'' \href{http://dx.doi.org/10.1007/JHEP07(2016)129}{{\em
  JHEP} {\bfseries 07} (2016) 129},
\href{http://arxiv.org/abs/1604.03110}{{\ttfamily arXiv:1604.03110 [hep-th]}}.

\bibitem{Hogervorst:2017sfd}
M.~Hogervorst and B.~C. van Rees, ``{Crossing symmetry in alpha space},''
  \href{http://dx.doi.org/10.1007/JHEP11(2017)193}{{\em JHEP} {\bfseries 11}
  (2017) 193},
\href{http://arxiv.org/abs/1702.08471}{{\ttfamily arXiv:1702.08471 [hep-th]}}.

\bibitem{Sleight:2018ryu}
C.~Sleight and M.~Taronna, ``{Anomalous Dimensions from Crossing Kernels},''
  \href{http://dx.doi.org/10.1007/JHEP11(2018)089}{{\em JHEP} {\bfseries 11}
  (2018) 089},
\href{http://arxiv.org/abs/1807.05941}{{\ttfamily arXiv:1807.05941 [hep-th]}}.

\bibitem{Gopakumar:2018xqi}
R.~Gopakumar and A.~Sinha, ``{On the Polyakov-Mellin bootstrap},''
  \href{http://dx.doi.org/10.1007/JHEP12(2018)040}{{\em JHEP} {\bfseries 12}
  (2018) 040},
\href{http://arxiv.org/abs/1809.10975}{{\ttfamily arXiv:1809.10975 [hep-th]}}.

\bibitem{Dobrev:1977qv}
V.~K. Dobrev, G.~Mack, V.~B. Petkova, S.~G. Petrova, and I.~T. Todorov,
  ``{Harmonic Analysis on the n-Dimensional Lorentz Group and Its Application
  to Conformal Quantum Field Theory},''
\href{http://dx.doi.org/10.1007/BFb0009678}{{\em Lect. Notes Phys.} {\bfseries
  63} (1977) 1--280}.

\bibitem{Zhou:2018sfz}
X.~Zhou, ``{Recursion Relations in Witten Diagrams and Conformal Partial
  Waves},'' \href{http://dx.doi.org/10.1007/JHEP05(2019)006}{{\em JHEP}
  {\bfseries 05} (2019) 006},
\href{http://arxiv.org/abs/1812.01006}{{\ttfamily arXiv:1812.01006 [hep-th]}}.

\bibitem{Freedman:1998tz}
D.~Z. Freedman, S.~D. Mathur, A.~Matusis, and L.~Rastelli, ``{Correlation
  functions in the CFT(d) / AdS(d+1) correspondence},''
  \href{http://dx.doi.org/10.1016/S0550-3213(99)00053-X}{{\em Nucl. Phys.}
  {\bfseries B546} (1999) 96--118},
\href{http://arxiv.org/abs/hep-th/9804058}{{\ttfamily arXiv:hep-th/9804058
  [hep-th]}}.

\bibitem{Freedman:1998bj}
D.~Z. Freedman, S.~D. Mathur, A.~Matusis, and L.~Rastelli, ``{Comments on 4
  point functions in the CFT / AdS correspondence},''
  \href{http://dx.doi.org/10.1016/S0370-2693(99)00229-4}{{\em Phys. Lett.}
  {\bfseries B452} (1999) 61--68},
\href{http://arxiv.org/abs/hep-th/9808006}{{\ttfamily arXiv:hep-th/9808006
  [hep-th]}}.

\bibitem{Heemskerk:2010ty}
I.~Heemskerk and J.~Sully, ``{More Holography from Conformal Field Theory},''
  \href{http://dx.doi.org/10.1007/JHEP09(2010)099}{{\em JHEP} {\bfseries 1009}
  (2010) 099},
\href{http://arxiv.org/abs/1006.0976}{{\ttfamily arXiv:1006.0976 [hep-th]}}.

\bibitem{Fitzpatrick:2012yx}
A.~L. Fitzpatrick, J.~Kaplan, D.~Poland, and D.~Simmons-Duffin, ``{The Analytic
  Bootstrap and AdS Superhorizon Locality},''
  \href{http://dx.doi.org/10.1007/JHEP12(2013)004}{{\em JHEP} {\bfseries 1312}
  (2013) 004},
\href{http://arxiv.org/abs/1212.3616}{{\ttfamily arXiv:1212.3616 [hep-th]}}.

\bibitem{Komargodski:2012ek}
Z.~Komargodski and A.~Zhiboedov, ``{Convexity and Liberation at Large Spin},''
  \href{http://dx.doi.org/10.1007/JHEP11(2013)140}{{\em JHEP} {\bfseries 1311}
  (2013) 140},
\href{http://arxiv.org/abs/1212.4103}{{\ttfamily arXiv:1212.4103 [hep-th]}}.

\bibitem{dsdi}
D.~Simmons-Duffin, ``{The Lightcone Bootstrap and the Spectrum of the 3d Ising
  CFT},'' \href{http://dx.doi.org/10.1007/JHEP03(2017)086}{{\em JHEP}
  {\bfseries 03} (2017) 086},
\href{http://arxiv.org/abs/1612.08471}{{\ttfamily arXiv:1612.08471 [hep-th]}}.

\bibitem{EllisPriv}
E.~Yuan, ``{Private Communication},''.

\bibitem{Fitzpatrick:2010zm}
A.~Fitzpatrick, E.~Katz, D.~Poland, and D.~Simmons-Duffin, ``{Effective
  Conformal Theory and the Flat-Space Limit of AdS},''
  \href{http://dx.doi.org/10.1007/JHEP07(2011)023}{{\em JHEP} {\bfseries 1107}
  (2011) 023},
\href{http://arxiv.org/abs/1007.2412}{{\ttfamily arXiv:1007.2412 [hep-th]}}.

\bibitem{Alday:2017gde}
L.~F. Alday, A.~Bissi, and E.~Perlmutter, ``{Holographic Reconstruction of AdS
  Exchanges from Crossing Symmetry},''
\href{http://arxiv.org/abs/1705.02318}{{\ttfamily arXiv:1705.02318 [hep-th]}}.

\bibitem{Rosenhaus:2018zqn}
V.~Rosenhaus, ``{Multipoint Conformal Blocks in the Comb Channel},''
\href{http://arxiv.org/abs/1810.03244}{{\ttfamily arXiv:1810.03244 [hep-th]}}.

\bibitem{Parikh:2019ygo}
S.~Parikh, ``{Holographic dual of the five-point conformal block},''
  \href{http://dx.doi.org/10.1007/JHEP05(2019)051}{{\em JHEP} {\bfseries 05}
  (2019) 051},
\href{http://arxiv.org/abs/1901.01267}{{\ttfamily arXiv:1901.01267 [hep-th]}}.

\bibitem{Jepsen:2019svc}
C.~B. Jepsen and S.~Parikh, ``{Propagator identities, holographic conformal
  blocks, and higher-point AdS diagrams},''
\href{http://arxiv.org/abs/1906.08405}{{\ttfamily arXiv:1906.08405 [hep-th]}}.

\bibitem{Parikh:2019dvm}
S.~Parikh, ``{A multipoint conformal block chain in $d$ dimensions},''
\href{http://arxiv.org/abs/1911.09190}{{\ttfamily arXiv:1911.09190 [hep-th]}}.

\bibitem{Fortin:2019zkm}
J.-F. Fortin, W.~Ma, and W.~Skiba, ``{Higher-Point Conformal Blocks in the Comb
  Channel},''
\href{http://arxiv.org/abs/1911.11046}{{\ttfamily arXiv:1911.11046 [hep-th]}}.

\bibitem{Hijano:2015zsa}
E.~Hijano, P.~Kraus, E.~Perlmutter, and R.~Snively, ``{Witten Diagrams
  Revisited: The AdS Geometry of Conformal Blocks},''
  \href{http://dx.doi.org/10.1007/JHEP01(2016)146}{{\em JHEP} {\bfseries 01}
  (2016) 146},
\href{http://arxiv.org/abs/1508.00501}{{\ttfamily arXiv:1508.00501 [hep-th]}}.

\bibitem{Goncalves:2019znr}
V.~Goncalves, R.~Pereira, and X.~Zhou, ``{$20'$ Five-Point Function from
  $AdS_5\times S^5$ Supergravity},''
\href{http://arxiv.org/abs/1906.05305}{{\ttfamily arXiv:1906.05305 [hep-th]}}.

\bibitem{Camanho:2014apa}
X.~O. Camanho, J.~D. Edelstein, J.~Maldacena, and A.~Zhiboedov, ``{Causality
  Constraints on Corrections to the Graviton Three-Point Coupling},''
  \href{http://dx.doi.org/10.1007/JHEP02(2016)020}{{\em JHEP} {\bfseries 02}
  (2016) 020},
\href{http://arxiv.org/abs/1407.5597}{{\ttfamily arXiv:1407.5597 [hep-th]}}.

\bibitem{Maldacena:2015iua}
J.~Maldacena, D.~Simmons-Duffin, and A.~Zhiboedov, ``{Looking for a bulk
  point},''
\href{http://arxiv.org/abs/1509.03612}{{\ttfamily arXiv:1509.03612 [hep-th]}}.

\bibitem{Hartman:2015lfa}
T.~Hartman, S.~Jain, and S.~Kundu, ``{Causality Constraints in Conformal Field
  Theory},'' \href{http://dx.doi.org/10.1007/JHEP05(2016)099}{{\em JHEP}
  {\bfseries 05} (2016) 099},
\href{http://arxiv.org/abs/1509.00014}{{\ttfamily arXiv:1509.00014 [hep-th]}}.

\bibitem{Afkhami-Jeddi:2016ntf}
N.~Afkhami-Jeddi, T.~Hartman, S.~Kundu, and A.~Tajdini, ``{Einstein gravity
  3-point functions from conformal field theory},''
\href{http://arxiv.org/abs/1610.09378}{{\ttfamily arXiv:1610.09378 [hep-th]}}.

\bibitem{Afkhami-Jeddi:2017rmx}
N.~Afkhami-Jeddi, T.~Hartman, S.~Kundu, and A.~Tajdini, ``{Shockwaves from the
  Operator Product Expansion},''
\href{http://arxiv.org/abs/1709.03597}{{\ttfamily arXiv:1709.03597 [hep-th]}}.

\bibitem{Afkhami-Jeddi:2018own}
N.~Afkhami-Jeddi, S.~Kundu, and A.~Tajdini, ``{A Conformal Collider for
  Holographic CFTs},''
\href{http://arxiv.org/abs/1805.07393}{{\ttfamily arXiv:1805.07393 [hep-th]}}.

\bibitem{KPZ2017}
M.~Kulaxizi, A.~Parnachev, and A.~Zhiboedov, ``{Bulk Phase Shift, CFT Regge
  Limit and Einstein Gravity},''
\href{http://arxiv.org/abs/1705.02934}{{\ttfamily arXiv:1705.02934 [hep-th]}}.

\bibitem{Costa:2017twz}
M.~S. Costa, T.~Hansen, and J.~Penedones, ``{Bounds for OPE coefficients on the
  Regge trajectory},''
\href{http://arxiv.org/abs/1707.07689}{{\ttfamily arXiv:1707.07689 [hep-th]}}.

\bibitem{Meltzer:2017rtf}
D.~Meltzer and E.~Perlmutter, ``{Beyond $a = c$: gravitational couplings to
  matter and the stress tensor OPE},''
  \href{http://dx.doi.org/10.1007/JHEP07(2018)157}{{\em JHEP} {\bfseries 07}
  (2018) 157},
\href{http://arxiv.org/abs/1712.04861}{{\ttfamily arXiv:1712.04861 [hep-th]}}.

\bibitem{Kologlu:2019bco}
M.~Kologlu, P.~Kravchuk, D.~Simmons-Duffin, and A.~Zhiboedov, ``{Shocks,
  Superconvergence, and a Stringy Equivalence Principle},''
\href{http://arxiv.org/abs/1904.05905}{{\ttfamily arXiv:1904.05905 [hep-th]}}.

\bibitem{Alday:2014tsa}
L.~F. Alday, A.~Bissi, and T.~Lukowski, ``{Lessons from crossing symmetry at
  large N},'' \href{http://dx.doi.org/10.1007/JHEP06(2015)074}{{\em JHEP}
  {\bfseries 06} (2015) 074},
\href{http://arxiv.org/abs/1410.4717}{{\ttfamily arXiv:1410.4717 [hep-th]}}.

\bibitem{Alday:2016htq}
L.~F. Alday and A.~Bissi, ``{Unitarity and positivity constraints for CFT at
  large central charge},''
\href{http://arxiv.org/abs/1606.09593}{{\ttfamily arXiv:1606.09593 [hep-th]}}.

\bibitem{Cornalba:2007fs}
L.~Cornalba, ``{Eikonal methods in AdS/CFT: Regge theory and multi-reggeon
  exchange},''
\href{http://arxiv.org/abs/0710.5480}{{\ttfamily arXiv:0710.5480 [hep-th]}}.

\bibitem{Cornalba:2009ax}
L.~Cornalba, M.~S. Costa, and J.~Penedones, ``{Deep Inelastic Scattering in
  Conformal QCD},'' \href{http://dx.doi.org/10.1007/JHEP03(2010)133}{{\em JHEP}
  {\bfseries 03} (2010) 133},
\href{http://arxiv.org/abs/0911.0043}{{\ttfamily arXiv:0911.0043 [hep-th]}}.

\bibitem{Li:2017lmh}
D.~Li, D.~Meltzer, and D.~Poland, ``{Conformal Bootstrap in the Regge Limit},''
\href{http://arxiv.org/abs/1705.03453}{{\ttfamily arXiv:1705.03453 [hep-th]}}.

\bibitem{Brower:2006ea}
R.~C. Brower, J.~Polchinski, M.~J. Strassler, and C.-I. Tan, ``{The Pomeron and
  gauge/string duality},''
  \href{http://dx.doi.org/10.1088/1126-6708/2007/12/005}{{\em JHEP} {\bfseries
  12} (2007) 005},
\href{http://arxiv.org/abs/hep-th/0603115}{{\ttfamily arXiv:hep-th/0603115
  [hep-th]}}.

\bibitem{Brower:2007xg}
R.~C. Brower, M.~J. Strassler, and C.-I. Tan, ``{On The Pomeron at Large 't
  Hooft Coupling},''
  \href{http://dx.doi.org/10.1088/1126-6708/2009/03/092}{{\em JHEP} {\bfseries
  03} (2009) 092},
\href{http://arxiv.org/abs/0710.4378}{{\ttfamily arXiv:0710.4378 [hep-th]}}.

\bibitem{Shenker:2014cwa}
S.~H. Shenker and D.~Stanford, ``{Stringy effects in scrambling},''
  \href{http://dx.doi.org/10.1007/JHEP05(2015)132}{{\em JHEP} {\bfseries 05}
  (2015) 132},
\href{http://arxiv.org/abs/1412.6087}{{\ttfamily arXiv:1412.6087 [hep-th]}}.

\bibitem{DO1}
F.~Dolan and H.~Osborn, ``{Conformal four point functions and the operator
  product expansion},''
  \href{http://dx.doi.org/10.1016/S0550-3213(01)00013-X}{{\em Nucl.Phys.}
  {\bfseries B599} (2001) 459--496},
\href{http://arxiv.org/abs/hep-th/0011040}{{\ttfamily arXiv:hep-th/0011040
  [hep-th]}}.

\bibitem{DO2}
F.~Dolan and H.~Osborn, ``{Conformal partial waves and the operator product
  expansion},'' \href{http://dx.doi.org/10.1016/j.nuclphysb.2003.11.016}{{\em
  Nucl.Phys.} {\bfseries B678} (2004) 491--507},
\href{http://arxiv.org/abs/hep-th/0309180}{{\ttfamily arXiv:hep-th/0309180
  [hep-th]}}.

\bibitem{DO3}
F.~Dolan and H.~Osborn, ``{Conformal Partial Waves: Further Mathematical
  Results},''
\href{http://arxiv.org/abs/1108.6194v2}{{\ttfamily arXiv:1108.6194v2
  [hep-th]}}.

\bibitem{Fitzpatrick:2015qma}
A.~L. Fitzpatrick, J.~Kaplan, M.~T. Walters, and J.~Wang, ``{Eikonalization of
  Conformal Blocks},'' \href{http://dx.doi.org/10.1007/JHEP09(2015)019}{{\em
  JHEP} {\bfseries 09} (2015) 019},
\href{http://arxiv.org/abs/1504.01737}{{\ttfamily arXiv:1504.01737 [hep-th]}}.

\end{thebibliography}\endgroup

\end{document}